%% file: hpdc17.tex
\documentclass[sigconf]{acmart}
\usepackage{listings}
\usepackage{gnuplot-lua-tikz}
\usepackage{subcaption}
\usepackage{algorithm}
\usepackage{algpseudocode}
\usepackage{enumitem}
\setcopyright{rightsretained}

\definecolor{lbcolor}{rgb}{0.9,0.9,0.9}

\newlength{\saveparindent}
\newlength{\saveparskip}
\algblock{ParFor}{EndParFor}
\algnewcommand\algorithmicparfor{\textbf{parfor}}
\algnewcommand\algorithmicpardo{\textbf{do}}
\algnewcommand\algorithmicendparfor{\textbf{end\ parfor}}
\algrenewtext{ParFor}[1]{\algorithmicparfor\ #1\ \algorithmicpardo}
\algrenewtext{EndParFor}{\algorithmicendparfor}
\algrenewcommand\algorithmicindent{1.0em}
\let\oldReturn\Return
\renewcommand{\Return}{\State\oldReturn}

\begin{document}

\copyrightyear{2017}
\acmYear{2017}
\setcopyright{acmlicensed}
\acmConference{HPDC '17}{June 26-30, 2017}{Washington , DC, USA}\acmPrice{15.00}\acmDOI{http://dx.doi.org/10.1145/3078597.3078607}
\acmISBN{978-1-4503-4699-3/17/06}

\title{knor: A NUMA-Optimized In-Memory, Distributed and \\ Semi-External-Memory k-means Library}


\author{Disa Mhembere}
\affiliation{
  \institution{Dept. of Computer Science, \\Johns Hopkins University}
  \city{Baltimore}
  \state{Maryland}
}

\author{Da Zheng}
\affiliation{
  \institution{Dept. of Computer Science, \\Johns Hopkins University}
  \city{Baltimore}
  \state{Maryland}
}

\author{Carey E. Priebe}
\affiliation{
  \institution{Dept. of Applied Math and Statistics, \\Johns Hopkins University}
  \city{Baltimore}
  \state{Maryland}
}

\author{Joshua T. Vogelstein}
\affiliation{
  \institution{Institute for Computational Medicine,
  Dept. of Biomedical Engineering, \\Johns Hopkins University}
  \city{Baltimore}
  \state{Maryland}
}

\author{Randal Burns}
\affiliation{
  \institution{Dept. of Computer Science, \\Johns Hopkins University}
  \city{Baltimore}
  \state{Maryland}
}

%


\begin{abstract}
k-means is one of the most influential and utilized
machine learning algorithms. Its computation limits the
performance and scalability of many statistical analysis and
machine learning tasks.
We rethink and optimize k-means in terms of modern NUMA architectures
to develop a novel parallelization scheme that delays and minimizes
synchronization barriers.
The \textit{k-means NUMA Optimized Routine} (\textsf{knor}) library has
(i) in-memory (\textsf{knori}), (ii) distributed memory (\textsf{knord}), and
(iii) semi-external memory (\textsf{knors})
modules that radically improve the performance
of k-means for varying memory and hardware budgets.
\textsf{knori} boosts performance for single machine datasets by an
order of magnitude or more. \textsf{knors} improves the scalability of k-means on a memory
budget using SSDs.
\textsf{knors} scales to billions of points on a single machine, using a fraction
of the resources that distributed in-memory systems require.
\textsf{knord} retains \textsf{knori}'s performance characteristics, while scaling in-memory
through distributed computation in the cloud. \textsf{knor} modifies Elkan's triangle inequality
pruning algorithm such that we utilize it on billion-point datasets without the
significant memory overhead of the original algorithm.
We demonstrate \textsf{knor}
outperforms distributed commercial products like H$_2$O, Turi (formerly Dato, GraphLab)
and Spark's MLlib by more than an order of magnitude for datasets of
$10^7$ to $10^9$ points.
\end{abstract}


\maketitle

\input{intro}

\input{relwork}

\input{theory}

\input{im-design}
\input{sem-design}
\input{dist-design}

\input{eval}

\input{future_discuss}

\input{conclusion}

\section{Acknowledgments}
This work is partially supported by DARPA GRAPHS N66001-14-1-4028
and DARPA SIMPLEX program through SPAWAR contract N66001-15-C-4041.

\bibliographystyle{abbrv}
\bibliography{hpdc17}

\end{document}

%% file: intro.tex
\section{Introduction}

Clustering data to maximize within-cluster similarity
and cross-cluster variance is highly desirable for the analysis
of big data. K-means is an intuitive and
highly popular method of clustering $n$ points in
$d$-dimensions into $k$ clusters.
A very popular synchronous variant of k-means
is Lloyd's algorithm \cite{lloyds}. Similar to Expectation
Maximization \cite{EM}, Lloyd's algorithm proceeds in
two phases. In phase one, we compute the distance from each
data point to each centroid (cluster mean). In phase two, we update the centroids
to be the mean of their membership. This proceeds until the centroids
no longer change from one iteration to the next. The algorithm locally
minimizes within-cluster \textit{distance}, for some distance metric
that often is Euclidean distance (Section \ref{algo}).
Despite k-means' popularity, state-of-the-art machine learning
libraries \cite{mllib, graphlab, h2o} experience many challenges scaling
performance well with respect to growing data sets. Furthermore,
these libraries place an emphasis on scaling-out computation
to the distributed setting, neglecting to fully utilize the resources
within each machine.

The decomposition of extremely large datasets into clusters of data points
that are similar is a topic of
great interest in industry and academia. For example, clustering
is the backbone upon which popular user recommendation systems at Netflix
\cite{netflixprize} are built. Furthermore, partitioning multi-billion
data points is essential
to targeted ad-driven organizations such as Google \cite{cfgoogle} and Facebook
\cite{facebookanatomy}.
In addition, clustering is highly applicable to neuroscience
and genetics research. Connectomics \cite{conn1, conn2, conn3},
uses clustering to group anatomical regions by structural, physiological, and
functional similarity, for the purposes of inference. Behavioromics
\cite{behavioromics} uses clustering to map neurons to distinct motor
patterns. In genetics, clustering is used to infer relationships between
genetically similar species \cite{genetic1, genetic2}.


The greatest challenges facing tool builders are \textit{(i)}
reducing the cost of the synchronization barrier between the
first and second phase of k-means, \textit{(ii)} mitigating the
latency of data movement through the memory hierarchy, and
\textit{(iii)} scaling to arbitrarily large datasets, while
maintaining performance. In addition,
fully asynchronous computation for Lloyd's algorithm is infeasible
because each iteration updates global state, membership and centroids.
The resulting global barriers pose a major challenge to the performance
and scalability of parallel and distributed implementations; this is especially
true for data that require large numbers of iterations to converge.

Popular frameworks \cite{graphlab, mahout, mllib}
have converged on scale-out, distributed processing in which data are partitioned
among cluster nodes, often randomly, and global updates are transmitted at the speed of the
interconnect. These frameworks are negatively affected by inefficient data
allocation, management, and task scheduling protocols with regards to k-means.
This design incurs heavy network traffic owing to data shuffling and centralized
master-worker designs.
Furthermore, such frameworks struggle to capitalize on potential gains from the use
of computation pruning techniques, such as Elkan's triangle inequality
algorithm (TI) \cite{triineq}.
Pruning introduces skew in which few workers have the bulk of the computation.
Skew degrades parallelism. While skew can be dealt with through dynamic scheduling,
this incurs data movement and message passing overheads.


In contrast, our k-means prefers scale-up computation on shared-memory
multicore machines in order to eliminate network traffic and perform
fine-grained synchronization. A current trend for hardware design scales up
a single machine, rather than scaling out to many networked machines,
integrating large memories and using solid-state storage devices (SSDs) to
 extend memory capacity.
This conforms to the node design for supercomputers \cite{Ang14}.
Recent findings, from Frank McSherry \cite{McSherry15,McSherryBlog} and our
prior work \cite{flashgraph},
show that the largest graph analytics tasks can be done on a small fraction of
the hardware, at less cost, as fast, and using less energy on a single shared-memory
node, rather than a distributed compute engine.  Our findings reveal
that clustering has the same structure.  We perform k-means on a single or few machines
to minimize network bottlenecks and find that even our single node performance (with SSDs)
outperforms our competitor's distributed performance on many instances.


Our approach advances Lloyd's algorithm for modern, multicore NUMA architectures
to achieve a high degree of parallelism by significantly merging the two phases
in Lloyd's. We present \textsf{knori}, a fast in-memory,
module that performs several orders of magnitude faster
than other state-of-the-art systems for datasets that fit into main memory.
We implement a practical modification to TI,
that we call the minimal triangle inequality (MTI).
TI incurs a memory increment of $\mathcal{O}(nd)$, limiting its scalability.
MTI requires an increase of only $\mathcal{O}(n)$ memory, drastically improving its
utility for large-scale datasets.
In practice, MTI outperforms TI because it requires significantly less data
structure maintenance, while still pruning computation comparably.
\textsf{knor} clusters data an order of magnitude faster than competitors.

The \textsf{knord} distributed module builds directly on \textsf{knori} and runs
across multiple machines using a decentralized driver.
It runs on larger datasets that fit in the aggregate memory of multiple nodes.

\textsf{knors} is a semi-external memory implementation that scales the computation
of a single node beyond memory bounds.
Semi-external memory (SEM) k-means
holds $\mathcal{O}(n)$ data in memory while streaming
$\mathcal{O}(nd)$ data from disk for a dataset, $\vec{V} \in \mathbb{R}^{n \texttt{x} d}$.
This notion of SEM is analogous to the
definition of SEM for graph algorithms \cite{Abello98, Pearce10}
in which vertex state is kept in memory and edge lists on disk.
We build
\textsf{knors} on a modified FlashGraph \cite{flashgraph} framework to access
asynchronous I/O and overlap I/O and computation.
\textsf{knors} uses a fraction of the memory of popular frameworks and outperforms
them by large factors using less hardware.

This work demonstrates that k-means on extremely large datasets can be
run on increasingly smaller/fewer machines than possible before;
creating reductions in monetary expense and power consumption. Furthermore,
our routines are highly portable. We simply require the C++11 standard
library be available, with thread-level parallelism is implemented
using the POSIX thread (p-threads) library. Distributed routines rely
on the Message Passing Library, MPI \cite{MPI}. The I/O components for
SEM routines are implemented using low-level Linux interfaces.

%% file: relwork.tex
\section{Related Work}

Zhao et al \cite{zhao2009parallel} developed a parallel k-means routine
on Hadoop!, an open source implementation of MapReduce \cite{mapreduce}.
Their implementation has much in common with Mahout
\cite{mahout}, a machine learning library for: Hadoop!.
The Map-Reduce paradigm consists of a \texttt{Map} phase, a synchronization barrier
in which data are shuffled to \textit{reducers}, then a \texttt{Reduce} phase.
For k-means one would perform distance computations in the \texttt{Map} phase
and build centroids to be used in the following iteration in \texttt{Reduce}
phase. The model allows for effortless
scalability and parallelism, but little flexibility in how to achieve either.
As such, the implementation is subject to skew within the reduce phase as
data points assigned to the same centroid end up at a single \textit{reducer}.

MLlib is a machine learning library for Spark \cite{spark}.
Spark imposes a functional paradigm to parallelism allowing for delayed computation
through the use of transformations that form a lineage. The lineage is then evaluated
and automatically parallelized. MLlib's performance is highly coupled with
Spark's ability to efficiently parallelize computation using the generic data
abstraction of resilient distributed datasets \cite{rdd}.

Other works focus on developing fast k-means approximations.
Sophia-ML uses a mini-batch algorithm that uses sampling to
reduce the cost of Lloyd's algorithm (also referred to as batched k-means)
and stochastic gradient descent k-means \cite{sculley2010web-scale}.
Sophia-ML's target application is online, real-time applications, which
differs from our goal of exact k-means on large scale data.
Shindler et al \cite{shindler2011fast} developed
a fast approximation that addresses scalability by streaming data from disk sequentially,
limiting the amount of memory necessary to iterate. This shares some similarity
with \textsf{knors}, but is designed for a single processor,
passing over the data just once and operating on medium-sized data.
We avoid approximations; they see little widespread use owing to
questions of cluster quality.

Elkan proposed the use of the triangle inequality (TI) with
bounds \cite{triineq}, to reduce the number of distance computations to fewer
than $\mathcal{O}(kn)$ per iteration. TI determines when
the distance of data point, $v_i$, that is assigned to a cluster, $c_i$,
is far enough from any other cluster, $c_x, x \in \{1..k\}-i$, so that no distance
computation is required between $v_i$ and $c_x$.
This method is extremely effective in pruning computation in real-world data,
i.e. data with multiple natural clusters.
The method relies on a sparse lower bound matrix of size $\mathcal{O}(nk)$.
Yinyang k-means \cite{ding2015yinyang} develop a competitor pruning
technique to TI that maintains a lower-bound matrix
of size $\mathcal{O}(nt)$, where $t$ is a parameter and $t = k/10$ is generally optimal.
Yingyang k-means outperforms TI by reducing the cost of maintenance of
their lower-bound matrix. Both Yinyang k-means and TI suffer from scalability
limitations because the lower-bound matrix increases in-memory state asymptotically.
We propose MTI for computation pruning. MTI costs a constant
$\mathcal{O}(n)$ more memory making it practical for use with big-data.

FlashGraph \cite{flashgraph} is a SEM graph computation framework that places
edge data on SSDs and allows user-defined vertex state to be held in memory.
FlashGraph partitions a graph then exposes a vertex-centric programming
interface that permits
users to define functions written from the perspective of a single vertex,
known as \textit{vertex programs}. Parallelization is obtained from running
multiple vertex programs concurrently. FlashGraph overlaps I/O with
computation to mask latency in data movement through the memory hierarchy.
FlashGraph is also tolerant to in-memory failures, allowing recovery in SEM
routines through lightweight checkpointing.

FlashGraph is built on top of a userspace filesystem called SAFS \cite{safs}. SAFS
provides a framework to perform high speed I/O from an array of SSDs. To facilitate
this, both SAFS and FlashGraph work to merge I/O requests for multiple
requests when requests are made for data located near one another on disk.
This I/O merging amortizes the cost of accesses to SSDs. SAFS creates and manages a
\textit{page cache} that pins frequently touched pages in memory. The page
cache reduces the number of actual I/O requests made to disk. Section
\ref{sec:fg-mods} discusses the modifications we make to FlashGraph to build
\textsf{knors} on top of FlashGraph.

%% file: theory.tex
\section{Nomenclature} \label{nomen}

We define notation that we use throughout the manuscript.
Let $\mathbb{N}$ be the set of all natural numbers. Let $\mathbb{R}$ be
the set of all real numbers. Let $\vec{v}$ be a $d$-dimension vector in
the dataset $\vec{V}$ with cardinality, $|\vec{V}| = n$.
Let $j$ be the number of iterations of Lloyd's algorithm we perform in a
single run of k-means. Let $t \in \{0...j\}$ be the current iteration
within a run of k-means.
Let $\vec{c}^{\,t}$ be a $d$-dimension vector representing
the mean of a cluster (i.e., a centroid), at iteration $t$.
Let $\vec{C}^t$ be the set of the $k$ centroids at iteration
$t$, with cardinality $|\vec{C}^t| = k$.
In a given iteration, $t$, we can cluster any point, $\vec{v}$ into
a cluster $\vec{c}^{\,t}$. We use Euclidean distance, denoted as
$\mathbf{d}$, as the dissimilarity metric between any $\vec{v}$ and
$\vec{c}^{\,t}$, such that $\mathbf{d}(\vec{v}, \vec{c}^{\,t}) =$ \newline
    $\sqrt{(\vec{v}_1 - \vec{c}^{\,t}_1)^2 + (\vec{v}_2 - \vec{c}^{\,t}_2)^2
        + ... + (\vec{v}_{d-1} - \vec{c}^{\,t}_{d-1})^2 +
    (\vec{v}_{d} - \vec{c}^{\,t}_{d})^2}$.

Let $f(\vec{c}^{\,t} | t > 0) = \mathbf{d}(\vec{c}^{\,t},
\vec{c}^{\,t-1})$. Finally, let $T$ be the number of threads of concurrent
execution, $P$ be the number of processing elements available (e.g. the number
of cores in the machine), and $N$ be the number of NUMA nodes.

\section{Parallel Lloyd's Algorithm}\label{algo}

Underlying further optimizations is our parallel version of Lloyd's algorithm
($||$Lloyd's) that boosts the performance of \textsf{knor} and reduces factors
limiting parallelism in a na\"{\i}ve parallel Lloyd's algorithm.
Traditionally Lloyd's operates in two-phases each separated by a global
barrier as follows:
\begin{enumerate}
    \itemsep0em
    \item Phase I: Compute the nearest centroid, $\vec{c\_nearest}^t$ to
        each data point, $\vec{v}$, at iteration $t$.
    \item Global barrier.
    \item Phase II: Update each centroid, for the next iteration,
        $\vec{c}^{\,t+1}$ to be the mean value of all points nearest to
        it in Phase I.
    \item Global barrier.
    \item Repeat until converged.
\end{enumerate}
Na\"{\i}ve Lloyd's uses two major data structures; A
read-only global centroids structure, $\vec{c}^{\,t}$, and a shared global
centroids for the next iteration, $\vec{c}^{\,t+1}$. Parallelism in Phase II
is limited to $k$ threads because $\vec{c}^{\,t+1}$ is shared. As such, Phase II
is plagued with substantial locking overhead because of the high likelihood
of data points concurrently attempting to update the the same nearest centroid.
Consequently, as $n$ gets larger with respect to $k$ this interference
worsens, further degrading performance.

$||$Lloyd's retains the read-only global centroid structure $\vec{c}^{\,t}$,
but provides each thread with its own local copy of the next iteration's
centroids. Thus we create $T$ copies of $\vec{c}^{\,t+1}$. Doing so means
$||$Lloyd's merges Phase I and II into a super-phase and eliminates the
barrier (Step 3 above). The super-phase concurrently computes the nearest
centroid to each point and updates a local version of the centroids to be used
in the following iteration. These local centroids can then be merged in parallel
through a reduction operation at the end of the iteration. $||$Lloyd's trades-off
increased parallelism for a slightly higher memory consumption by a factor of
$\mathcal{O}(T)$ over Lloyd's. This algorithm design naturally leads to lock-free
routines that require fewer synchronization barriers as we show in
Algorithm \ref{alg:merge}.

\begin{algorithm}
\caption{$||$ Lloyd's algorithm}
\label{alg:merge}
\begin{algorithmic}[1]
	\Procedure{$||$means}{$\vec{V}$, $\vec{C}^t$, $k$}
		\State$\vec{ptC^t}$ \Comment{Per-thread centroids}
		\State$\vec{clusterAssignment^t}$ \Comment{Shared, no conflict}
        \State tid \Comment{Current thread ID}
		\ParFor{$\vec{v} \in \vec{V}$}
			\State $dist = \infty$
			\State $\vec{c\_nearest}^t$ = INVALID
			\For{$\vec{c}^{\,t} \in \vec{C}^t$}
				\If {$\mathbf{d}(\vec{v}, \vec{c^{\,t}}) < dist$}
				\State $dist = \mathbf{d}(\vec{v}, \vec{c^{\,t}})$
				\State $\vec{c\_nearest}^t = \vec{c}^{\,t}$
				\EndIf
			\EndFor
		\State$\vec{ptC^t}[tid][\vec{c\_nearest}^t$] += $\vec{v}$
		\EndParFor
        \State clusterMeans = \textsc{mergePtStructs}($\vec{ptC^t}$)
	\EndProcedure

	\item[]
	\Procedure{mergePtStructs}{$\vec{vectors}$}
		\While{$|\vec{vectors}| > 1$}
        \State PAR\_MERGE($\vec{vectors}$) \Comment{$\mathcal{O}(T log n)$}
		\EndWhile
		\Return vectors[0]
	\EndProcedure
\end{algorithmic}
\end{algorithm}

\subsection*{Minimal Triangle Inequality (MTI) Pruning}\label{sec:mte}

We simplify Elkan's Algorithm for triangle inequality pruning (TI)
\cite{triineq} by removing the the necessity for the lower
bound matrix of size $\mathcal{O}(nd)$. Omitting the lower
bound matrix means we forego
the opportunity to prune certain computations; we accept this tradeoff
in order to limit main memory consumption and prioritize usability.
Due to space limitations, we omit experiments exhibiting that in practice
pruning benefits of maintaining a lower bound matrix are minimal.
With $\mathcal{O}(n)$, memory we implement three of the four \cite{triineq}
pruning clauses invoked for
each data point in an iteration of a \textsf{knor} module
with pruning \textit{enabled}.
Let $u^t = \mathbf{d}(\vec{v}, \vec{c\_nearest}^t) + f(\vec{c\_nearest}^t)$,
be the upper bound
of the distance of a sample, $\vec{v}$, in iteration $t$
from its assigned cluster $\vec{c\_nearest}^t$.
Finally, we define $U$ to be an update function such that
$U(u^t)$ fully tightens the upper bound of $u^t$.

\begin{itemize}[label={},leftmargin=*]
\item\textbf{Clause 1:} if $u^t \leq \min \mathbf{d}(\vec{c\_nearest}^t,
    \vec{c}^{\,t} \, \forall \, \vec{c}^{\,t} \in \vec{C}^t)$,
		then $\vec{v}$ remains in the same cluster for the current iteration.
		For \textsf{knors}, this is extremely significant because no I/O
		request is made for data.

\item\textbf{Clause 2:} if $u^t \leq \mathbf{d}(\vec{c\_nearest}^t,
		\vec{c}^{\,t} \, \forall \, \vec{c}^{\,t} \in \vec{C}^t)$,
		then the distance computation between data point $\vec{v}$ and
		centroid $\vec{c}^{\,t}$ is pruned.

\item\textbf{Clause 3:} if $U(u^t) \leq \mathbf{d}(\vec{c\_nearest}^t,
		\vec{c}^{\,t} \, \forall \, \vec{c}^{\,t} \in \vec{C}^t)$,
		then the distance computation between data point $\vec{v}$ and
		centroid $\vec{c}^{\,t}$ is pruned.
\end{itemize}

%% file: im-design.tex
\section {In-memory design} \label{im-design}

We prioritize practical performance when we implement \textsf{knori}
optimizations. We make design tradeoffs to balance
the opposing forces of minimizing memory usage and maximizing
CPU cycles spent on parallel computing. Section \ref{im-design} chronicles
the memory bounds that we achieve and optimizations that we apply.

\subsection{Asymptotic Memory Consumption}

\textsf{knori} with MTI \textit{disabled}, which we call \textsf{knori-},
retains the computation complexity of a serial
routine, i.e., $\mathcal{O}(ndk)$, but has a memory bound of
$\mathcal{O}(nd + Tkd)$ as compared to the original $\mathcal{O}(nd + kd)$. The factor of $T$
arises from the per-thread centroids we maintain.
\textsf{knori} (MTI \textit{enabled}) uses additional memory $\mathcal{O}(k^2 + n)$,
which does not increase the asymptotic bound.
The $\mathcal{O}(k^2)$ comes from maintaining an
upper/lower triangular centroid-to-centroid distance matrix and $\mathcal{O}(n)$ comes from
maintaining the upper bound of each data point's distance to any centroids. The
$\mathcal{O}(n)$ in practice adds between $6$-$10$ Bytes per data point or
$\leq 1GB$ when $n = 100$ million and $d$ is unrestricted. We justify the
tradeoff of slightly higher memory consumption for an improvement in performance
in Section \ref{sec:eval}. A complete summary of \textsf{knor} routine memory
bounds is shown in Table \ref{tbl:complexity}.

\subsection{In-memory optimizations}

The following design principles and optimizations improve
the performance of Algorithm \ref{alg:merge}.

\textbf{Prioritize data locality for NUMA}:
Non-uniform memory
access (NUMA) architectures are characterized by groups of processors that have
affinity to a local memory bank via a shared local bus. Other non-local
memory banks must be accessed through a globally shared interconnect.
The effect is low latency accesses with high throughput to local memory banks,
and conversely higher latency and lower throughput
for remote memory accesses to non-local memory.

\begin{figure}[t]
\centering
\includegraphics[width=.8\linewidth]{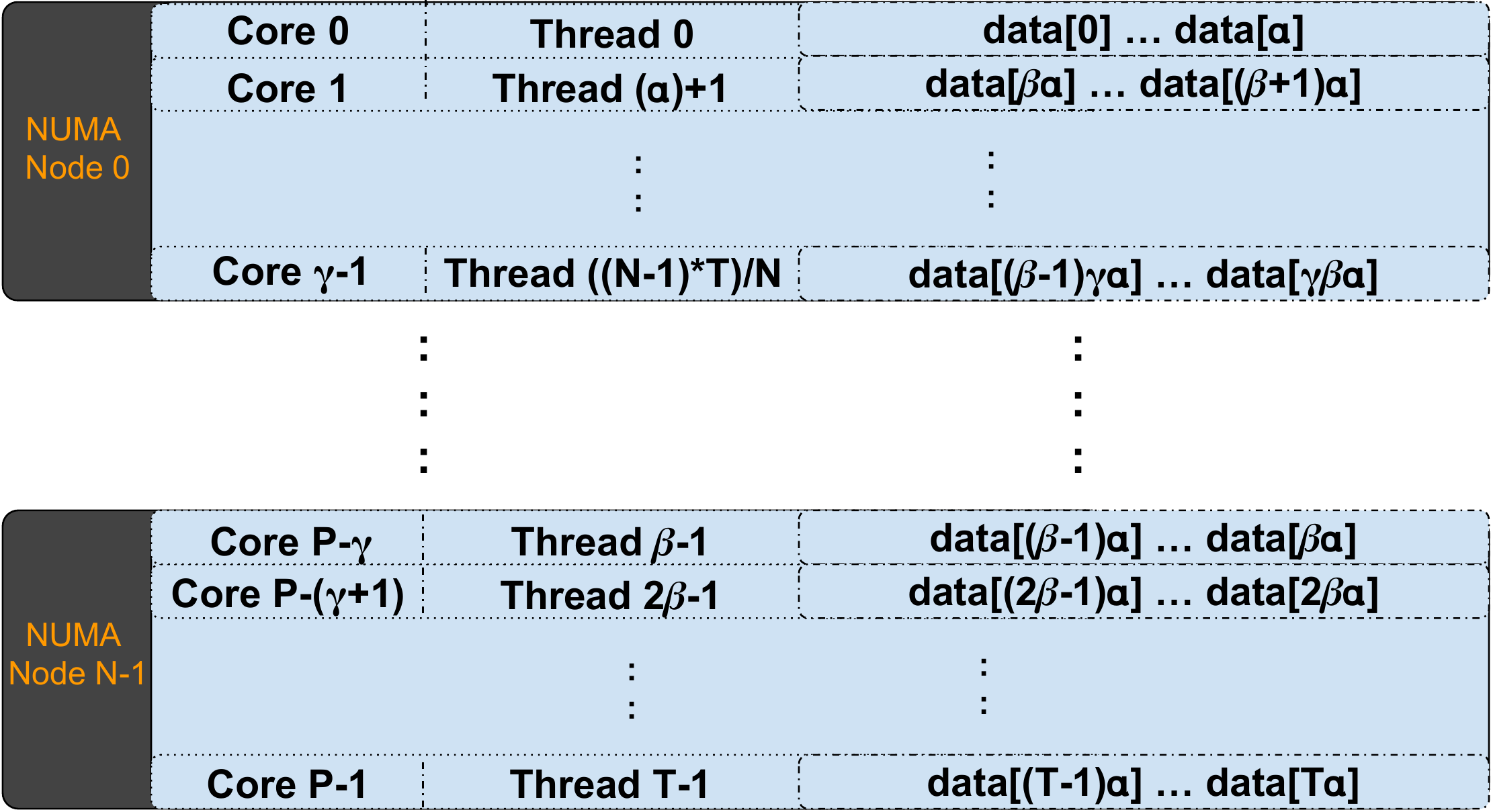}
\caption{The memory allocation and thread assignment scheme we utilize
for \textsf{knori} and \textsf{knord} on each machine. $\alpha = n/T$ is
the amount of data per thread, $\beta = T/N$ is the number of threads
per NUMA node, and $\gamma = P/N$ is the number of physical processors per NUMA node.
Distributing memory across NUMA nodes maximizes memory throughput
while binding threads to NUMA nodes reduces remote memory accesses.}
\label{fig:numa-mem}
\end{figure}

To minimize remote memory accesses, we bind every thread to a single NUMA node,
equally partition the dataset across NUMA nodes, and sequentially allocate data
structures to the local NUMA node's memory. Every thread works
independently.  Threads only communicate or share data to aggregate per-thread results.
Figure \ref{fig:numa-mem} shows the data allocation and access scheme we employ. We
bind threads to NUMA nodes rather than specific CPU cores because the latter
is too restrictive to the OS scheduler. CPU thread-binding
may cause performance degradation if the number of worker threads exceeds
the number of physical cores.

\begin{figure}[t]
\centering
\includegraphics[width=.8\linewidth]{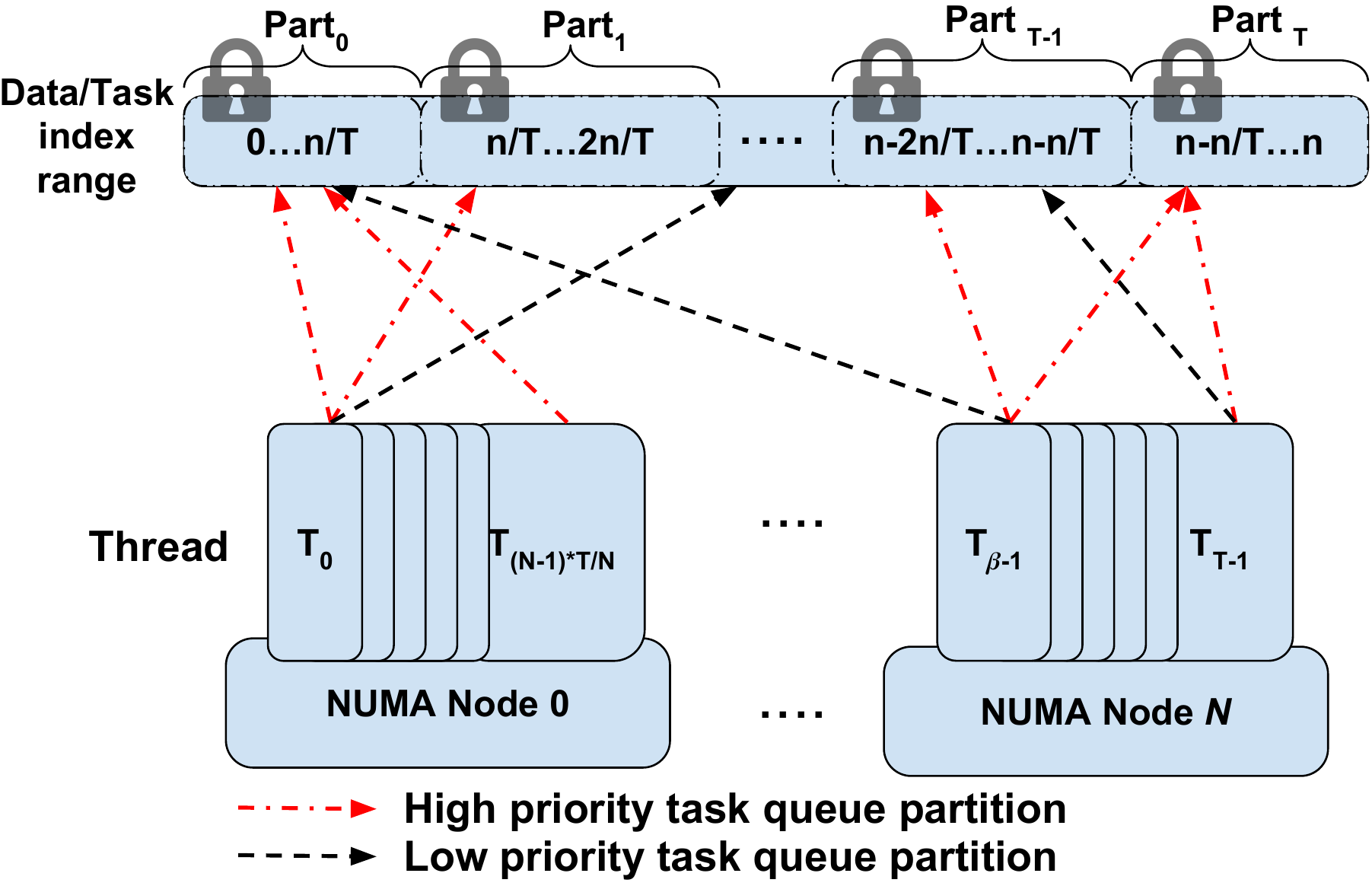}
\caption{The NUMA-aware partitioned task scheduler we utilize
for \textsf{knori} and \textsf{knord} on each machine. The
scheduler minimizes task queue lock contention and remote memory
accesses by prioritizing tasks with data in the local NUMA memory
bank.}
\label{fig:task-queue}
\vspace{-10pt}
\end{figure}

\textbf{Dynamic Scheduling and Work Stealing}: To achieve optimal performance
when MTI pruning is \textit{disabled}, statically scheduling thread tasks to
locally allocated data partitions is sufficient.
When MTI is \textit{enabled}, we see worker
skew at the thread level. In response, we develop a NUMA-aware
partitioned priority task queue, (Figure \ref{fig:task-queue}),
to feed worker threads, prioritizing tasks that maximize local memory access
and thus limit remote memory accesses.

The task queue enables idle threads to \textit{steal} work from threads bound to the
same NUMA node first, minimizing remote memory accesses. The queue is
partitioned into $T$ parts, each with a lock required for access.
We allow a thread to cycle through the task queue once looking for high priority
tasks before settling on another, possibly lower priority task.
This tradeoff avoids starvation and ensures threads are idle for negligible
periods of time. The result is good load balancing when pruning in addition to
optimized memory access patterns.

\textbf{Avoid interference and delay the synchronization barrier}: We
employ per-thread local centroids and write-conflict free shared data structures to
eliminate interference. Local centroids are un-finalized running totals of
global centroids used in the following iteration for distance computations.
Local centroids are concurrently updated. Finally, we require only a single global
barrier prior to merging local clusters in a parallel funnelsort-like \cite{cache-obl}
reduction routine for use in the following iteration.

\textbf{Effective data layout for CPU cache exploitation}: Both per-thread and
global data structures are contiguously allocated chunks of memory.
Contiguous data organization and sequential access patterns
when computing cluster-to-centroid distances maximizes both
prefetching and CPU caching opportunities.

%% file: sem-design.tex
\section{Semi-external Memory Design}

We design a highly optimized semi-external memory module, \textsf{knors}, that targets
the stand-alone server environment and when data are too large to reside
fully in main memory. \textsf{knors} extends its in-memory counterpart from the
perspective of scalability by placing data on SSDs and requesting data as
necessary. Data requests are performed asynchronously allowing the overlap
of I/O and computation. The SEM model allows us to reduce the asymptotic memory
bounds so as to scale to larger datasets. A SEM routine uses $\mathcal{O}(n)$
memory for a dataset, $\vec{V} \in \mathbb{R}^{n \texttt{x} d}$ that when processed
completely in memory would require $\mathcal{O}(nd)$ memory. For completeness, we summarize all \textsf{knor} memory
asymptotic bounds in Table \ref{tbl:complexity}.

\subsection{FlashGraph modifications} \label{sec:fg-mods}

We modify FlashGraph to support matrix-like computations. FlashGraph's primitive
data type is the \texttt{page\_vertex} that is interpreted as a vertex with an index
to the edge list of the \texttt{page\_vertex} on SSDs.
We define a \textit{row} of data to be equivalent to a $d$-dimension
data point, $\vec{v}_i$. Each row is composed of a unique
identifier, \textit{row-ID}, and $d$-dimension data vector, \textit{row-data}.
We add a \texttt{page\_row} data type to FlashGraph and modify FlashGraph's
I/O layer to support reading floating point
row-data from SSD rather than the numeric data type associated with edge lists.
The \texttt{page\_row} type computes it row-ID and row-data location on disk meaning
only user-defined state is stored in-memory.
The \texttt{page\_row} reduces the memory necessary to use FlashGraph by $\mathcal{O}(n)$
because it does not store a row-data index to data on SSDs unlike a
\texttt{page\_vertex}. This allows \textsf{knors} to scale to larger
datasets than possible before on a single machine.

\subsection{Semi-external Memory Asymptotic \newline Analysis}

SEM implementations do not alter the computation bounds, whereas they do lower
the memory bounds for k-means to $\mathcal{O}(n + Tkd)$. This improves on the
$\mathcal{O}(nd + Tkd)$ bound of \textsf{knori} and the $\mathcal{O}(nd + kd)$ bound of Lloyd's
original algorithm. In practice, the disk I/O bound of \textsf{knors}
is much lower than the worst case of $\mathcal{O}(nd)$ obtained when MTI pruning is
\textit{disabled} (\textsf{knors-}), especially for data with natural clusters.

\subsubsection{I/O minimization} \label{sec:io-min-eff}

I/O bounds the performance of k-means in the SEM model.
Accordingly, we reduce the number of data-rows that need
to be brought into memory each iteration. Only Clause 1
of MTI (Section \ref{sec:mte}) facilitates the skipping of all
distance computations for a data point. In this case we do not
issue an I/O request for the data point's row-data.
This results in a reduction in I/O, but because data are pruned in a near-random
fashion, we retrieve significantly more data than necessary from SSDs due to
fragmentation. Reducing the filesystem \textit{page size}, i.e. minimum read size
from SSDs alleviates this to an extent, but a small page size can lead to
higher amounts of I/O requests, offsetting any gains achieved by the
reduction in fragmentation.
We utilize a minimum read size of $4KB$; even with this relatively
small value we still receive significantly more data from disk than we request
(Figure \ref{fig:tot-io}). To address this, we develop a lazily-updated
partitioned \textit{row cache} described in Section \ref{sec:row-cache},
that drastically reduces the amount of data brought into memory
as shown in Figure \ref{fig:io-by-iter}.

\subsubsection{Partitioned Row Cache (RC)} \label{sec:row-cache}

\begin{figure}[t]
\centering
\includegraphics[width=.8\linewidth]{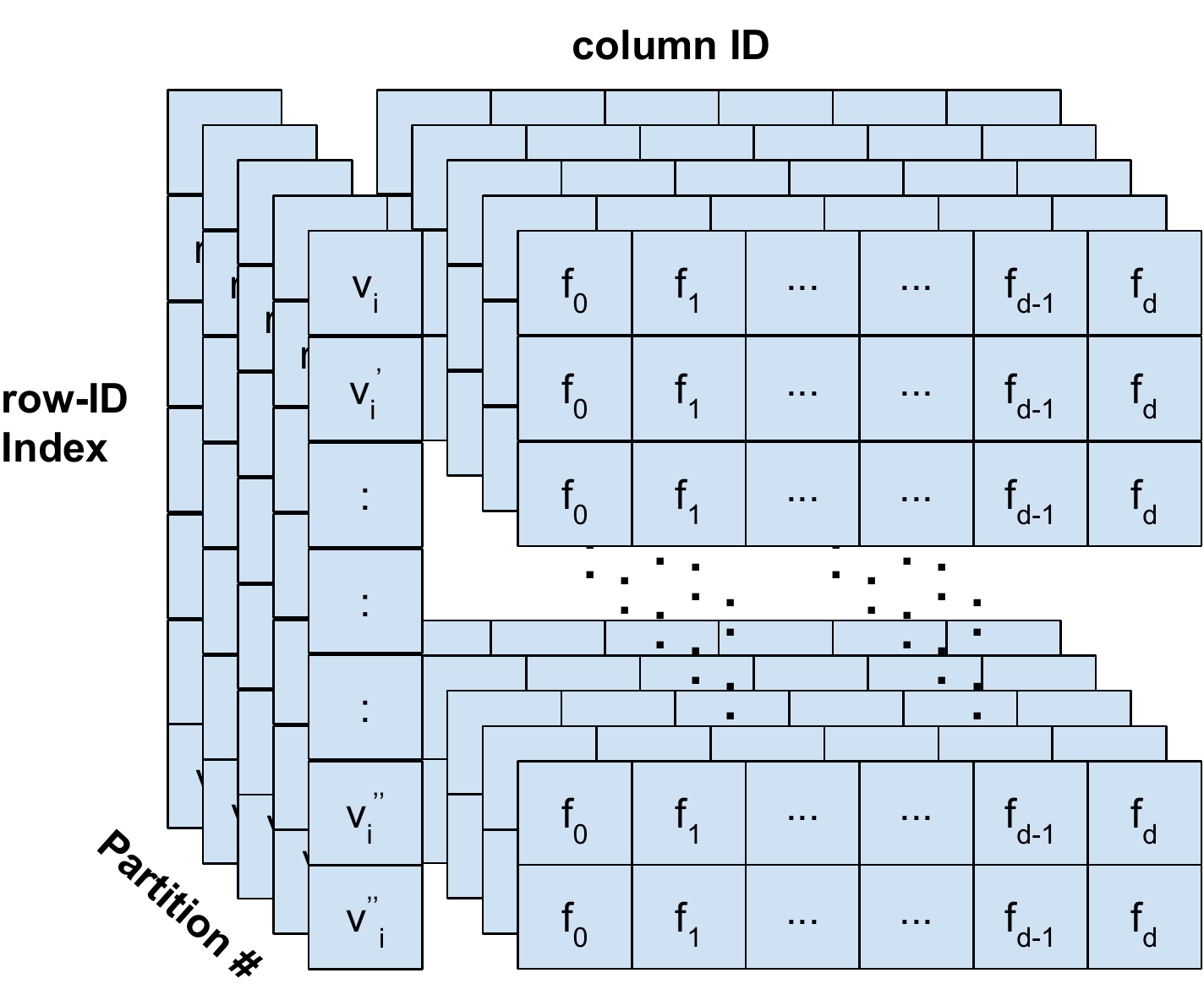}
\caption{The structure of the row cache we utilize for \textsf{knors}.
Partitioning the row cache eliminates the need for locking during cache population.}
\label{fig:row-cache}
\vspace{-10pt}
\end{figure}

We add a layer to the memory hierarchy for SEM applications by designing a
lazily-updated row cache (Figure \ref{fig:row-cache}).
The row cache improves performance
by reducing I/O and minimizing I/O request merging and page
caching overhead in FlashGraph. A row is \textit{active} when
it performs an I/O request in the current iteration for its row-data.
The row cache pins active rows to memory at the granularity of a row,
rather than a page, improving its effectiveness in reducing I/O compared
to a page cache.

The row cache lazily updates at certain iterations of the k-means algorithm
based on a user defined \textit{cache update interval} ($I_{cache}$).
The cache updates/refreshes at iteration $I_{cache}$ then the update
frequency increases exponentially such that the next RC update is
performed after $2I_{cache}$ iterations and so forth. This means that
row-data in the RC remains static for several iterations before
the RC is flushed then repopulated. We justify lazy updates by observing that k-means,
especially on real-world data, follows predictable row activation patterns.
In early iterations the cache's utility is of minimal benefit as the row
activation pattern is near-random. As the algorithm progresses, data points that are active tend to stay active for many iterations as they
are near more than one established centroid.
This means the cache can remain static for longer periods of time
while achieving very high cache hit rates. We set $I_{cache}$
to $5$ for all experiments in the evaluation (Section \ref{sec:eval}).
The design trade-off is cache freshness for reduced cache maintenance
overhead. We demonstrate the efficacy of this design in
Figure \ref{fig:cache-hits-by-iter}.

We partition the row cache into as many partitions as FlashGraph creates for the
underlying matrix, generally equal to the number of threads of execution.
Each partition is updated locally in a lock-free caching structure.
This vastly reduces the cache maintenance overhead, keeping the RC lightweight.
At the completion of an iteration in which the RC
refreshes, each partition updates a global map that stores pointers
to the actual data. The size of the cache is user-defined, but
$1GB$ is sufficient to significantly improve the performance of
billion-point datasets.

%% file: dist-design.tex
\section{Distributed Design}

\textsf{knord} scales to distributed clusters through the Message Passing Interface (MPI).
We employ modular design principles and build our distributed routines as a
layer above our parallel in-memory routines. The implication is that each
machine maintains a decentralized \textit{driver} (MPI) process that
launches \textit{worker} (pthread) threads that retain the NUMA performance
optimizations we develop for \textsf{knori} in Section \ref{im-design}.

We do not address load balancing between machines in the cluster.
We recognize that in some
cases it may be beneficial to dynamically dispatch tasks, but we argue that this
would negatively affect the performance enhancing NUMA polices we implement. We
further argue that the gains in performance of our data partitioning
scheme (shown in Figure \ref{fig:numa-mem}) outweigh the effects of skew in
this setting. We validate these assertions empirically in
Section \ref{sec:dist-eval}.

%% file: eval.tex
\section{Experimental Evaluation} \label{sec:eval}

We evaluate \textsf{knor} by benchmarking optimizations in addition to comparing
its performance against other state-of-the-art frameworks.
In Section \ref{sec:baseline} we evaluate the performance of our baseline single
threaded implementation to ensure all speedup experiments are relative to a
state-of-the-art baseline performance.
Sections \ref{sec:im-opt-eval} and \ref{sec:sem-opt-eval} evaluate the effect
of specific optimizations on our in-memory and semi-external memory tools
respectively. Section \ref{sec:vs-others} evaluates the performance of
\textsf{knori} and \textsf{knors} relative
to other popular frameworks from the perspective of time and resource
consumption. Section \ref{sec:dist-eval} specifically performs comparison
between \textsf{knord} and MLlib in a cluster.

We evaluate \textsf{knor} optimizations on the Friendster top-8 and top-32
eigenvector datasets, because the Friendster dataset represents
real-world machine learning data. The Friendster dataset is derived
from a graph that follows a power law distribution of edges. As such, the
resulting eigenvectors contain natural clusters with well defined centroids, which
makes MTI pruning effective, because many data points fall into
strongly rooted clusters and do not change membership. These trends hold
true for other large scale datasets, albeit to a lesser extent on uniformly
random generated data (Section \ref{sec:vs-others}). The datasets we use for performance
and scalability evaluation are shown in Table \ref{tbl:matrices}.

We use the following notation throughout the evaluation:
\begin{itemize}
    \item \textbf{\textsf{knori-}} is \textsf{knori} with MTI pruning \textit{disabled}.
    \item \textbf{\textsf{knors-}} is \textsf{knors} with MTI pruning \textit{disabled}.
    \item \textbf{\textsf{knors-{}-}} is \textsf{knors} with both MTI pruning and the row cache
        (RC) \textit{disabled}.
    \item \textbf{\textsf{knord-}} is \textsf{knord} with MTI pruning \textit{disabled}.
    \item \textbf{MLlib-EC$2$} is MLlib's k-means routine running on Amazon EC$2$
        instances \cite{aws}.
    \item \textbf{MPI} is a pure MPI \cite{MPI} distributed implementation of our $||$Lloyd's
    algorithm (Section \ref{algo}) with MTI pruning. We develop this in order to compare
			its performance to \textsf{knord}.
    \item \textbf{MPI-} is a pure MPI distributed implementation of our $||$Lloyd's
        algorithm with MTI pruning \textit{disabled}.
\end{itemize}

\begin{table}[!htb]
\caption{Asymptotic memory complexity of \textsf{knor} routines.}
\vspace{-10pt}
\begin{center}
\footnotesize
\begin{tabular}{|c|c|c|}
\hline
Module / Routine & Memory complexity \\
\hline
Na\"{\i}ve Lloyd's & $\mathcal{O} (nd + kd)$\\
\hline 

\textsf{knors-}, \textsf{knors-{}-} & $\mathcal{O} (n + Tkd)$\\
\hline

\textsf{knors} & $\mathcal{O} (2n + Tkd + k^2)$\\
\hline

\textsf{knori-}, \textsf{knord-} & $\mathcal{O} (nd + Tkd) $\\
\hline

\textsf{knori}, \textsf{knord} & $\mathcal{O}(nd + Tkd + n + k^2)$\\
\hline

\end{tabular}
\normalsize
\end{center}
\label{tbl:complexity}
\end{table}

\begin{table}[!htb]
\caption{The datasets under evaluation in this study.}
\vspace{-10pt}
\begin{center}
\footnotesize
\begin{tabular}{|c|c|c|c|}
\hline
\textbf{Data Matrix} & $n$ & $d$ & \textbf{Size}\\
\hline
Friendster-8 \cite{friendster} eigenvectors & $66$M &
  $8$ & $4$GB\\
\hline
Friendster-32 \cite{friendster} eigenvectors & $66$M &
  $32$ & $16$GB\\
\hline
Rand-Multivariate (RM$_{856M}$)& $856$M & 16 & $103$GB\\
\hline
Rand-Multivariate (RM$_{1B}$)& $1.1$B & 32 & $251$GB\\
\hline
Rand-Univariate (RU$_{2B}$)& $2.1$B & 64 & $1.1$TB\\
\hline
\end{tabular}
\normalsize
\end{center}
\label{tbl:matrices}
\end{table}

For completeness we note versions of all frameworks and libraries we use for comparison
in this study; Spark v2.0.1 for MLlib, H$_2$O v3.7, Turi v2.1, R v3.3.1, MATLAB R2016b,
BLAS v3.7.0, Scikit-learn v0.18, MLpack v2.1.0.

\subsection{Single Node Evaluation Hardware}

We perform single node experiments relating to \textsf{knori}, \textsf{knors}
on a NUMA server with
four Intel Xeon E$7$-$4860$ processors clocked at $2.6$ GHz and $1$TB
of DDR$3$-$1600$ memory. Each processor has $12$ cores. The machine has three LSI
SAS $9300$-$8$e host bus adapters (HBA) connected to a SuperMicro storage
chassis, in which $24$ OCZ Intrepid $3000$ SSDs are installed. The machine
runs Linux kernel v$3.13.0$. The C++ code is compiled using mpicxx.mpich2 version
$4.8.4$ with the -O$3$ flag.

\subsection{Cluster Evaluation Hardware}

We perform distributed memory experiments relating to \textsf{knord} on
Amazon EC$2$ compute optimized instances of type c$4.8$xlarge with $60$GB of
DDR$3$-$1600$ memory, running Linux kernel v$3.13.0$-$91$.
Each machine has $36$ vCPUS, corresponding to $18$ physical
Intel Xeon E$5$-$2666$ v$3$ processors,
clocking $2.9$ GHz, sitting on $2$ independent sockets.
We allow no more that $18$ independent MPI processes or equivalently
$18$ Spark workers to exist on any single machine. We constrain the cluster to a single
availability zone, subnet and placement group, maximizing cluster-wide data locality
and minimizing network latency on the 10 Gigabit interconnect. We measure all
experiments from the point when all data is in RAM on all machines. For MLlib we
ensure that the Spark engine is configured to use the maximum available memory and
does not perform any checkpointing or I/O during computation.

\input{im-opt-eval}
\input{sem-opt-eval}

\subsection{MTI Evaluation}

Figures \ref{fig:f8-per-iter-perf-knor} and \ref{fig:f32-per-iter-perf-knor}
highlight the performance improvement of
\textsf{knor} modules with MTI \textit{enabled} over MTI \textit{disabled}
counterparts. We show that MTI provides a few factors of improvement in
time even without some of the pruning ability Elkan's TI \cite{triineq} provides.
Figure \ref{fig:mem-knor} highlights that MTI increases the memory load by
negligible amounts compared to non-pruning modules.
We conclude that MTI (unlike TI) is a viable optimization for large-scale datasets.

\begin{figure}[!htb]
\centering
\footnotesize
\vspace{-15pt}
\begin{subfigure}{.5\textwidth}
		\include{./charts/perf.iter.friendster8.knor}
\vspace{-15pt}
\caption{The Friendster graph top-8 eigenvector dataset.}
\label{fig:f8-per-iter-perf-knor}
\end{subfigure}

\begin{subfigure}{.5\textwidth}
\include{./charts/perf.iter.friendster32.knor}
\vspace{-15pt}
\caption{The Friendster graph top-32 eigenvector dataset.}
\label{fig:f32-per-iter-perf-knor}
\end{subfigure}

\begin{subfigure}{.5\textwidth}
\include{./charts/mem.friendster.knor}
\vspace{-15pt}
\caption{Memory comparison of fully optimized \textsf{knor} routines
(\textsf{knori}, \textsf{knors}) compared to more vanilla \textsf{knor}
routines (\textsf{knori-}, \textsf{knors-{}-})}
\label{fig:mem-knor}
\end{subfigure}

\caption{Performance and memory usage comparison of \textsf{knor} modules
		on matrices from the Friendster graph top-8 and top-32 eigenvectors.}
\label{fig:knor-comp}
\vspace{-10pt}
\end{figure}

\subsection{Comparison with Other Frameworks} \label{sec:vs-others}

We evaluate the performance of our routines in comparison with other
frameworks on the datasets in Table \ref{tbl:matrices}. We show that \textsf{knori}
achieves greater than an order of magnitude improvement over other state-of-the-art
frameworks. Finally, we demonstrate \textsf{knors} outperforms other state-of-the-art
frameworks by several factors.

Both our in-memory and semi-external memory modules incur
little memory overhead when compared with other frameworks.
Figure \ref{fig:mem} shows memory consumption.
We note that MLlib requires the placement of temporary
Spark block manager files. Because the block manager cannot be disabled,
we provide an in-memory RAM-disk so as to not influence MLlib's performance
negatively. We configure MLlib, H$_2$O and Turi to
use the minimum amount of memory necessary to achieve their highest
performance. We acknowledge that a reduction in memory for these frameworks is possible,
but would degrade computation time and lead to unfair comparisons.
All measurements are an average of $10$ runs; we drop all caches between runs.

\begin{figure}[!htb]
\centering
\footnotesize
\vspace{-5pt}
\begin{subfigure}{.5\textwidth}
		\include{./charts/perf.iter.friendster8}
\vspace{-15pt}
\caption{The Friendster graph top-8 eigenvector dataset.}
\label{fig:f8-per-iter-perf}
\end{subfigure}

\begin{subfigure}{.5\textwidth}
\include{./charts/perf.iter.friendster32}
\vspace{-15pt}
\caption{The Friendster graph top-32 eigenvector dataset.}
\label{fig:f32-per-iter-perf}
\end{subfigure}

\begin{subfigure}{.5\textwidth}
\include{./charts/mem.friendster}
\vspace{-15pt}
\caption{Peak memory consumption on the Friendster eigenvectors dataset,
    with $k=10$. Row cache size = $512$MB, page cache size = $1$GB.
}
\label{fig:mem}
\end{subfigure}

\caption{Performance comparison on matrices from the Friendster
    \cite{friendster} graph top-8 and top-32 eigenvectors.}
\label{fig:per-iter-perf}
\end{figure}

We demonstrate that \textsf{knori}
is no less than an order of magnitude faster than all
competitor frameworks (Figure \ref{fig:per-iter-perf}).
\textsf{knori} is often hundreds of
times faster than Turi. Furthermore, \textsf{knors} is consistently
twice as fast as competitor in-memory frameworks.
We further demonstrate performance improvements over competitor
frameworks on algorithmically identical implementations by \textit{disabling} MTI.
\textsf{knori-} is nearly $10$x
faster than competitor solutions, whereas \textsf{knors-}
is comparable and often faster than competitor in-memory solutions.
We attribute our performance gains over other frameworks when MTI is \textit{disabled}
to our parallelization scheme for Lloyd's (Algorithm \ref{alg:merge}).
Lastly, Figure \ref{fig:knor-comp} demonstrates a
consistent $30\%$ improvement in \textsf{knors} when we utilize the row cache. This is
evidence that the design of our lazily updated row cache provides a performance boost.

Finally, comparing \textsf{knori-} and \textsf{knors-{}-} to MLlib, H$_2$O and Turi
(Figures \ref{fig:knor-comp} and \ref{fig:per-iter-perf}) reveals \textsf{knor}
to be several times faster and to use significantly less memory.
This is relevant because \textsf{knori-} and \textsf{knors-{}-} are
algorithmically identical to k-means within MLlib, Turi and H$_2$O.

\subsection{Single-node Scalability Evaluation}

To demonstrate scalability, we compare performance on synthetic datasets drawn from
random distributions that contain hundreds of
millions to billions of data points. Uniformly random data are typically
the worst case scenario for the convergence of k-means, because many
data points tend to be near several centroids.

Both in-memory and SEM modules outperform popular
frameworks on $100$GB+ datasets. We achieve $7$-$20$x improvement when
in-memory and $3$-$6$x improvement in SEM when compared to MLlib, H$_2$O and Turi.
As data increases in size, the performance difference between \textsf{knori} and
\textsf{knors} narrows since there is now enough data to mask I/O latency and to turn
\textsf{knors} from an being I/O bound to being computation bound. We observe
\textsf{knors} is only $3$-$4$x slower than its in-memory counterpart in such cases.

\begin{figure}[!htb]
\centering
\footnotesize
\vspace{-5pt}
\begin{subfigure}{.5\textwidth}
		\include{./charts/perf.iter.rmvm}
\vspace{-15pt}
\caption{Per iteration time elapsed of each routine.}
\label{fig:perf-rmvm}
\end{subfigure}

\begin{subfigure}{.5\textwidth}
\include{./charts/mem.rmvm}
\vspace{-15pt}
\caption{Memory consumption of each routine.}
\label{fig:mem-rmvm}
\end{subfigure}
\caption{Performance comparison on RM$_{856M}$ and RM$_{1B}$ datasets.
  Turi is unable to run on RM$_{1B}$ on our machine and only SEM routines
  are able to run on RU$_{2B}$ on our machine. Page cache size = $4$GB,
  Row cache size = $2$GB.}
\label{fig:rmvm}
\end{figure}

Memory capacity limits the scalability of k-means and semi-external memory
allows algorithms to scale well beyond the limits of physical memory.
The 1B point matrix (RM$_{1B}$) is the largest that fits in 1TB of memory
on our machine. At 2B points (RU$_{2B}$), semi-external memory algorithms
continue to execute proportionally and all other algorithms fail.

\subsection{Distributed Comparison vs.
        \newline Other Frameworks} \label{sec:dist-eval}

We analyze performance of \textsf{knord} and \textsf{knord-} on Amazon's EC$2$ cloud in
comparison to that of (i) MLlib (\textbf{MLlib-EC$2$}), (ii) a
pure MPI implementation of our $||$Lloyd's algorithm with
MTI pruning (\textbf{MPI}), and (iii) a pure MPI implementation of $||$Lloyd's algorithm with
pruning \textit{disabled} (\textbf{MPI-}).
Note that H$_2$O has no distributed memory implementation and Turi
discontinued their distributed memory interface prior to our experiments.

\begin{figure}[!htb]
    \centering
    \footnotesize
    \vspace{-5pt}
    \begin{subfigure}{.5\textwidth}
        \include{./charts/numa.speedup.dist.friendster32}
        \vspace{-15pt}
        \caption {Distributed speedup comparison on the Friendster-32 dataset.}
        \label{fig:dist-speedupfr32}
    \end{subfigure}

    \begin{subfigure}{.5\textwidth}
        \include{./charts/numa.speedup.dist.rm1b}
        \vspace{-15pt}
        \caption{Distributed speedup comparison on the RM$_{1B}$ dataset.}
        \label{fig:dist-speeduprm1b}
    \end{subfigure}

    \caption {Speedup experiments are normalized to each implementation's
			serial performance. Each machine has 18 physical cores with 1 thread per core.}
		\label{fig:dist-speedup}
        \vspace{-5pt}
\end{figure}

\begin{figure}[!htb]
    \centering
    \begin{subfigure}{.5\textwidth}
        \vspace{-15pt}
        \begin{center}
            \include{./charts/titles}
        \end{center}
        \vspace{-15pt}
    \end{subfigure}

    \begin{subfigure}{.5\textwidth}
        \include{./charts/perf.iter.dist.friendsterX}
        \vspace{-10pt}
        \caption{Friendster8 (left) and Friendster32 (right) datasets
        computation time per iteration for $k=100$.}
		\label{fig:dist-perf-frX}
	\end{subfigure}

    \begin{subfigure}{.5\textwidth}
        \include{./charts/perf.iter.dist.rand}
        \vspace{-10pt}
        \caption{RM$_{856M}$ (left) and RM$_{1B}$ (right) datasets
        computation time per iteration for $k=10$.}
		\label{fig:dist-perf-rmX}
	\end{subfigure}

	\caption {Distributed performance comparison of \textsf{knord}, MPI and MLlib
		on Amazon's EC$2$ cloud. Each machine has 18 physical cores with 1 thread per core.}
   \label{fig:dist-perf}
\end{figure}

Figures \ref{fig:dist-speedup} and \ref{fig:dist-perf} reveal several fundamental
and important results. Figure \ref{fig:dist-speedup} shows that \textsf{knord} scales
well to very large numbers of machines, performing within a constant factor of
linear performance. This is a necessity today as many organization push big-data
computation to the cloud. Figure \ref{fig:dist-perf} shows that in a cluster,
\textsf{knord}, even with TI \textit{disabled}, outperforms MLlib by a factor of
$5$ or more. This means we can often use fractions of the hardware required
by MLlib to perform equivalent tasks. Figure \ref{fig:dist-perf} demonstrates
that \textsf{knord} also benefits from our in-memory NUMA optimizations as
we outperform a NUMA-oblivious MPI routine by $20$-$50$\%,
depending on the dataset. Finally, Figure \ref{fig:dist-perf} shows that
MTI remains a low-overhead, effective method to reduce computation
even in the distributed setting.

\subsubsection{Semi-External Memory in the Cloud}

We conclude our experiments by measuring the performance of \textsf{knors} on a
single 32 core i3.16xlarge machine with 8 SSDs on Amazon EC2 compared
to \textsf{knord}, MLlib and an optimized MPI routine running in a cluster.
We run \textsf{knors} with 48 threads, with extra parallelism coming
from symmetric multiprocessing.
We run all other implementations with the same number of processes/threads as physical cores.

Figure \ref{fig:sem-ec2} highlights that \textsf{knors} often outperforms MLlib
even when MLLib runs in a cluster that contains more physical CPU cores.
\textsf{knors} has comparable performance to both MPI and \textsf{knord},
leading to our assertion that the SEM scale-up model should be considered
prior to moving to the distributed setting.

\begin{figure}[!htb]
\centering
\footnotesize
\include{./charts/perf.iter.sem-ec2}
\vspace{-15pt}
\caption{Performance comparison of \textsf{knors} to distributed packages.
\textsf{knors} uses one i3.16xlarge machine with 32 physical cores.
\textsf{knord}, MLlib-EC2 and MPI use 3 c4.8xlarge with a total of 48 physical
cores for all datasets other than RU$_{1B}$ where they use 8 c4.8xlarge with a
total of 128 physical cores.}
\label{fig:sem-ec2}
\vspace{-5pt}
\end{figure}

%% file: im-opt-eval.tex
\subsection{Baseline Single-thread Performance} \label{sec:baseline}

\textsf{knori}, even with MTI pruning \textit{disabled},
performs on par with state-of-the-art implementations of Lloyd's
algorithm. This is true for implementations that utilize generalized
matrix multiplication (GEMM) techniques and vectorized operations, such as
MATLAB \cite{matlab} and BLAS \cite{BLAS}.
We find the same to be true of popular statistics
packages and frameworks such as MLpack \cite{mlpack},
Scikit-learn \cite{sklearn} and R
\cite{rpackage} all of which use highly optimized C/C++ code,
although some use scripting language wrappers.
Table \ref{tbl:baseline} shows performance at 1 thread. Table
\ref{tbl:baseline} provides credence to our speedup results since our baseline
single threaded performance tops other state-of-the-art serial routines.

\begin{table}[!htb]
\caption{Serial performance of popular, optimized k-means routines,
		all using Lloyd's algorithm, on the Friendster-8 dataset.
		For fairness all implementations perform all distance computations.
        The \textbf{Language} column refers to the underlying language of
        implementation and not any user-facing higher level wrapper.}
\vspace{-10pt}
\begin{center}
\footnotesize
\resizebox{.475\textwidth}{!}{\begin{tabular}{|c|c|c|c|}
\hline
\textbf{Implementation} & \textbf{Type} & \textbf{Language} & \textbf{Time/iter (sec)}\\
\hline
\textbf{\textsf{knori}} & \textbf{Iterative} & \textbf{C++} & \textbf{7.49} \\
\hline
MATLAB & GEMM & C++ & 20.68 \\
\hline
BLAS & GEMM & C++ & 20.7 \\
\hline
R & Iterative & C & 8.63 \\
\hline
Scikit-learn & Iterative & Cython & 12.84 \\
\hline
MLpack & Iterative & C++ & 13.09 \\
\hline
\end{tabular}}
\normalsize
\end{center}
\label{tbl:baseline}
\end{table}

\subsection{In-memory Optimization Evaluation} \label{sec:im-opt-eval}

We show NUMA-node thread binding, maintaining
NUMA memory locality, and NUMA-aware task scheduling for \textsf{knori} is
highly effective in improving performance.
We achieve near-linear speedup (Figure \ref{fig:numa-mem-eval}).
Because the machine has $48$ physical cores, speedup degrades slightly at 64 cores;
additional speedup beyond 48 cores comes from simultaneous multithreading (hyperthreading).
The NUMA-aware implementation is nearly $6$x faster at $64$ threads
compared to a routine containing no NUMA optimizations, henceforth referred to as
\textit{NUMA-oblivious}. The NUMA-oblivious routine relies on the OS
to determine memory allocation, thread scheduling, and load
balancing policies.

We further show that although both the NUMA-oblivious and NUMA-aware
implementation speedup linearly, the NUMA-oblivious routine has a lower linear
constant when compared with a NUMA-aware implementation (Figure \ref{fig:numa-mem-eval}).

Increased parallelism amplifies the performance degradation of
the NUMA-oblivious implementation. We identify the following as the greatest
contributors:

\begin{itemize}
\item the NUMA-oblivious allocation policies of traditional memory
allocators, such as \texttt{malloc}, place data in a contiguous
chunk within a single NUMA memory bank whenever possible. This
leads to a large number of threads performing remote memory accesses
as $T$ increases;
\item a dynamic NUMA-oblivious task scheduler may give tasks to threads
that cause worker threads to perform many more remote memory accesses than
necessary when thread-binding and static scheduling are employed.
\end{itemize}

\begin{figure}[!htb]
	\begin{center}
		\small
		\vspace{-15pt}
		\include{./charts/numa.speedup.friendster-8}
		\vspace{-15pt}
		\caption{Speedup of \textsf{knori} (which is NUMA-aware) vs.
			a NUMA-oblivious routine for on the Friendster top-8
			eigenvector dataset, with $k=10$.}
		\label{fig:numa-mem-eval}
	\end{center}
\end{figure}

We demonstrate the effectiveness of a
NUMA-aware partitioned task scheduler for pruned computations
via \textsf{knori} (Figure \ref{fig:sched-eval}).
We define a \textit{task} as a block of
data points in contiguous memory given to a thread for computation.
We set a minimum \textit{task size}, i.e. the number of data points in the block,
to $8192$. We empirically determine that this task size is small enough
to not artificially introduce skew in billion-point datasets.
We compare against a static and a first in, first out (FIFO) task scheduler.
The static scheduler preassigns $n/T$ rows
to each worker thread. The FIFO scheduler first assigns
threads to tasks that are local to the thread's partition
of data, then allows threads to steal tasks from straggler
threads whose data resides on any NUMA node.

We observe that as $k$ increases, so
does the potential for skew. When $k=10$, the NUMA-aware
scheduler performs negligibly worse than both FIFO and static
static scheduling, but as $k$, increases the NUMA-aware scheduler improves
performance---by more than $40\%$ when $k = 100$.
We observe similar trends in other datasets; we omit these
results for space reasons and to avoid redundancy.

\begin{figure}[!htb]
	\begin{center}
		\small
		\vspace{-15pt}
		\include{./charts/sched.friendster8}
		\vspace{-15pt}
		\caption{Performance of the partitioned NUMA-aware
            scheduler (default for \textsf{knori}) vs. FIFO and static scheduling
            for \textsf{knori} on the Friendster-8 dataset.}
		\label{fig:sched-eval}
	\end{center}
    \vspace{-15pt}
\end{figure}

%% file: charts/numa.speedup.friendster-8.tex
\begin{tikzpicture}[gnuplot]
\path (0.000,0.000) rectangle (8.382,5.334);
\gpcolor{color=gp lt color border}
\gpsetlinetype{gp lt border}
\gpsetlinewidth{1.00}
\draw[gp path] (1.320,0.985)--(1.500,0.985);
\draw[gp path] (7.829,0.985)--(7.649,0.985);
\node[gp node right] at (1.136,0.985) { 1};
\draw[gp path] (1.320,1.494)--(1.500,1.494);
\draw[gp path] (7.829,1.494)--(7.649,1.494);
\node[gp node right] at (1.136,1.494) { 2};
\draw[gp path] (1.320,2.004)--(1.500,2.004);
\draw[gp path] (7.829,2.004)--(7.649,2.004);
\node[gp node right] at (1.136,2.004) { 4};
\draw[gp path] (1.320,2.513)--(1.500,2.513);
\draw[gp path] (7.829,2.513)--(7.649,2.513);
\node[gp node right] at (1.136,2.513) { 8};
\draw[gp path] (1.320,3.022)--(1.500,3.022);
\draw[gp path] (7.829,3.022)--(7.649,3.022);
\node[gp node right] at (1.136,3.022) { 16};
\draw[gp path] (1.320,3.532)--(1.500,3.532);
\draw[gp path] (7.829,3.532)--(7.649,3.532);
\node[gp node right] at (1.136,3.532) { 32};
\draw[gp path] (1.320,4.041)--(1.500,4.041);
\draw[gp path] (7.829,4.041)--(7.649,4.041);
\node[gp node right] at (1.136,4.041) { 64};
\draw[gp path] (1.320,0.985)--(1.320,1.165);
\draw[gp path] (1.320,4.041)--(1.320,3.861);
\node[gp node center] at (1.320,0.677) {1};
\draw[gp path] (2.405,0.985)--(2.405,1.165);
\draw[gp path] (2.405,4.041)--(2.405,3.861);
\node[gp node center] at (2.405,0.677) {2};
\draw[gp path] (3.490,0.985)--(3.490,1.165);
\draw[gp path] (3.490,4.041)--(3.490,3.861);
\node[gp node center] at (3.490,0.677) {4};
\draw[gp path] (4.575,0.985)--(4.575,1.165);
\draw[gp path] (4.575,4.041)--(4.575,3.861);
\node[gp node center] at (4.575,0.677) {8};
\draw[gp path] (5.659,0.985)--(5.659,1.165);
\draw[gp path] (5.659,4.041)--(5.659,3.861);
\node[gp node center] at (5.659,0.677) {16};
\draw[gp path] (6.744,0.985)--(6.744,1.165);
\draw[gp path] (6.744,4.041)--(6.744,3.861);
\node[gp node center] at (6.744,0.677) {32};
\draw[gp path] (7.829,0.985)--(7.829,1.165);
\draw[gp path] (7.829,4.041)--(7.829,3.861);
\node[gp node center] at (7.829,0.677) {64};
\draw[gp path] (1.320,4.041)--(1.320,0.985)--(7.829,0.985)--(7.829,4.041)--cycle;
\node[gp node center,rotate=-270] at (0.246,2.513) {Relative Performance};
\node[gp node center] at (4.574,0.215) {No. of Threads};
\node[gp node right] at (3.290,5.000) {\textsf{knori}};
\gpcolor{rgb color={0.000,1.000,0.000}}
\gpsetlinetype{gp lt plot 0}
\draw[gp path] (3.474,5.000)--(4.390,5.000);
\draw[gp path] (1.320,0.985)--(2.405,1.495)--(3.490,1.865)--(4.575,2.335)--(5.659,2.885)%
  --(6.744,3.318)--(7.829,3.634);
\gpsetpointsize{4.00}
\gppoint{gp mark 1}{(1.320,0.985)}
\gppoint{gp mark 1}{(2.405,1.495)}
\gppoint{gp mark 1}{(3.490,1.865)}
\gppoint{gp mark 1}{(4.575,2.335)}
\gppoint{gp mark 1}{(5.659,2.885)}
\gppoint{gp mark 1}{(6.744,3.318)}
\gppoint{gp mark 1}{(7.829,3.634)}
\gppoint{gp mark 1}{(3.932,5.000)}
\gpcolor{color=gp lt color border}
\node[gp node right] at (3.290,4.692) {NUMA-oblivious};
\gpcolor{color=gp lt color 3}
\draw[gp path] (3.474,4.692)--(4.390,4.692);
\draw[gp path] (1.320,0.985)--(2.405,1.023)--(3.490,1.198)--(4.575,1.408)--(5.659,1.489)%
  --(6.744,2.102)--(7.829,2.463);
\gppoint{gp mark 1}{(1.320,0.985)}
\gppoint{gp mark 1}{(2.405,1.023)}
\gppoint{gp mark 1}{(3.490,1.198)}
\gppoint{gp mark 1}{(4.575,1.408)}
\gppoint{gp mark 1}{(5.659,1.489)}
\gppoint{gp mark 1}{(6.744,2.102)}
\gppoint{gp mark 1}{(7.829,2.463)}
\gppoint{gp mark 1}{(3.932,4.692)}
\gpcolor{color=gp lt color border}
\node[gp node right] at (7.150,5.000) {Linear (Ideal)};
\gpcolor{color=gp lt color 4}
\gpsetlinetype{gp lt plot 1}
\draw[gp path] (7.334,5.000)--(8.250,5.000);
\draw[gp path] (1.320,0.985)--(2.405,1.494)--(3.490,2.004)--(4.575,2.513)--(5.659,3.022)%
  --(6.744,3.532)--(7.829,4.041);
\gppoint{gp mark 2}{(1.320,0.985)}
\gppoint{gp mark 2}{(2.405,1.494)}
\gppoint{gp mark 2}{(3.490,2.004)}
\gppoint{gp mark 2}{(4.575,2.513)}
\gppoint{gp mark 2}{(5.659,3.022)}
\gppoint{gp mark 2}{(6.744,3.532)}
\gppoint{gp mark 2}{(7.829,4.041)}
\gppoint{gp mark 2}{(7.792,5.000)}
\gpcolor{color=gp lt color border}
\gpsetlinetype{gp lt border}
\draw[gp path] (1.320,4.041)--(1.320,0.985)--(7.829,0.985)--(7.829,4.041)--cycle;
\gpdefrectangularnode{gp plot 1}{\pgfpoint{1.320cm}{0.985cm}}{\pgfpoint{7.829cm}{4.041cm}}
\end{tikzpicture}

%% file: charts/sched.friendster8.tex
\begin{tikzpicture}[gnuplot]
\path (0.000,0.000) rectangle (8.382,4.826);
\gpcolor{color=gp lt color border}
\gpsetlinetype{gp lt border}
\gpsetlinewidth{1.00}
\draw[gp path] (1.504,0.616)--(1.684,0.616);
\draw[gp path] (7.829,0.616)--(7.649,0.616);
\node[gp node right] at (1.320,0.616) { 60};
\draw[gp path] (1.504,0.909)--(1.684,0.909);
\draw[gp path] (7.829,0.909)--(7.649,0.909);
\node[gp node right] at (1.320,0.909) { 80};
\draw[gp path] (1.504,1.202)--(1.684,1.202);
\draw[gp path] (7.829,1.202)--(7.649,1.202);
\node[gp node right] at (1.320,1.202) { 100};
\draw[gp path] (1.504,1.496)--(1.684,1.496);
\draw[gp path] (7.829,1.496)--(7.649,1.496);
\node[gp node right] at (1.320,1.496) { 120};
\draw[gp path] (1.504,1.789)--(1.684,1.789);
\draw[gp path] (7.829,1.789)--(7.649,1.789);
\node[gp node right] at (1.320,1.789) { 140};
\draw[gp path] (1.504,2.082)--(1.684,2.082);
\draw[gp path] (7.829,2.082)--(7.649,2.082);
\node[gp node right] at (1.320,2.082) { 160};
\draw[gp path] (1.504,2.375)--(1.684,2.375);
\draw[gp path] (7.829,2.375)--(7.649,2.375);
\node[gp node right] at (1.320,2.375) { 180};
\draw[gp path] (1.504,2.668)--(1.684,2.668);
\draw[gp path] (7.829,2.668)--(7.649,2.668);
\node[gp node right] at (1.320,2.668) { 200};
\draw[gp path] (1.504,2.961)--(1.684,2.961);
\draw[gp path] (7.829,2.961)--(7.649,2.961);
\node[gp node right] at (1.320,2.961) { 220};
\draw[gp path] (1.504,3.255)--(1.684,3.255);
\draw[gp path] (7.829,3.255)--(7.649,3.255);
\node[gp node right] at (1.320,3.255) { 240};
\draw[gp path] (1.504,3.548)--(1.684,3.548);
\draw[gp path] (7.829,3.548)--(7.649,3.548);
\node[gp node right] at (1.320,3.548) { 260};
\draw[gp path] (1.504,3.841)--(1.684,3.841);
\draw[gp path] (7.829,3.841)--(7.649,3.841);
\node[gp node right] at (1.320,3.841) { 280};
\draw[gp path] (2.769,0.616)--(2.769,0.796);
\draw[gp path] (2.769,3.841)--(2.769,3.661);
\node[gp node center] at (2.769,0.308) {k=10};
\draw[gp path] (4.034,0.616)--(4.034,0.796);
\draw[gp path] (4.034,3.841)--(4.034,3.661);
\node[gp node center] at (4.034,0.308) {k=20};
\draw[gp path] (5.299,0.616)--(5.299,0.796);
\draw[gp path] (5.299,3.841)--(5.299,3.661);
\node[gp node center] at (5.299,0.308) {k=50};
\draw[gp path] (6.564,0.616)--(6.564,0.796);
\draw[gp path] (6.564,3.841)--(6.564,3.661);
\node[gp node center] at (6.564,0.308) {k=100};
\draw[gp path] (1.504,3.841)--(1.504,0.616)--(7.829,0.616)--(7.829,3.841)--cycle;
\node[gp node center,rotate=-270] at (0.246,2.228) {Time/iter (msec)};
\node[gp node right] at (2.188,4.492) {\textsf{knori}};
\gpfill{rgb color={0.000,1.000,0.000},color=.!50} (2.372,4.415)--(3.288,4.415)--(3.288,4.569)--(2.372,4.569)--cycle;
\gpfill{rgb color={0.000,1.000,0.000},color=.!50} (2.516,0.616)--(2.770,0.616)--(2.770,1.020)--(2.516,1.020)--cycle;
\gpfill{rgb color={0.000,1.000,0.000},color=.!50} (3.781,0.616)--(4.035,0.616)--(4.035,0.872)--(3.781,0.872)--cycle;
\gpfill{rgb color={0.000,1.000,0.000},color=.!50} (5.046,0.616)--(5.300,0.616)--(5.300,1.747)--(5.046,1.747)--cycle;
\gpfill{rgb color={0.000,1.000,0.000},color=.!50} (6.311,0.616)--(6.565,0.616)--(6.565,2.480)--(6.311,2.480)--cycle;
\node[gp node right] at (4.576,4.492) {FIFO};
\def\gpfillpath{(4.760,4.415)--(5.676,4.415)--(5.676,4.569)--(4.760,4.569)--cycle}
\gpfill{color=gpbgfillcolor} \gpfillpath;
\gpfill{rgb color={0.000,0.000,1.000},gp pattern 1,pattern color=.} \gpfillpath;
\gpcolor{rgb color={0.000,0.000,1.000}}
\gpsetlinetype{gp lt plot 1}
\draw[gp path] (4.760,4.415)--(5.676,4.415)--(5.676,4.569)--(4.760,4.569)--cycle;
\def\gpfillpath{(2.769,0.616)--(3.023,0.616)--(3.023,0.918)--(2.769,0.918)--cycle}
\gpfill{color=gpbgfillcolor} \gpfillpath;
\gpfill{rgb color={0.000,0.000,1.000},gp pattern 1,pattern color=.} \gpfillpath;
\draw[gp path] (2.769,0.616)--(2.769,0.917)--(3.022,0.917)--(3.022,0.616)--cycle;
\def\gpfillpath{(4.034,0.616)--(4.288,0.616)--(4.288,1.099)--(4.034,1.099)--cycle}
\gpfill{color=gpbgfillcolor} \gpfillpath;
\gpfill{rgb color={0.000,0.000,1.000},gp pattern 1,pattern color=.} \gpfillpath;
\draw[gp path] (4.034,0.616)--(4.034,1.098)--(4.287,1.098)--(4.287,0.616)--cycle;
\def\gpfillpath{(5.299,0.616)--(5.553,0.616)--(5.553,2.230)--(5.299,2.230)--cycle}
\gpfill{color=gpbgfillcolor} \gpfillpath;
\gpfill{rgb color={0.000,0.000,1.000},gp pattern 1,pattern color=.} \gpfillpath;
\draw[gp path] (5.299,0.616)--(5.299,2.229)--(5.552,2.229)--(5.552,0.616)--cycle;
\def\gpfillpath{(6.564,0.616)--(6.818,0.616)--(6.818,2.974)--(6.564,2.974)--cycle}
\gpfill{color=gpbgfillcolor} \gpfillpath;
\gpfill{rgb color={0.000,0.000,1.000},gp pattern 1,pattern color=.} \gpfillpath;
\draw[gp path] (6.564,0.616)--(6.564,2.973)--(6.817,2.973)--(6.817,0.616)--cycle;
\gpcolor{color=gp lt color border}
\node[gp node right] at (6.964,4.492) {Static};
\def\gpfillpath{(7.148,4.415)--(8.064,4.415)--(8.064,4.569)--(7.148,4.569)--cycle}
\gpfill{color=gpbgfillcolor} \gpfillpath;
\gpfill{rgb color={1.000,0.000,0.000},gp pattern 2,pattern color=.} \gpfillpath;
\gpcolor{rgb color={1.000,0.000,0.000}}
\gpsetlinetype{gp lt plot 2}
\draw[gp path] (7.148,4.415)--(8.064,4.415)--(8.064,4.569)--(7.148,4.569)--cycle;
\def\gpfillpath{(3.022,0.616)--(3.276,0.616)--(3.276,0.910)--(3.022,0.910)--cycle}
\gpfill{color=gpbgfillcolor} \gpfillpath;
\gpfill{rgb color={1.000,0.000,0.000},gp pattern 2,pattern color=.} \gpfillpath;
\draw[gp path] (3.022,0.616)--(3.022,0.909)--(3.275,0.909)--(3.275,0.616)--cycle;
\def\gpfillpath{(4.287,0.616)--(4.541,0.616)--(4.541,1.158)--(4.287,1.158)--cycle}
\gpfill{color=gpbgfillcolor} \gpfillpath;
\gpfill{rgb color={1.000,0.000,0.000},gp pattern 2,pattern color=.} \gpfillpath;
\draw[gp path] (4.287,0.616)--(4.287,1.157)--(4.540,1.157)--(4.540,0.616)--cycle;
\def\gpfillpath{(5.552,0.616)--(5.806,0.616)--(5.806,2.551)--(5.552,2.551)--cycle}
\gpfill{color=gpbgfillcolor} \gpfillpath;
\gpfill{rgb color={1.000,0.000,0.000},gp pattern 2,pattern color=.} \gpfillpath;
\draw[gp path] (5.552,0.616)--(5.552,2.550)--(5.805,2.550)--(5.805,0.616)--cycle;
\def\gpfillpath{(6.817,0.616)--(7.071,0.616)--(7.071,3.643)--(6.817,3.643)--cycle}
\gpfill{color=gpbgfillcolor} \gpfillpath;
\gpfill{rgb color={1.000,0.000,0.000},gp pattern 2,pattern color=.} \gpfillpath;
\draw[gp path] (6.817,0.616)--(6.817,3.642)--(7.070,3.642)--(7.070,0.616)--cycle;
\gpcolor{color=gp lt color border}
\gpsetlinetype{gp lt border}
\draw[gp path] (1.504,3.841)--(1.504,0.616)--(7.829,0.616)--(7.829,3.841)--cycle;
\gpdefrectangularnode{gp plot 1}{\pgfpoint{1.504cm}{0.616cm}}{\pgfpoint{7.829cm}{3.841cm}}
\end{tikzpicture}

%% file: sem-opt-eval.tex
\subsection{Semi-external Memory Evaluation}
\label{sec:sem-opt-eval}

We evaluate \textsf{knors} optimizations, performance and scalability.
\textsf{knors} utilizes a $4$KB FlashGraph \textit{page cache} size, minimizing the
amount of superfluous data read from disk due to data fragmentation.
Additionally, we disable checkpoint failure recovery during
performance evaluation for both our routines and those of our competitors.

\begin{figure}[!htb]
	\centering
	\footnotesize
	\begin{subfigure}{.5\textwidth}
		\include{./charts/io.per.iter}
		\caption{\textsf{knors} data requested (req) from SSDs vs. data read (read)
        from SSDs each iteration when the row cache (RC) was \textit{enabled}
        or \textit{disabled}. Because of MTI pruning, algorithms may request
        only a few points from any block, but the entire block must still
        be read from SSD.}
		\label{fig:io-by-iter}
	\end{subfigure}

\begin{subfigure}{.5\textwidth}
	\include{./charts/tot.io}
    \caption{Total data requested (req) vs. data read from SSDs when
	(i) both MTI and RC are \textit{disabled} (\textsf{knors}-{}-),
	(ii) Only MTI is \textit{enabled} (\textsf{knors}-),
	(iii) both MTI and RC are \textit{enabled} (\textsf{knors}).
		Without pruning, all data are requested and read.}
	\label{fig:tot-io}
\end{subfigure}

\caption{The effect of the row cache and MTI on I/O for the Friendster
    top-32 eigenvectors dataset. Row cache size = $512$MB, page
    cache size = $1$GB, $k=10$.}
\label{fig:io}
\end{figure}

We drastically reduce the amount of data read from SSDs by utilizing
the row cache.
Figure \ref{fig:io-by-iter} shows that as the number of iterations increase,
the row cache's ability to reduce I/O and improve speed also
increases because most rows that are active are pinned in memory.
Figure \ref{fig:tot-io} contrasts the total amount of data that an implementation
requests from SSDs with the amount of data SAFS actually reads and
transports into memory. When \textsf{knors} \textit{disables} both
MTI pruning and the row cache (i.e., \textsf{knors-{}-}, every
request issued to FlashGraph for row-data
is either served by FlashGraph's page cache or
read from SSDs. When \textsf{knors} \textit{enables} MTI pruning,
but \textit{disables} the row cache (i.e., \textsf{knors-}, we read
an order of magnitude more data from SSDs than
when we \textit{enable} the row cache.
Figure \ref{fig:io} demonstrates that a page cache is \textbf{not}
sufficient for k-means
and that caching at the granularity of row-data is necessary to achieve
significant reductions in I/O and improvements in performance for real-world
datasets.

\begin{figure}[!htb]
	\begin{center}
		\small
		\vspace{-15pt}
		\include{./charts/cache.hits}
		\vspace{-15pt}
    \caption{Row cache hits per iteration contrasted
      with the maximum achievable number of hits on
        the Friendster top-32 eigenvectors dataset.}
		\label{fig:cache-hits-by-iter}
	\end{center}
\end{figure}

Lazy row cache updates reduce I/O significantly.
Figure \ref{fig:cache-hits-by-iter} justifies our design decision
for a lazily updated row cache. As the algorithm progresses,
we obtain nearly a $100\%$ cache hit rate, meaning that \textsf{knors} operates
at in-memory speeds for the vast majority of iterations.

%% file: charts/io.per.iter.tex
\begin{tikzpicture}[gnuplot]
\path (0.000,0.000) rectangle (8.382,4.572);
\gpcolor{color=gp lt color border}
\gpsetlinetype{gp lt border}
\gpsetlinewidth{1.00}
\draw[gp path] (1.688,1.025)--(1.868,1.025);
\draw[gp path] (7.829,1.025)--(7.649,1.025);
\node[gp node right] at (1.504,1.025) {0.016};
\draw[gp path] (1.688,1.251)--(1.778,1.251);
\draw[gp path] (7.829,1.251)--(7.739,1.251);
\draw[gp path] (1.688,1.476)--(1.868,1.476);
\draw[gp path] (7.829,1.476)--(7.649,1.476);
\node[gp node right] at (1.504,1.476) {0.062};
\draw[gp path] (1.688,1.701)--(1.778,1.701);
\draw[gp path] (7.829,1.701)--(7.739,1.701);
\draw[gp path] (1.688,1.927)--(1.868,1.927);
\draw[gp path] (7.829,1.927)--(7.649,1.927);
\node[gp node right] at (1.504,1.927) {0.25};
\draw[gp path] (1.688,2.152)--(1.778,2.152);
\draw[gp path] (7.829,2.152)--(7.739,2.152);
\draw[gp path] (1.688,2.378)--(1.868,2.378);
\draw[gp path] (7.829,2.378)--(7.649,2.378);
\node[gp node right] at (1.504,2.378) {1};
\draw[gp path] (1.688,2.603)--(1.778,2.603);
\draw[gp path] (7.829,2.603)--(7.739,2.603);
\draw[gp path] (1.688,2.828)--(1.868,2.828);
\draw[gp path] (7.829,2.828)--(7.649,2.828);
\node[gp node right] at (1.504,2.828) {4};
\draw[gp path] (1.688,3.054)--(1.778,3.054);
\draw[gp path] (7.829,3.054)--(7.739,3.054);
\draw[gp path] (1.688,3.279)--(1.868,3.279);
\draw[gp path] (7.829,3.279)--(7.649,3.279);
\node[gp node right] at (1.504,3.279) {16};
\draw[gp path] (1.688,0.985)--(1.688,1.165);
\draw[gp path] (1.688,3.279)--(1.688,3.099);
\node[gp node center] at (1.688,0.677) { 0};
\draw[gp path] (2.892,0.985)--(2.892,1.165);
\draw[gp path] (2.892,3.279)--(2.892,3.099);
\node[gp node center] at (2.892,0.677) { 20};
\draw[gp path] (4.096,0.985)--(4.096,1.165);
\draw[gp path] (4.096,3.279)--(4.096,3.099);
\node[gp node center] at (4.096,0.677) { 40};
\draw[gp path] (5.300,0.985)--(5.300,1.165);
\draw[gp path] (5.300,3.279)--(5.300,3.099);
\node[gp node center] at (5.300,0.677) { 60};
\draw[gp path] (6.504,0.985)--(6.504,1.165);
\draw[gp path] (6.504,3.279)--(6.504,3.099);
\node[gp node center] at (6.504,0.677) { 80};
\draw[gp path] (7.709,0.985)--(7.709,1.165);
\draw[gp path] (7.709,3.279)--(7.709,3.099);
\node[gp node center] at (7.709,0.677) { 100};
\draw[gp path] (1.688,3.279)--(1.688,0.985)--(7.829,0.985)--(7.829,3.279)--cycle;
\node[gp node center,rotate=-270] at (0.246,2.132) {Data (GB)};
\node[gp node center] at (4.758,0.215) {Iteration No.};
\node[gp node right] at (3.474,4.238) {No RC req};
\gpcolor{rgb color={1.000,0.000,0.000}}
\gpsetlinetype{gp lt plot 2}
\draw[gp path] (3.658,4.238)--(4.574,4.238);
\draw[gp path] (1.748,3.272)--(1.808,2.586)--(1.869,2.422)--(1.929,2.338)--(1.989,2.274)%
  --(2.049,2.231)--(2.109,2.229)--(2.170,2.225)--(2.230,2.217)--(2.290,2.206)--(2.350,2.194)%
  --(2.410,2.181)--(2.471,2.169)--(2.531,2.158)--(2.591,2.147)--(2.651,2.136)--(2.712,2.125)%
  --(2.772,2.131)--(2.832,2.137)--(2.892,2.143)--(2.952,2.149)--(3.013,2.154)--(3.073,2.160)%
  --(3.133,2.165)--(3.193,2.169)--(3.253,2.173)--(3.314,2.176)--(3.374,2.178)--(3.434,2.180)%
  --(3.494,2.181)--(3.554,2.182)--(3.615,2.184)--(3.675,2.185)--(3.735,2.186)--(3.795,2.187)%
  --(3.855,2.188)--(3.916,2.188)--(3.976,2.189)--(4.036,2.190)--(4.096,2.190)--(4.156,2.191)%
  --(4.217,2.191)--(4.277,2.192)--(4.337,2.192)--(4.397,2.192)--(4.457,2.192)--(4.518,2.192)%
  --(4.578,2.192)--(4.638,2.192)--(4.698,2.192)--(4.759,2.192)--(4.819,2.192)--(4.879,2.192)%
  --(4.939,2.192)--(4.999,2.192)--(5.060,2.191)--(5.120,2.191)--(5.180,2.191)--(5.240,2.191)%
  --(5.300,2.191)--(5.361,2.191)--(5.421,2.191)--(5.481,2.191)--(5.541,2.191)--(5.601,2.191)%
  --(5.662,2.191)--(5.722,2.191)--(5.782,2.190)--(5.842,2.190)--(5.902,2.190)--(5.963,2.190)%
  --(6.023,2.190)--(6.083,2.190)--(6.143,2.190)--(6.203,2.190)--(6.264,2.190)--(6.324,2.190)%
  --(6.384,2.190)--(6.444,2.190)--(6.504,2.190)--(6.565,2.190)--(6.625,2.189)--(6.685,2.189)%
  --(6.745,2.189)--(6.806,2.189)--(6.866,2.189)--(6.926,2.189)--(6.986,2.189)--(7.046,2.189)%
  --(7.107,2.189)--(7.167,2.189)--(7.227,2.189)--(7.287,2.189)--(7.347,2.189)--(7.408,2.189)%
  --(7.468,2.189)--(7.528,2.189)--(7.588,2.189)--(7.648,2.189)--(7.709,2.189)--(7.769,2.189)%
  --(7.829,2.189);
\gpcolor{color=gp lt color border}
\node[gp node right] at (3.474,3.930) {No RC read};
\gpcolor{rgb color={1.000,0.000,0.000}}
\gpsetlinetype{gp lt plot 0}
\draw[gp path] (3.658,3.930)--(4.574,3.930);
\draw[gp path] (1.748,3.272)--(1.808,3.180)--(1.869,3.066)--(1.929,2.984)--(1.989,2.914)%
  --(2.049,2.864)--(2.109,2.861)--(2.170,2.856)--(2.230,2.846)--(2.290,2.833)--(2.350,2.818)%
  --(2.410,2.802)--(2.471,2.787)--(2.531,2.772)--(2.591,2.757)--(2.651,2.743)--(2.712,2.729)%
  --(2.772,2.737)--(2.832,2.744)--(2.892,2.752)--(2.952,2.760)--(3.013,2.768)--(3.073,2.775)%
  --(3.133,2.781)--(3.193,2.786)--(3.253,2.791)--(3.314,2.795)--(3.374,2.798)--(3.434,2.800)%
  --(3.494,2.802)--(3.554,2.804)--(3.615,2.805)--(3.675,2.807)--(3.735,2.807)--(3.795,2.809)%
  --(3.855,2.810)--(3.916,2.811)--(3.976,2.812)--(4.036,2.812)--(4.096,2.813)--(4.156,2.814)%
  --(4.217,2.814)--(4.277,2.815)--(4.337,2.815)--(4.397,2.816)--(4.457,2.816)--(4.518,2.816)%
  --(4.578,2.816)--(4.638,2.815)--(4.698,2.815)--(4.759,2.815)--(4.819,2.815)--(4.879,2.815)%
  --(4.939,2.815)--(4.999,2.815)--(5.060,2.815)--(5.120,2.815)--(5.180,2.815)--(5.240,2.814)%
  --(5.300,2.815)--(5.361,2.814)--(5.421,2.814)--(5.481,2.814)--(5.541,2.814)--(5.601,2.814)%
  --(5.662,2.814)--(5.722,2.814)--(5.782,2.814)--(5.842,2.813)--(5.902,2.813)--(5.963,2.814)%
  --(6.023,2.813)--(6.083,2.813)--(6.143,2.813)--(6.203,2.813)--(6.264,2.813)--(6.324,2.813)%
  --(6.384,2.813)--(6.444,2.813)--(6.504,2.813)--(6.565,2.813)--(6.625,2.812)--(6.685,2.812)%
  --(6.745,2.812)--(6.806,2.812)--(6.866,2.812)--(6.926,2.812)--(6.986,2.812)--(7.046,2.812)%
  --(7.107,2.812)--(7.167,2.812)--(7.227,2.811)--(7.287,2.812)--(7.347,2.812)--(7.408,2.811)%
  --(7.468,2.811)--(7.528,2.811)--(7.588,2.811)--(7.648,2.812)--(7.709,2.812)--(7.769,2.811)%
  --(7.829,2.812);
\gpcolor{color=gp lt color border}
\node[gp node right] at (6.782,4.238) {\textsf{knors} req};
\gpcolor{rgb color={0.000,1.000,0.000}}
\gpsetlinetype{gp lt plot 2}
\draw[gp path] (6.966,4.238)--(7.882,4.238);
\draw[gp path] (1.748,3.271)--(1.808,2.577)--(1.869,2.406)--(1.929,2.318)--(1.989,2.249)%
  --(2.049,2.203)--(2.109,2.201)--(2.170,2.225)--(2.230,1.382)--(2.290,1.367)--(2.350,1.354)%
  --(2.410,1.339)--(2.471,1.324)--(2.531,1.310)--(2.591,1.297)--(2.651,1.284)--(2.712,1.270)%
  --(2.772,1.277)--(2.832,1.285)--(2.892,1.292)--(2.952,1.300)--(3.013,1.307)--(3.073,1.314)%
  --(3.133,1.320)--(3.193,1.325)--(3.253,1.330)--(3.314,1.333)--(3.374,1.336)--(3.434,1.338)%
  --(3.494,1.340)--(3.554,1.341)--(3.615,1.343)--(3.675,1.345)--(3.735,1.346)--(3.795,1.347)%
  --(3.855,1.348)--(3.916,1.349)--(3.976,2.189)--(4.036,1.023)--(4.096,1.039)--(4.156,1.060)%
  --(4.217,1.058)--(4.277,1.067)--(4.337,1.071)--(4.397,1.077)--(4.457,1.079)--(4.518,1.079)%
  --(4.578,1.076)--(4.638,1.072)--(4.698,1.069)--(4.759,1.069)--(4.819,1.069)--(4.879,1.067)%
  --(4.939,1.065)--(4.999,1.063)--(5.060,1.061)--(5.120,1.059)--(5.180,1.057)--(5.240,1.056)%
  --(5.300,1.054)--(5.361,1.052)--(5.421,1.050)--(5.481,1.049)--(5.541,1.046)--(5.601,1.045)%
  --(5.662,1.044)--(5.722,1.042)--(5.782,1.041)--(5.842,1.040)--(5.902,1.038)--(5.963,1.036)%
  --(6.023,1.035)--(6.083,1.033)--(6.143,1.031)--(6.203,1.029)--(6.264,1.028)--(6.324,1.026)%
  --(6.384,1.024)--(6.444,1.022)--(6.504,1.020)--(6.565,1.018)--(6.625,1.015)--(6.685,1.012)%
  --(6.745,1.010)--(6.806,1.008)--(6.866,1.007)--(6.926,1.005)--(6.986,1.003)--(7.046,1.001)%
  --(7.107,0.999)--(7.167,0.997)--(7.227,0.996)--(7.287,0.995)--(7.347,0.994)--(7.408,0.992)%
  --(7.468,0.991)--(7.528,0.990)--(7.588,0.990)--(7.648,0.988)--(7.709,0.987)--(7.769,0.986)%
  --(7.829,0.985);
\gpcolor{color=gp lt color border}
\node[gp node right] at (6.782,3.930) {\textsf{knors} read};
\gpcolor{rgb color={0.000,1.000,0.000}}
\gpsetlinetype{gp lt plot 0}
\draw[gp path] (6.966,3.930)--(7.882,3.930);
\draw[gp path] (1.748,3.272)--(1.808,3.176)--(1.869,3.053)--(1.929,2.964)--(1.989,2.886)%
  --(2.049,2.830)--(2.109,2.827)--(2.170,2.859)--(2.230,1.996)--(2.290,1.980)--(2.350,1.962)%
  --(2.410,1.942)--(2.471,1.923)--(2.531,1.904)--(2.591,1.885)--(2.651,1.867)--(2.712,1.849)%
  --(2.772,1.859)--(2.832,1.869)--(2.892,1.879)--(2.952,1.889)--(3.013,1.899)--(3.073,1.909)%
  --(3.133,1.917)--(3.193,1.924)--(3.253,1.929)--(3.314,1.935)--(3.374,1.938)--(3.434,1.941)%
  --(3.494,1.943)--(3.554,1.945)--(3.615,1.947)--(3.675,1.949)--(3.735,1.950)--(3.795,1.951)%
  --(3.855,1.953)--(3.916,1.954)--(3.976,2.813)--(4.036,1.565)--(4.096,1.569)--(4.156,1.575)%
  --(4.217,1.575)--(4.277,1.578)--(4.337,1.580)--(4.397,1.582)--(4.457,1.582)--(4.518,1.582)%
  --(4.578,1.581)--(4.638,1.580)--(4.698,1.579)--(4.759,1.579)--(4.819,1.579)--(4.879,1.579)%
  --(4.939,1.578)--(4.999,1.577)--(5.060,1.577)--(5.120,1.576)--(5.180,1.575)--(5.240,1.575)%
  --(5.300,1.575)--(5.361,1.574)--(5.421,1.573)--(5.481,1.572)--(5.541,1.572)--(5.601,1.571)%
  --(5.662,1.571)--(5.722,1.571)--(5.782,1.570)--(5.842,1.570)--(5.902,1.569)--(5.963,1.569)%
  --(6.023,1.569)--(6.083,1.568)--(6.143,1.567)--(6.203,1.567)--(6.264,1.566)--(6.324,1.566)%
  --(6.384,1.566)--(6.444,1.565)--(6.504,1.564)--(6.565,1.564)--(6.625,1.563)--(6.685,1.562)%
  --(6.745,1.562)--(6.806,1.562)--(6.866,1.561)--(6.926,1.561)--(6.986,1.560)--(7.046,1.559)%
  --(7.107,1.559)--(7.167,1.559)--(7.227,1.558)--(7.287,1.558)--(7.347,1.558)--(7.408,1.558)%
  --(7.468,1.558)--(7.528,1.557)--(7.588,1.557)--(7.648,1.557)--(7.709,1.557)--(7.769,1.556)%
  --(7.829,1.556);
\gpcolor{color=gp lt color border}
\gpsetlinetype{gp lt border}
\draw[gp path] (1.688,3.279)--(1.688,0.985)--(7.829,0.985)--(7.829,3.279)--cycle;
\gpdefrectangularnode{gp plot 1}{\pgfpoint{1.688cm}{0.985cm}}{\pgfpoint{7.829cm}{3.279cm}}
\end{tikzpicture}

%% file: charts/tot.io.tex
\begin{tikzpicture}[gnuplot]
\path (0.000,0.000) rectangle (8.382,4.826);
\gpcolor{color=gp lt color border}
\gpsetlinetype{gp lt border}
\gpsetlinewidth{1.00}
\draw[gp path] (1.688,0.616)--(1.868,0.616);
\draw[gp path] (7.829,0.616)--(7.649,0.616);
\node[gp node right] at (1.504,0.616) { 10};
\draw[gp path] (1.688,1.055)--(1.778,1.055);
\draw[gp path] (7.829,1.055)--(7.739,1.055);
\draw[gp path] (1.688,1.312)--(1.778,1.312);
\draw[gp path] (7.829,1.312)--(7.739,1.312);
\draw[gp path] (1.688,1.494)--(1.778,1.494);
\draw[gp path] (7.829,1.494)--(7.739,1.494);
\draw[gp path] (1.688,1.635)--(1.778,1.635);
\draw[gp path] (7.829,1.635)--(7.739,1.635);
\draw[gp path] (1.688,1.751)--(1.778,1.751);
\draw[gp path] (7.829,1.751)--(7.739,1.751);
\draw[gp path] (1.688,1.849)--(1.778,1.849);
\draw[gp path] (7.829,1.849)--(7.739,1.849);
\draw[gp path] (1.688,1.933)--(1.778,1.933);
\draw[gp path] (7.829,1.933)--(7.739,1.933);
\draw[gp path] (1.688,2.008)--(1.778,2.008);
\draw[gp path] (7.829,2.008)--(7.739,2.008);
\draw[gp path] (1.688,2.075)--(1.868,2.075);
\draw[gp path] (7.829,2.075)--(7.649,2.075);
\node[gp node right] at (1.504,2.075) { 100};
\draw[gp path] (1.688,2.514)--(1.778,2.514);
\draw[gp path] (7.829,2.514)--(7.739,2.514);
\draw[gp path] (1.688,2.770)--(1.778,2.770);
\draw[gp path] (7.829,2.770)--(7.739,2.770);
\draw[gp path] (1.688,2.953)--(1.778,2.953);
\draw[gp path] (7.829,2.953)--(7.739,2.953);
\draw[gp path] (1.688,3.094)--(1.778,3.094);
\draw[gp path] (7.829,3.094)--(7.739,3.094);
\draw[gp path] (1.688,3.209)--(1.778,3.209);
\draw[gp path] (7.829,3.209)--(7.739,3.209);
\draw[gp path] (1.688,3.307)--(1.778,3.307);
\draw[gp path] (7.829,3.307)--(7.739,3.307);
\draw[gp path] (1.688,3.392)--(1.778,3.392);
\draw[gp path] (7.829,3.392)--(7.739,3.392);
\draw[gp path] (1.688,3.466)--(1.778,3.466);
\draw[gp path] (7.829,3.466)--(7.739,3.466);
\draw[gp path] (1.688,3.533)--(1.868,3.533);
\draw[gp path] (7.829,3.533)--(7.649,3.533);
\node[gp node right] at (1.504,3.533) { 1000};
\draw[gp path] (3.735,0.616)--(3.735,0.796);
\draw[gp path] (3.735,3.533)--(3.735,3.353);
\node[gp node center] at (3.735,0.308) {Req I/O};
\draw[gp path] (5.782,0.616)--(5.782,0.796);
\draw[gp path] (5.782,3.533)--(5.782,3.353);
\node[gp node center] at (5.782,0.308) {Read from SSD};
\draw[gp path] (1.688,3.533)--(1.688,0.616)--(7.829,0.616)--(7.829,3.533)--cycle;
\node[gp node center,rotate=-270] at (0.246,2.074) {Log Scale Data (GB)};
\node[gp node right] at (3.474,4.492) {\textsf{knors}};
\gpfill{rgb color={0.000,1.000,0.000},color=.!50} (3.658,4.415)--(4.574,4.415)--(4.574,4.569)--(3.658,4.569)--cycle;
\gpfill{rgb color={0.000,1.000,0.000},color=.!50} (3.326,0.616)--(3.736,0.616)--(3.736,1.186)--(3.326,1.186)--cycle;
\gpfill{rgb color={0.000,1.000,0.000},color=.!50} (5.373,0.616)--(5.783,0.616)--(5.783,1.892)--(5.373,1.892)--cycle;
\node[gp node right] at (3.474,4.184) {\textsf{knors-}};
\def\gpfillpath{(3.658,4.107)--(4.574,4.107)--(4.574,4.261)--(3.658,4.261)--cycle}
\gpfill{color=gpbgfillcolor} \gpfillpath;
\gpfill{rgb color={0.000,0.000,1.000},gp pattern 1,pattern color=.} \gpfillpath;
\gpcolor{rgb color={0.000,0.000,1.000}}
\gpsetlinetype{gp lt plot 1}
\draw[gp path] (3.658,4.107)--(4.574,4.107)--(4.574,4.261)--(3.658,4.261)--cycle;
\def\gpfillpath{(3.735,0.616)--(4.145,0.616)--(4.145,1.885)--(3.735,1.885)--cycle}
\gpfill{color=gpbgfillcolor} \gpfillpath;
\gpfill{rgb color={0.000,0.000,1.000},gp pattern 1,pattern color=.} \gpfillpath;
\draw[gp path] (3.735,0.616)--(3.735,1.884)--(4.144,1.884)--(4.144,0.616)--cycle;
\def\gpfillpath{(5.782,0.616)--(6.192,0.616)--(6.192,2.972)--(5.782,2.972)--cycle}
\gpfill{color=gpbgfillcolor} \gpfillpath;
\gpfill{rgb color={0.000,0.000,1.000},gp pattern 1,pattern color=.} \gpfillpath;
\draw[gp path] (5.782,0.616)--(5.782,2.971)--(6.191,2.971)--(6.191,0.616)--cycle;
\gpcolor{color=gp lt color border}
\node[gp node right] at (6.230,4.492) {\textsf{knors-{}-}};
\def\gpfillpath{(6.414,4.415)--(7.330,4.415)--(7.330,4.569)--(6.414,4.569)--cycle}
\gpfill{color=gpbgfillcolor} \gpfillpath;
\gpfill{rgb color={1.000,0.000,0.000},gp pattern 2,pattern color=.} \gpfillpath;
\gpcolor{rgb color={1.000,0.000,0.000}}
\gpsetlinetype{gp lt plot 2}
\draw[gp path] (6.414,4.415)--(7.330,4.415)--(7.330,4.569)--(6.414,4.569)--cycle;
\def\gpfillpath{(4.144,0.616)--(4.555,0.616)--(4.555,3.227)--(4.144,3.227)--cycle}
\gpfill{color=gpbgfillcolor} \gpfillpath;
\gpfill{rgb color={1.000,0.000,0.000},gp pattern 2,pattern color=.} \gpfillpath;
\draw[gp path] (4.144,0.616)--(4.144,3.226)--(4.554,3.226)--(4.554,0.616)--cycle;
\def\gpfillpath{(6.191,0.616)--(6.602,0.616)--(6.602,3.226)--(6.191,3.226)--cycle}
\gpfill{color=gpbgfillcolor} \gpfillpath;
\gpfill{rgb color={1.000,0.000,0.000},gp pattern 2,pattern color=.} \gpfillpath;
\draw[gp path] (6.191,0.616)--(6.191,3.225)--(6.601,3.225)--(6.601,0.616)--cycle;
\gpcolor{color=gp lt color border}
\gpsetlinetype{gp lt border}
\draw[gp path] (1.688,3.533)--(1.688,0.616)--(7.829,0.616)--(7.829,3.533)--cycle;
\gpdefrectangularnode{gp plot 1}{\pgfpoint{1.688cm}{0.616cm}}{\pgfpoint{7.829cm}{3.533cm}}
\end{tikzpicture}

%% file: charts/cache.hits.tex
\begin{tikzpicture}[gnuplot]
\path (0.000,0.000) rectangle (8.382,4.572);
\gpcolor{color=gp lt color border}
\gpsetlinetype{gp lt border}
\gpsetlinewidth{1.00}
\draw[gp path] (1.136,0.985)--(1.316,0.985);
\draw[gp path] (7.829,0.985)--(7.649,0.985);
\node[gp node right] at (0.952,0.985) { 0};
\draw[gp path] (1.136,1.636)--(1.316,1.636);
\draw[gp path] (7.829,1.636)--(7.649,1.636);
\node[gp node right] at (0.952,1.636) { 2};
\draw[gp path] (1.136,2.286)--(1.316,2.286);
\draw[gp path] (7.829,2.286)--(7.649,2.286);
\node[gp node right] at (0.952,2.286) { 4};
\draw[gp path] (1.136,2.937)--(1.316,2.937);
\draw[gp path] (7.829,2.937)--(7.649,2.937);
\node[gp node right] at (0.952,2.937) { 6};
\draw[gp path] (1.136,3.587)--(1.316,3.587);
\draw[gp path] (7.829,3.587)--(7.649,3.587);
\node[gp node right] at (0.952,3.587) { 8};
\draw[gp path] (1.136,0.985)--(1.136,1.165);
\draw[gp path] (1.136,3.587)--(1.136,3.407);
\node[gp node center] at (1.136,0.677) { 0};
\draw[gp path] (2.448,0.985)--(2.448,1.165);
\draw[gp path] (2.448,3.587)--(2.448,3.407);
\node[gp node center] at (2.448,0.677) { 20};
\draw[gp path] (3.761,0.985)--(3.761,1.165);
\draw[gp path] (3.761,3.587)--(3.761,3.407);
\node[gp node center] at (3.761,0.677) { 40};
\draw[gp path] (5.073,0.985)--(5.073,1.165);
\draw[gp path] (5.073,3.587)--(5.073,3.407);
\node[gp node center] at (5.073,0.677) { 60};
\draw[gp path] (6.385,0.985)--(6.385,1.165);
\draw[gp path] (6.385,3.587)--(6.385,3.407);
\node[gp node center] at (6.385,0.677) { 80};
\draw[gp path] (7.698,0.985)--(7.698,1.165);
\draw[gp path] (7.698,3.587)--(7.698,3.407);
\node[gp node center] at (7.698,0.677) { 100};
\draw[gp path] (1.136,3.587)--(1.136,0.985)--(7.829,0.985)--(7.829,3.587)--cycle;
\node[gp node center,rotate=-270] at (0.246,2.286) {No. of points x $10^6$};
\node[gp node center] at (4.482,0.215) {Iteration No.};
\node[gp node right] at (3.198,4.238) {Cache hits};
\gpcolor{rgb color={1.000,0.647,0.000}}
\gpsetlinetype{gp lt plot 0}
\draw[gp path] (3.382,4.238)--(4.298,4.238);
\draw[gp path] (1.267,1.059)--(1.333,1.058)--(1.398,1.057)--(1.464,1.057)--(1.530,1.057)%
  --(1.595,1.056)--(1.661,0.985)--(1.727,1.753)--(1.792,1.729)--(1.858,1.702)--(1.923,1.676)%
  --(1.989,1.651)--(2.055,1.628)--(2.120,1.606)--(2.186,1.587)--(2.252,1.568)--(2.317,1.578)%
  --(2.383,1.588)--(2.448,1.599)--(2.514,1.610)--(2.580,1.621)--(2.645,1.632)--(2.711,1.642)%
  --(2.776,1.650)--(2.842,1.658)--(2.908,1.663)--(2.973,1.668)--(3.039,1.672)--(3.105,1.675)%
  --(3.170,1.677)--(3.236,1.680)--(3.301,1.683)--(3.367,1.685)--(3.433,1.686)--(3.498,1.688)%
  --(3.564,1.690)--(3.629,0.985)--(3.695,1.729)--(3.761,1.729)--(3.826,1.730)--(3.892,1.730)%
  --(3.958,1.731)--(4.023,1.731)--(4.089,1.731)--(4.154,1.731)--(4.220,1.731)--(4.286,1.731)%
  --(4.351,1.731)--(4.417,1.731)--(4.483,1.731)--(4.548,1.731)--(4.614,1.731)--(4.679,1.731)%
  --(4.745,1.731)--(4.811,1.731)--(4.876,1.731)--(4.942,1.731)--(5.007,1.731)--(5.073,1.731)%
  --(5.139,1.731)--(5.204,1.731)--(5.270,1.730)--(5.336,1.730)--(5.401,1.730)--(5.467,1.730)%
  --(5.532,1.730)--(5.598,1.730)--(5.664,1.730)--(5.729,1.730)--(5.795,1.730)--(5.860,1.730)%
  --(5.926,1.730)--(5.992,1.730)--(6.057,1.730)--(6.123,1.730)--(6.189,1.730)--(6.254,1.729)%
  --(6.320,1.729)--(6.385,1.729)--(6.451,1.729)--(6.517,1.729)--(6.582,1.729)--(6.648,1.729)%
  --(6.714,1.729)--(6.779,1.729)--(6.845,1.729)--(6.910,1.729)--(6.976,1.729)--(7.042,1.729)%
  --(7.107,1.729)--(7.173,1.729)--(7.238,1.729)--(7.304,1.729)--(7.370,1.729)--(7.435,1.729)%
  --(7.501,1.729)--(7.567,1.729)--(7.632,1.729)--(7.698,1.729)--(7.763,1.729)--(7.829,1.729);
\gpcolor{color=gp lt color border}
\node[gp node right] at (6.874,4.238) {Active points};
\gpcolor{rgb color={0.000,0.000,1.000}}
\gpsetlinetype{gp lt plot 1}
\draw[gp path] (7.058,4.238)--(7.974,4.238);
\draw[gp path] (1.267,3.576)--(1.333,2.549)--(1.398,2.192)--(1.464,1.977)--(1.530,1.855)%
  --(1.595,1.849)--(1.661,1.837)--(1.727,1.817)--(1.792,1.790)--(1.858,1.760)--(1.923,1.731)%
  --(1.989,1.704)--(2.055,1.679)--(2.120,1.656)--(2.186,1.634)--(2.252,1.613)--(2.317,1.624)%
  --(2.383,1.636)--(2.448,1.647)--(2.514,1.660)--(2.580,1.672)--(2.645,1.684)--(2.711,1.694)%
  --(2.776,1.704)--(2.842,1.712)--(2.908,1.718)--(2.973,1.724)--(3.039,1.728)--(3.105,1.731)%
  --(3.170,1.734)--(3.236,1.737)--(3.301,1.739)--(3.367,1.742)--(3.433,1.744)--(3.498,1.745)%
  --(3.564,1.747)--(3.629,1.749)--(3.695,1.750)--(3.761,1.752)--(3.826,1.754)--(3.892,1.754)%
  --(3.958,1.755)--(4.023,1.756)--(4.089,1.756)--(4.154,1.756)--(4.220,1.756)--(4.286,1.756)%
  --(4.351,1.756)--(4.417,1.755)--(4.483,1.755)--(4.548,1.755)--(4.614,1.755)--(4.679,1.755)%
  --(4.745,1.755)--(4.811,1.755)--(4.876,1.754)--(4.942,1.754)--(5.007,1.754)--(5.073,1.754)%
  --(5.139,1.754)--(5.204,1.754)--(5.270,1.753)--(5.336,1.753)--(5.401,1.753)--(5.467,1.753)%
  --(5.532,1.753)--(5.598,1.752)--(5.664,1.752)--(5.729,1.752)--(5.795,1.752)--(5.860,1.752)%
  --(5.926,1.752)--(5.992,1.751)--(6.057,1.751)--(6.123,1.751)--(6.189,1.751)--(6.254,1.751)%
  --(6.320,1.751)--(6.385,1.750)--(6.451,1.750)--(6.517,1.750)--(6.582,1.750)--(6.648,1.750)%
  --(6.714,1.750)--(6.779,1.749)--(6.845,1.749)--(6.910,1.749)--(6.976,1.749)--(7.042,1.749)%
  --(7.107,1.749)--(7.173,1.749)--(7.238,1.749)--(7.304,1.749)--(7.370,1.749)--(7.435,1.749)%
  --(7.501,1.748)--(7.567,1.748)--(7.632,1.748)--(7.698,1.748)--(7.763,1.748)--(7.829,1.748);
\gpcolor{color=gp lt color border}
\gpsetlinetype{gp lt border}
\draw[gp path] (1.136,3.587)--(1.136,0.985)--(7.829,0.985)--(7.829,3.587)--cycle;
\gpdefrectangularnode{gp plot 1}{\pgfpoint{1.136cm}{0.985cm}}{\pgfpoint{7.829cm}{3.587cm}}
\end{tikzpicture}

%% file: charts/perf.iter.friendster8.knor.tex
\begin{tikzpicture}[gnuplot]
\path (0.000,0.000) rectangle (8.382,4.826);
\gpcolor{color=gp lt color border}
\gpsetlinetype{gp lt border}
\gpsetlinewidth{1.00}
\draw[gp path] (1.504,0.616)--(1.684,0.616);
\draw[gp path] (7.829,0.616)--(7.649,0.616);
\node[gp node right] at (1.320,0.616) {0.01};
\draw[gp path] (1.504,0.909)--(1.594,0.909);
\draw[gp path] (7.829,0.909)--(7.739,0.909);
\draw[gp path] (1.504,1.080)--(1.594,1.080);
\draw[gp path] (7.829,1.080)--(7.739,1.080);
\draw[gp path] (1.504,1.201)--(1.594,1.201);
\draw[gp path] (7.829,1.201)--(7.739,1.201);
\draw[gp path] (1.504,1.296)--(1.594,1.296);
\draw[gp path] (7.829,1.296)--(7.739,1.296);
\draw[gp path] (1.504,1.373)--(1.594,1.373);
\draw[gp path] (7.829,1.373)--(7.739,1.373);
\draw[gp path] (1.504,1.438)--(1.594,1.438);
\draw[gp path] (7.829,1.438)--(7.739,1.438);
\draw[gp path] (1.504,1.494)--(1.594,1.494);
\draw[gp path] (7.829,1.494)--(7.739,1.494);
\draw[gp path] (1.504,1.544)--(1.594,1.544);
\draw[gp path] (7.829,1.544)--(7.739,1.544);
\draw[gp path] (1.504,1.588)--(1.684,1.588);
\draw[gp path] (7.829,1.588)--(7.649,1.588);
\node[gp node right] at (1.320,1.588) {0.1};
\draw[gp path] (1.504,1.881)--(1.594,1.881);
\draw[gp path] (7.829,1.881)--(7.739,1.881);
\draw[gp path] (1.504,2.052)--(1.594,2.052);
\draw[gp path] (7.829,2.052)--(7.739,2.052);
\draw[gp path] (1.504,2.174)--(1.594,2.174);
\draw[gp path] (7.829,2.174)--(7.739,2.174);
\draw[gp path] (1.504,2.268)--(1.594,2.268);
\draw[gp path] (7.829,2.268)--(7.739,2.268);
\draw[gp path] (1.504,2.345)--(1.594,2.345);
\draw[gp path] (7.829,2.345)--(7.739,2.345);
\draw[gp path] (1.504,2.410)--(1.594,2.410);
\draw[gp path] (7.829,2.410)--(7.739,2.410);
\draw[gp path] (1.504,2.466)--(1.594,2.466);
\draw[gp path] (7.829,2.466)--(7.739,2.466);
\draw[gp path] (1.504,2.516)--(1.594,2.516);
\draw[gp path] (7.829,2.516)--(7.739,2.516);
\draw[gp path] (1.504,2.561)--(1.684,2.561);
\draw[gp path] (7.829,2.561)--(7.649,2.561);
\node[gp node right] at (1.320,2.561) {1};
\draw[gp path] (1.504,2.853)--(1.594,2.853);
\draw[gp path] (7.829,2.853)--(7.739,2.853);
\draw[gp path] (1.504,3.025)--(1.594,3.025);
\draw[gp path] (7.829,3.025)--(7.739,3.025);
\draw[gp path] (1.504,3.146)--(1.594,3.146);
\draw[gp path] (7.829,3.146)--(7.739,3.146);
\draw[gp path] (1.504,3.240)--(1.594,3.240);
\draw[gp path] (7.829,3.240)--(7.739,3.240);
\draw[gp path] (1.504,3.317)--(1.594,3.317);
\draw[gp path] (7.829,3.317)--(7.739,3.317);
\draw[gp path] (1.504,3.382)--(1.594,3.382);
\draw[gp path] (7.829,3.382)--(7.739,3.382);
\draw[gp path] (1.504,3.439)--(1.594,3.439);
\draw[gp path] (7.829,3.439)--(7.739,3.439);
\draw[gp path] (1.504,3.489)--(1.594,3.489);
\draw[gp path] (7.829,3.489)--(7.739,3.489);
\draw[gp path] (1.504,3.533)--(1.684,3.533);
\draw[gp path] (7.829,3.533)--(7.649,3.533);
\node[gp node right] at (1.320,3.533) {10};
\draw[gp path] (2.769,0.616)--(2.769,0.796);
\draw[gp path] (2.769,3.533)--(2.769,3.353);
\node[gp node center] at (2.769,0.308) {k=10};
\draw[gp path] (4.034,0.616)--(4.034,0.796);
\draw[gp path] (4.034,3.533)--(4.034,3.353);
\node[gp node center] at (4.034,0.308) {k=20};
\draw[gp path] (5.299,0.616)--(5.299,0.796);
\draw[gp path] (5.299,3.533)--(5.299,3.353);
\node[gp node center] at (5.299,0.308) {k=50};
\draw[gp path] (6.564,0.616)--(6.564,0.796);
\draw[gp path] (6.564,3.533)--(6.564,3.353);
\node[gp node center] at (6.564,0.308) {k=100};
\draw[gp path] (1.504,3.533)--(1.504,0.616)--(7.829,0.616)--(7.829,3.533)--cycle;
\node[gp node center,rotate=-270] at (0.246,2.074) {Log Scale Time/iter (sec)};
\node[gp node right] at (3.382,4.492) {\textsf{knori}};
\gpfill{rgb color={0.000,1.000,0.000},color=.!50} (3.566,4.415)--(4.482,4.415)--(4.482,4.569)--(3.566,4.569)--cycle;
\gpfill{rgb color={0.000,1.000,0.000},color=.!50} (2.453,0.616)--(2.665,0.616)--(2.665,1.533)--(2.453,1.533)--cycle;
\gpfill{rgb color={0.000,1.000,0.000},color=.!50} (3.718,0.616)--(3.930,0.616)--(3.930,1.481)--(3.718,1.481)--cycle;
\gpfill{rgb color={0.000,1.000,0.000},color=.!50} (4.983,0.616)--(5.195,0.616)--(5.195,1.564)--(4.983,1.564)--cycle;
\gpfill{rgb color={0.000,1.000,0.000},color=.!50} (6.248,0.616)--(6.460,0.616)--(6.460,1.724)--(6.248,1.724)--cycle;
\node[gp node right] at (3.382,4.184) {\textsf{knori-}};
\def\gpfillpath{(3.566,4.107)--(4.482,4.107)--(4.482,4.261)--(3.566,4.261)--cycle}
\gpfill{color=gpbgfillcolor} \gpfillpath;
\gpfill{rgb color={0.498,1.000,0.831},gp pattern 3,pattern color=.} \gpfillpath;
\gpcolor{rgb color={0.498,1.000,0.831}}
\gpsetlinetype{gp lt plot 1}
\draw[gp path] (3.566,4.107)--(4.482,4.107)--(4.482,4.261)--(3.566,4.261)--cycle;
\def\gpfillpath{(2.664,0.616)--(2.875,0.616)--(2.875,1.810)--(2.664,1.810)--cycle}
\gpfill{color=gpbgfillcolor} \gpfillpath;
\gpfill{rgb color={0.498,1.000,0.831},gp pattern 3,pattern color=.} \gpfillpath;
\draw[gp path] (2.664,0.616)--(2.664,1.809)--(2.874,1.809)--(2.874,0.616)--cycle;
\def\gpfillpath{(3.929,0.616)--(4.140,0.616)--(4.140,2.057)--(3.929,2.057)--cycle}
\gpfill{color=gpbgfillcolor} \gpfillpath;
\gpfill{rgb color={0.498,1.000,0.831},gp pattern 3,pattern color=.} \gpfillpath;
\draw[gp path] (3.929,0.616)--(3.929,2.056)--(4.139,2.056)--(4.139,0.616)--cycle;
\def\gpfillpath{(5.194,0.616)--(5.405,0.616)--(5.405,2.421)--(5.194,2.421)--cycle}
\gpfill{color=gpbgfillcolor} \gpfillpath;
\gpfill{rgb color={0.498,1.000,0.831},gp pattern 3,pattern color=.} \gpfillpath;
\draw[gp path] (5.194,0.616)--(5.194,2.420)--(5.404,2.420)--(5.404,0.616)--cycle;
\def\gpfillpath{(6.459,0.616)--(6.670,0.616)--(6.670,2.706)--(6.459,2.706)--cycle}
\gpfill{color=gpbgfillcolor} \gpfillpath;
\gpfill{rgb color={0.498,1.000,0.831},gp pattern 3,pattern color=.} \gpfillpath;
\draw[gp path] (6.459,0.616)--(6.459,2.705)--(6.669,2.705)--(6.669,0.616)--cycle;
\gpcolor{color=gp lt color border}
\node[gp node right] at (6.138,4.492) {\textsf{knors}};
\def\gpfillpath{(6.322,4.415)--(7.238,4.415)--(7.238,4.569)--(6.322,4.569)--cycle}
\gpfill{color=gpbgfillcolor} \gpfillpath;
\gpfill{rgb color={0.373,0.620,0.627},gp pattern 4,pattern color=.} \gpfillpath;
\gpcolor{rgb color={0.373,0.620,0.627}}
\gpsetlinetype{gp lt plot 2}
\draw[gp path] (6.322,4.415)--(7.238,4.415)--(7.238,4.569)--(6.322,4.569)--cycle;
\def\gpfillpath{(2.874,0.616)--(3.086,0.616)--(3.086,2.195)--(2.874,2.195)--cycle}
\gpfill{color=gpbgfillcolor} \gpfillpath;
\gpfill{rgb color={0.373,0.620,0.627},gp pattern 4,pattern color=.} \gpfillpath;
\draw[gp path] (2.874,0.616)--(2.874,2.194)--(3.085,2.194)--(3.085,0.616)--cycle;
\def\gpfillpath{(4.139,0.616)--(4.351,0.616)--(4.351,2.324)--(4.139,2.324)--cycle}
\gpfill{color=gpbgfillcolor} \gpfillpath;
\gpfill{rgb color={0.373,0.620,0.627},gp pattern 4,pattern color=.} \gpfillpath;
\draw[gp path] (4.139,0.616)--(4.139,2.323)--(4.350,2.323)--(4.350,0.616)--cycle;
\def\gpfillpath{(5.404,0.616)--(5.616,0.616)--(5.616,2.369)--(5.404,2.369)--cycle}
\gpfill{color=gpbgfillcolor} \gpfillpath;
\gpfill{rgb color={0.373,0.620,0.627},gp pattern 4,pattern color=.} \gpfillpath;
\draw[gp path] (5.404,0.616)--(5.404,2.368)--(5.615,2.368)--(5.615,0.616)--cycle;
\def\gpfillpath{(6.669,0.616)--(6.881,0.616)--(6.881,2.544)--(6.669,2.544)--cycle}
\gpfill{color=gpbgfillcolor} \gpfillpath;
\gpfill{rgb color={0.373,0.620,0.627},gp pattern 4,pattern color=.} \gpfillpath;
\draw[gp path] (6.669,0.616)--(6.669,2.543)--(6.880,2.543)--(6.880,0.616)--cycle;
\gpcolor{color=gp lt color border}
\node[gp node right] at (6.138,4.184) {\textsf{knors-{}-}};
\def\gpfillpath{(6.322,4.107)--(7.238,4.107)--(7.238,4.261)--(6.322,4.261)--cycle}
\gpfill{color=gpbgfillcolor} \gpfillpath;
\gpfill{rgb color={0.392,0.584,0.929},gp pattern 5,pattern color=.} \gpfillpath;
\gpcolor{rgb color={0.392,0.584,0.929}}
\gpsetlinetype{gp lt plot 3}
\draw[gp path] (6.322,4.107)--(7.238,4.107)--(7.238,4.261)--(6.322,4.261)--cycle;
\def\gpfillpath{(3.085,0.616)--(3.297,0.616)--(3.297,2.500)--(3.085,2.500)--cycle}
\gpfill{color=gpbgfillcolor} \gpfillpath;
\gpfill{rgb color={0.392,0.584,0.929},gp pattern 5,pattern color=.} \gpfillpath;
\draw[gp path] (3.085,0.616)--(3.085,2.499)--(3.296,2.499)--(3.296,0.616)--cycle;
\def\gpfillpath{(4.350,0.616)--(4.562,0.616)--(4.562,2.694)--(4.350,2.694)--cycle}
\gpfill{color=gpbgfillcolor} \gpfillpath;
\gpfill{rgb color={0.392,0.584,0.929},gp pattern 5,pattern color=.} \gpfillpath;
\draw[gp path] (4.350,0.616)--(4.350,2.693)--(4.561,2.693)--(4.561,0.616)--cycle;
\def\gpfillpath{(5.615,0.616)--(5.827,0.616)--(5.827,3.011)--(5.615,3.011)--cycle}
\gpfill{color=gpbgfillcolor} \gpfillpath;
\gpfill{rgb color={0.392,0.584,0.929},gp pattern 5,pattern color=.} \gpfillpath;
\draw[gp path] (5.615,0.616)--(5.615,3.010)--(5.826,3.010)--(5.826,0.616)--cycle;
\def\gpfillpath{(6.880,0.616)--(7.092,0.616)--(7.092,3.268)--(6.880,3.268)--cycle}
\gpfill{color=gpbgfillcolor} \gpfillpath;
\gpfill{rgb color={0.392,0.584,0.929},gp pattern 5,pattern color=.} \gpfillpath;
\draw[gp path] (6.880,0.616)--(6.880,3.267)--(7.091,3.267)--(7.091,0.616)--cycle;
\gpcolor{color=gp lt color border}
\gpsetlinetype{gp lt border}
\draw[gp path] (1.504,3.533)--(1.504,0.616)--(7.829,0.616)--(7.829,3.533)--cycle;
\gpdefrectangularnode{gp plot 1}{\pgfpoint{1.504cm}{0.616cm}}{\pgfpoint{7.829cm}{3.533cm}}
\end{tikzpicture}

%% file: charts/perf.iter.friendster32.knor.tex
\begin{tikzpicture}[gnuplot]
\path (0.000,0.000) rectangle (8.382,4.826);
\gpcolor{color=gp lt color border}
\gpsetlinetype{gp lt border}
\gpsetlinewidth{1.00}
\draw[gp path] (1.320,0.616)--(1.500,0.616);
\draw[gp path] (7.829,0.616)--(7.649,0.616);
\node[gp node right] at (1.136,0.616) {0.1};
\draw[gp path] (1.320,1.194)--(1.410,1.194);
\draw[gp path] (7.829,1.194)--(7.739,1.194);
\draw[gp path] (1.320,1.532)--(1.410,1.532);
\draw[gp path] (7.829,1.532)--(7.739,1.532);
\draw[gp path] (1.320,1.772)--(1.410,1.772);
\draw[gp path] (7.829,1.772)--(7.739,1.772);
\draw[gp path] (1.320,1.958)--(1.410,1.958);
\draw[gp path] (7.829,1.958)--(7.739,1.958);
\draw[gp path] (1.320,2.110)--(1.410,2.110);
\draw[gp path] (7.829,2.110)--(7.739,2.110);
\draw[gp path] (1.320,2.239)--(1.410,2.239);
\draw[gp path] (7.829,2.239)--(7.739,2.239);
\draw[gp path] (1.320,2.350)--(1.410,2.350);
\draw[gp path] (7.829,2.350)--(7.739,2.350);
\draw[gp path] (1.320,2.449)--(1.410,2.449);
\draw[gp path] (7.829,2.449)--(7.739,2.449);
\draw[gp path] (1.320,2.537)--(1.500,2.537);
\draw[gp path] (7.829,2.537)--(7.649,2.537);
\node[gp node right] at (1.136,2.537) {1};
\draw[gp path] (1.320,3.115)--(1.410,3.115);
\draw[gp path] (7.829,3.115)--(7.739,3.115);
\draw[gp path] (1.320,3.453)--(1.410,3.453);
\draw[gp path] (7.829,3.453)--(7.739,3.453);
\draw[gp path] (1.320,3.693)--(1.410,3.693);
\draw[gp path] (7.829,3.693)--(7.739,3.693);
\draw[gp path] (1.320,3.879)--(1.410,3.879);
\draw[gp path] (7.829,3.879)--(7.739,3.879);
\draw[gp path] (1.320,4.031)--(1.410,4.031);
\draw[gp path] (7.829,4.031)--(7.739,4.031);
\draw[gp path] (1.320,4.160)--(1.410,4.160);
\draw[gp path] (7.829,4.160)--(7.739,4.160);
\draw[gp path] (1.320,4.271)--(1.410,4.271);
\draw[gp path] (7.829,4.271)--(7.739,4.271);
\draw[gp path] (1.320,4.369)--(1.410,4.369);
\draw[gp path] (7.829,4.369)--(7.739,4.369);
\draw[gp path] (1.320,4.457)--(1.500,4.457);
\draw[gp path] (7.829,4.457)--(7.649,4.457);
\node[gp node right] at (1.136,4.457) {10};
\draw[gp path] (2.622,0.616)--(2.622,0.796);
\draw[gp path] (2.622,4.457)--(2.622,4.277);
\node[gp node center] at (2.622,0.308) {k=10};
\draw[gp path] (3.924,0.616)--(3.924,0.796);
\draw[gp path] (3.924,4.457)--(3.924,4.277);
\node[gp node center] at (3.924,0.308) {k=20};
\draw[gp path] (5.225,0.616)--(5.225,0.796);
\draw[gp path] (5.225,4.457)--(5.225,4.277);
\node[gp node center] at (5.225,0.308) {k=50};
\draw[gp path] (6.527,0.616)--(6.527,0.796);
\draw[gp path] (6.527,4.457)--(6.527,4.277);
\node[gp node center] at (6.527,0.308) {k=100};
\draw[gp path] (1.320,4.457)--(1.320,0.616)--(7.829,0.616)--(7.829,4.457)--cycle;
\node[gp node center,rotate=-270] at (0.246,2.536) {Log Scale Time/iter (sec)};
\gpfill{rgb color={0.000,1.000,0.000},color=.!50} (2.296,0.616)--(2.514,0.616)--(2.514,0.735)--(2.296,0.735)--cycle;
\gpfill{rgb color={0.000,1.000,0.000},color=.!50} (3.598,0.616)--(3.816,0.616)--(3.816,1.217)--(3.598,1.217)--cycle;
\gpfill{rgb color={0.000,1.000,0.000},color=.!50} (4.900,0.616)--(5.118,0.616)--(5.118,1.805)--(4.900,1.805)--cycle;
\gpfill{rgb color={0.000,1.000,0.000},color=.!50} (6.202,0.616)--(6.420,0.616)--(6.420,2.258)--(6.202,2.258)--cycle;
\def\gpfillpath{(2.513,0.616)--(2.731,0.616)--(2.731,1.943)--(2.513,1.943)--cycle}
\gpfill{color=gpbgfillcolor} \gpfillpath;
\gpfill{rgb color={0.498,1.000,0.831},gp pattern 3,pattern color=.} \gpfillpath;
\gpcolor{rgb color={0.498,1.000,0.831}}
\gpsetlinetype{gp lt plot 1}
\draw[gp path] (2.513,0.616)--(2.513,1.942)--(2.730,1.942)--(2.730,0.616)--cycle;
\def\gpfillpath{(3.815,0.616)--(4.033,0.616)--(4.033,2.373)--(3.815,2.373)--cycle}
\gpfill{color=gpbgfillcolor} \gpfillpath;
\gpfill{rgb color={0.498,1.000,0.831},gp pattern 3,pattern color=.} \gpfillpath;
\draw[gp path] (3.815,0.616)--(3.815,2.372)--(4.032,2.372)--(4.032,0.616)--cycle;
\def\gpfillpath{(5.117,0.616)--(5.335,0.616)--(5.335,3.085)--(5.117,3.085)--cycle}
\gpfill{color=gpbgfillcolor} \gpfillpath;
\gpfill{rgb color={0.498,1.000,0.831},gp pattern 3,pattern color=.} \gpfillpath;
\draw[gp path] (5.117,0.616)--(5.117,3.084)--(5.334,3.084)--(5.334,0.616)--cycle;
\def\gpfillpath{(6.419,0.616)--(6.637,0.616)--(6.637,3.651)--(6.419,3.651)--cycle}
\gpfill{color=gpbgfillcolor} \gpfillpath;
\gpfill{rgb color={0.498,1.000,0.831},gp pattern 3,pattern color=.} \gpfillpath;
\draw[gp path] (6.419,0.616)--(6.419,3.650)--(6.636,3.650)--(6.636,0.616)--cycle;
\def\gpfillpath{(2.730,0.616)--(2.948,0.616)--(2.948,3.192)--(2.730,3.192)--cycle}
\gpfill{color=gpbgfillcolor} \gpfillpath;
\gpfill{rgb color={0.373,0.620,0.627},gp pattern 4,pattern color=.} \gpfillpath;
\gpcolor{rgb color={0.373,0.620,0.627}}
\gpsetlinetype{gp lt plot 2}
\draw[gp path] (2.730,0.616)--(2.730,3.191)--(2.947,3.191)--(2.947,0.616)--cycle;
\def\gpfillpath{(4.032,0.616)--(4.250,0.616)--(4.250,3.222)--(4.032,3.222)--cycle}
\gpfill{color=gpbgfillcolor} \gpfillpath;
\gpfill{rgb color={0.373,0.620,0.627},gp pattern 4,pattern color=.} \gpfillpath;
\draw[gp path] (4.032,0.616)--(4.032,3.221)--(4.249,3.221)--(4.249,0.616)--cycle;
\def\gpfillpath{(5.334,0.616)--(5.552,0.616)--(5.552,3.700)--(5.334,3.700)--cycle}
\gpfill{color=gpbgfillcolor} \gpfillpath;
\gpfill{rgb color={0.373,0.620,0.627},gp pattern 4,pattern color=.} \gpfillpath;
\draw[gp path] (5.334,0.616)--(5.334,3.699)--(5.551,3.699)--(5.551,0.616)--cycle;
\def\gpfillpath{(6.636,0.616)--(6.854,0.616)--(6.854,3.982)--(6.636,3.982)--cycle}
\gpfill{color=gpbgfillcolor} \gpfillpath;
\gpfill{rgb color={0.373,0.620,0.627},gp pattern 4,pattern color=.} \gpfillpath;
\draw[gp path] (6.636,0.616)--(6.636,3.981)--(6.853,3.981)--(6.853,0.616)--cycle;
\def\gpfillpath{(2.947,0.616)--(3.165,0.616)--(3.165,3.376)--(2.947,3.376)--cycle}
\gpfill{color=gpbgfillcolor} \gpfillpath;
\gpfill{rgb color={0.392,0.584,0.929},gp pattern 5,pattern color=.} \gpfillpath;
\gpcolor{rgb color={0.392,0.584,0.929}}
\gpsetlinetype{gp lt plot 3}
\draw[gp path] (2.947,0.616)--(2.947,3.375)--(3.164,3.375)--(3.164,0.616)--cycle;
\def\gpfillpath{(4.249,0.616)--(4.467,0.616)--(4.467,3.389)--(4.249,3.389)--cycle}
\gpfill{color=gpbgfillcolor} \gpfillpath;
\gpfill{rgb color={0.392,0.584,0.929},gp pattern 5,pattern color=.} \gpfillpath;
\draw[gp path] (4.249,0.616)--(4.249,3.388)--(4.466,3.388)--(4.466,0.616)--cycle;
\def\gpfillpath{(5.551,0.616)--(5.769,0.616)--(5.769,3.783)--(5.551,3.783)--cycle}
\gpfill{color=gpbgfillcolor} \gpfillpath;
\gpfill{rgb color={0.392,0.584,0.929},gp pattern 5,pattern color=.} \gpfillpath;
\draw[gp path] (5.551,0.616)--(5.551,3.782)--(5.768,3.782)--(5.768,0.616)--cycle;
\def\gpfillpath{(6.853,0.616)--(7.071,0.616)--(7.071,4.286)--(6.853,4.286)--cycle}
\gpfill{color=gpbgfillcolor} \gpfillpath;
\gpfill{rgb color={0.392,0.584,0.929},gp pattern 5,pattern color=.} \gpfillpath;
\draw[gp path] (6.853,0.616)--(6.853,4.285)--(7.070,4.285)--(7.070,0.616)--cycle;
\gpcolor{color=gp lt color border}
\gpsetlinetype{gp lt border}
\draw[gp path] (1.320,4.457)--(1.320,0.616)--(7.829,0.616)--(7.829,4.457)--cycle;
\gpdefrectangularnode{gp plot 1}{\pgfpoint{1.320cm}{0.616cm}}{\pgfpoint{7.829cm}{4.457cm}}
\end{tikzpicture}

%% file: charts/mem.friendster.knor.tex
\begin{tikzpicture}[gnuplot]
\path (0.000,0.000) rectangle (8.382,4.826);
\gpcolor{color=gp lt color border}
\gpsetlinetype{gp lt border}
\gpsetlinewidth{1.00}
\draw[gp path] (1.320,1.425)--(1.500,1.425);
\draw[gp path] (7.829,1.425)--(7.649,1.425);
\node[gp node right] at (1.136,1.425) { 5};
\draw[gp path] (1.320,2.435)--(1.500,2.435);
\draw[gp path] (7.829,2.435)--(7.649,2.435);
\node[gp node right] at (1.136,2.435) { 10};
\draw[gp path] (1.320,3.446)--(1.500,3.446);
\draw[gp path] (7.829,3.446)--(7.649,3.446);
\node[gp node right] at (1.136,3.446) { 15};
\draw[gp path] (1.320,4.457)--(1.500,4.457);
\draw[gp path] (7.829,4.457)--(7.649,4.457);
\node[gp node right] at (1.136,4.457) { 20};
\draw[gp path] (3.490,0.616)--(3.490,0.796);
\draw[gp path] (3.490,4.457)--(3.490,4.277);
\node[gp node center] at (3.490,0.308) {Friendster-8};
\draw[gp path] (5.659,0.616)--(5.659,0.796);
\draw[gp path] (5.659,4.457)--(5.659,4.277);
\node[gp node center] at (5.659,0.308) {Friendster-32};
\draw[gp path] (1.320,4.457)--(1.320,0.616)--(7.829,0.616)--(7.829,4.457)--cycle;
\node[gp node center,rotate=-270] at (0.246,2.536) {Memory (GB)};
\gpfill{rgb color={0.000,1.000,0.000},color=.!50} (2.947,0.616)--(3.310,0.616)--(3.310,1.466)--(2.947,1.466)--cycle;
\gpfill{rgb color={0.000,1.000,0.000},color=.!50} (5.117,0.616)--(5.480,0.616)--(5.480,3.690)--(5.117,3.690)--cycle;
\def\gpfillpath{(3.309,0.616)--(3.671,0.616)--(3.671,1.345)--(3.309,1.345)--cycle}
\gpfill{color=gpbgfillcolor} \gpfillpath;
\gpfill{rgb color={0.498,1.000,0.831},gp pattern 3,pattern color=.} \gpfillpath;
\gpcolor{rgb color={0.498,1.000,0.831}}
\gpsetlinetype{gp lt plot 1}
\draw[gp path] (3.309,0.616)--(3.309,1.344)--(3.670,1.344)--(3.670,0.616)--cycle;
\def\gpfillpath{(5.479,0.616)--(5.841,0.616)--(5.841,3.649)--(5.479,3.649)--cycle}
\gpfill{color=gpbgfillcolor} \gpfillpath;
\gpfill{rgb color={0.498,1.000,0.831},gp pattern 3,pattern color=.} \gpfillpath;
\draw[gp path] (5.479,0.616)--(5.479,3.648)--(5.840,3.648)--(5.840,0.616)--cycle;
\def\gpfillpath{(3.670,0.616)--(4.033,0.616)--(4.033,1.304)--(3.670,1.304)--cycle}
\gpfill{color=gpbgfillcolor} \gpfillpath;
\gpfill{rgb color={0.373,0.620,0.627},gp pattern 4,pattern color=.} \gpfillpath;
\gpcolor{rgb color={0.373,0.620,0.627}}
\gpsetlinetype{gp lt plot 2}
\draw[gp path] (3.670,0.616)--(3.670,1.303)--(4.032,1.303)--(4.032,0.616)--cycle;
\def\gpfillpath{(5.840,0.616)--(6.203,0.616)--(6.203,1.365)--(5.840,1.365)--cycle}
\gpfill{color=gpbgfillcolor} \gpfillpath;
\gpfill{rgb color={0.373,0.620,0.627},gp pattern 4,pattern color=.} \gpfillpath;
\draw[gp path] (5.840,0.616)--(5.840,1.364)--(6.202,1.364)--(6.202,0.616)--cycle;
\def\gpfillpath{(4.032,0.616)--(4.395,0.616)--(4.395,1.122)--(4.032,1.122)--cycle}
\gpfill{color=gpbgfillcolor} \gpfillpath;
\gpfill{rgb color={0.392,0.584,0.929},gp pattern 5,pattern color=.} \gpfillpath;
\gpcolor{rgb color={0.392,0.584,0.929}}
\gpsetlinetype{gp lt plot 3}
\draw[gp path] (4.032,0.616)--(4.032,1.121)--(4.394,1.121)--(4.394,0.616)--cycle;
\def\gpfillpath{(6.202,0.616)--(6.564,0.616)--(6.564,1.264)--(6.202,1.264)--cycle}
\gpfill{color=gpbgfillcolor} \gpfillpath;
\gpfill{rgb color={0.392,0.584,0.929},gp pattern 5,pattern color=.} \gpfillpath;
\draw[gp path] (6.202,0.616)--(6.202,1.263)--(6.563,1.263)--(6.563,0.616)--cycle;
\gpcolor{color=gp lt color border}
\gpsetlinetype{gp lt border}
\draw[gp path] (1.320,4.457)--(1.320,0.616)--(7.829,0.616)--(7.829,4.457)--cycle;
\gpdefrectangularnode{gp plot 1}{\pgfpoint{1.320cm}{0.616cm}}{\pgfpoint{7.829cm}{4.457cm}}
\end{tikzpicture}

%% file: charts/perf.iter.friendster8.tex
\begin{tikzpicture}[gnuplot]
\path (0.000,0.000) rectangle (8.382,4.826);
\gpcolor{color=gp lt color border}
\gpsetlinetype{gp lt border}
\gpsetlinewidth{1.00}
\draw[gp path] (1.504,0.616)--(1.684,0.616);
\draw[gp path] (7.829,0.616)--(7.649,0.616);
\node[gp node right] at (1.320,0.616) {0.01};
\draw[gp path] (1.504,0.836)--(1.594,0.836);
\draw[gp path] (7.829,0.836)--(7.739,0.836);
\draw[gp path] (1.504,0.964)--(1.594,0.964);
\draw[gp path] (7.829,0.964)--(7.739,0.964);
\draw[gp path] (1.504,1.055)--(1.594,1.055);
\draw[gp path] (7.829,1.055)--(7.739,1.055);
\draw[gp path] (1.504,1.126)--(1.594,1.126);
\draw[gp path] (7.829,1.126)--(7.739,1.126);
\draw[gp path] (1.504,1.183)--(1.594,1.183);
\draw[gp path] (7.829,1.183)--(7.739,1.183);
\draw[gp path] (1.504,1.232)--(1.594,1.232);
\draw[gp path] (7.829,1.232)--(7.739,1.232);
\draw[gp path] (1.504,1.275)--(1.594,1.275);
\draw[gp path] (7.829,1.275)--(7.739,1.275);
\draw[gp path] (1.504,1.312)--(1.594,1.312);
\draw[gp path] (7.829,1.312)--(7.739,1.312);
\draw[gp path] (1.504,1.345)--(1.684,1.345);
\draw[gp path] (7.829,1.345)--(7.649,1.345);
\node[gp node right] at (1.320,1.345) {0.1};
\draw[gp path] (1.504,1.565)--(1.594,1.565);
\draw[gp path] (7.829,1.565)--(7.739,1.565);
\draw[gp path] (1.504,1.693)--(1.594,1.693);
\draw[gp path] (7.829,1.693)--(7.739,1.693);
\draw[gp path] (1.504,1.784)--(1.594,1.784);
\draw[gp path] (7.829,1.784)--(7.739,1.784);
\draw[gp path] (1.504,1.855)--(1.594,1.855);
\draw[gp path] (7.829,1.855)--(7.739,1.855);
\draw[gp path] (1.504,1.913)--(1.594,1.913);
\draw[gp path] (7.829,1.913)--(7.739,1.913);
\draw[gp path] (1.504,1.962)--(1.594,1.962);
\draw[gp path] (7.829,1.962)--(7.739,1.962);
\draw[gp path] (1.504,2.004)--(1.594,2.004);
\draw[gp path] (7.829,2.004)--(7.739,2.004);
\draw[gp path] (1.504,2.041)--(1.594,2.041);
\draw[gp path] (7.829,2.041)--(7.739,2.041);
\draw[gp path] (1.504,2.075)--(1.684,2.075);
\draw[gp path] (7.829,2.075)--(7.649,2.075);
\node[gp node right] at (1.320,2.075) {1};
\draw[gp path] (1.504,2.294)--(1.594,2.294);
\draw[gp path] (7.829,2.294)--(7.739,2.294);
\draw[gp path] (1.504,2.422)--(1.594,2.422);
\draw[gp path] (7.829,2.422)--(7.739,2.422);
\draw[gp path] (1.504,2.514)--(1.594,2.514);
\draw[gp path] (7.829,2.514)--(7.739,2.514);
\draw[gp path] (1.504,2.584)--(1.594,2.584);
\draw[gp path] (7.829,2.584)--(7.739,2.584);
\draw[gp path] (1.504,2.642)--(1.594,2.642);
\draw[gp path] (7.829,2.642)--(7.739,2.642);
\draw[gp path] (1.504,2.691)--(1.594,2.691);
\draw[gp path] (7.829,2.691)--(7.739,2.691);
\draw[gp path] (1.504,2.733)--(1.594,2.733);
\draw[gp path] (7.829,2.733)--(7.739,2.733);
\draw[gp path] (1.504,2.770)--(1.594,2.770);
\draw[gp path] (7.829,2.770)--(7.739,2.770);
\draw[gp path] (1.504,2.804)--(1.684,2.804);
\draw[gp path] (7.829,2.804)--(7.649,2.804);
\node[gp node right] at (1.320,2.804) {10};
\draw[gp path] (1.504,3.023)--(1.594,3.023);
\draw[gp path] (7.829,3.023)--(7.739,3.023);
\draw[gp path] (1.504,3.152)--(1.594,3.152);
\draw[gp path] (7.829,3.152)--(7.739,3.152);
\draw[gp path] (1.504,3.243)--(1.594,3.243);
\draw[gp path] (7.829,3.243)--(7.739,3.243);
\draw[gp path] (1.504,3.313)--(1.594,3.313);
\draw[gp path] (7.829,3.313)--(7.739,3.313);
\draw[gp path] (1.504,3.371)--(1.594,3.371);
\draw[gp path] (7.829,3.371)--(7.739,3.371);
\draw[gp path] (1.504,3.420)--(1.594,3.420);
\draw[gp path] (7.829,3.420)--(7.739,3.420);
\draw[gp path] (1.504,3.462)--(1.594,3.462);
\draw[gp path] (7.829,3.462)--(7.739,3.462);
\draw[gp path] (1.504,3.500)--(1.594,3.500);
\draw[gp path] (7.829,3.500)--(7.739,3.500);
\draw[gp path] (1.504,3.533)--(1.684,3.533);
\draw[gp path] (7.829,3.533)--(7.649,3.533);
\node[gp node right] at (1.320,3.533) {100};
\draw[gp path] (2.769,0.616)--(2.769,0.796);
\draw[gp path] (2.769,3.533)--(2.769,3.353);
\node[gp node center] at (2.769,0.308) {k=10};
\draw[gp path] (4.034,0.616)--(4.034,0.796);
\draw[gp path] (4.034,3.533)--(4.034,3.353);
\node[gp node center] at (4.034,0.308) {k=20};
\draw[gp path] (5.299,0.616)--(5.299,0.796);
\draw[gp path] (5.299,3.533)--(5.299,3.353);
\node[gp node center] at (5.299,0.308) {k=50};
\draw[gp path] (6.564,0.616)--(6.564,0.796);
\draw[gp path] (6.564,3.533)--(6.564,3.353);
\node[gp node center] at (6.564,0.308) {k=100};
\draw[gp path] (1.504,3.533)--(1.504,0.616)--(7.829,0.616)--(7.829,3.533)--cycle;
\node[gp node center,rotate=-270] at (0.246,2.074) {Log Scale Time/iter (sec)};
\node[gp node right] at (2.188,4.492) {\textsf{knori}};
\gpfill{rgb color={0.000,1.000,0.000},color=.!50} (2.372,4.415)--(3.288,4.415)--(3.288,4.569)--(2.372,4.569)--cycle;
\gpfill{rgb color={0.000,1.000,0.000},color=.!50} (2.408,0.616)--(2.589,0.616)--(2.589,1.304)--(2.408,1.304)--cycle;
\gpfill{rgb color={0.000,1.000,0.000},color=.!50} (3.673,0.616)--(3.854,0.616)--(3.854,1.265)--(3.673,1.265)--cycle;
\gpfill{rgb color={0.000,1.000,0.000},color=.!50} (4.938,0.616)--(5.119,0.616)--(5.119,1.327)--(4.938,1.327)--cycle;
\gpfill{rgb color={0.000,1.000,0.000},color=.!50} (6.203,0.616)--(6.384,0.616)--(6.384,1.448)--(6.203,1.448)--cycle;
\node[gp node right] at (2.188,4.184) {\textsf{knors}};
\def\gpfillpath{(2.372,4.107)--(3.288,4.107)--(3.288,4.261)--(2.372,4.261)--cycle}
\gpfill{color=gpbgfillcolor} \gpfillpath;
\gpfill{rgb color={0.373,0.620,0.627},gp pattern 4,pattern color=.} \gpfillpath;
\gpcolor{rgb color={0.373,0.620,0.627}}
\gpsetlinetype{gp lt plot 1}
\draw[gp path] (2.372,4.107)--(3.288,4.107)--(3.288,4.261)--(2.372,4.261)--cycle;
\def\gpfillpath{(2.588,0.616)--(2.770,0.616)--(2.770,1.800)--(2.588,1.800)--cycle}
\gpfill{color=gpbgfillcolor} \gpfillpath;
\gpfill{rgb color={0.373,0.620,0.627},gp pattern 4,pattern color=.} \gpfillpath;
\draw[gp path] (2.588,0.616)--(2.588,1.799)--(2.769,1.799)--(2.769,0.616)--cycle;
\def\gpfillpath{(3.853,0.616)--(4.035,0.616)--(4.035,1.897)--(3.853,1.897)--cycle}
\gpfill{color=gpbgfillcolor} \gpfillpath;
\gpfill{rgb color={0.373,0.620,0.627},gp pattern 4,pattern color=.} \gpfillpath;
\draw[gp path] (3.853,0.616)--(3.853,1.896)--(4.034,1.896)--(4.034,0.616)--cycle;
\def\gpfillpath{(5.118,0.616)--(5.300,0.616)--(5.300,1.931)--(5.118,1.931)--cycle}
\gpfill{color=gpbgfillcolor} \gpfillpath;
\gpfill{rgb color={0.373,0.620,0.627},gp pattern 4,pattern color=.} \gpfillpath;
\draw[gp path] (5.118,0.616)--(5.118,1.930)--(5.299,1.930)--(5.299,0.616)--cycle;
\def\gpfillpath{(6.383,0.616)--(6.565,0.616)--(6.565,2.062)--(6.383,2.062)--cycle}
\gpfill{color=gpbgfillcolor} \gpfillpath;
\gpfill{rgb color={0.373,0.620,0.627},gp pattern 4,pattern color=.} \gpfillpath;
\draw[gp path] (6.383,0.616)--(6.383,2.061)--(6.564,2.061)--(6.564,0.616)--cycle;
\gpcolor{color=gp lt color border}
\node[gp node right] at (4.576,4.492) {H$_2$O};
\def\gpfillpath{(4.760,4.415)--(5.676,4.415)--(5.676,4.569)--(4.760,4.569)--cycle}
\gpfill{color=gpbgfillcolor} \gpfillpath;
\gpfill{rgb color={0.498,0.498,0.498},gp pattern 7,pattern color=.} \gpfillpath;
\gpcolor{rgb color={0.498,0.498,0.498}}
\gpsetlinetype{gp lt plot 2}
\draw[gp path] (4.760,4.415)--(5.676,4.415)--(5.676,4.569)--(4.760,4.569)--cycle;
\def\gpfillpath{(2.769,0.616)--(2.951,0.616)--(2.951,2.073)--(2.769,2.073)--cycle}
\gpfill{color=gpbgfillcolor} \gpfillpath;
\gpfill{rgb color={0.498,0.498,0.498},gp pattern 7,pattern color=.} \gpfillpath;
\draw[gp path] (2.769,0.616)--(2.769,2.072)--(2.950,2.072)--(2.950,0.616)--cycle;
\def\gpfillpath{(4.034,0.616)--(4.216,0.616)--(4.216,2.171)--(4.034,2.171)--cycle}
\gpfill{color=gpbgfillcolor} \gpfillpath;
\gpfill{rgb color={0.498,0.498,0.498},gp pattern 7,pattern color=.} \gpfillpath;
\draw[gp path] (4.034,0.616)--(4.034,2.170)--(4.215,2.170)--(4.215,0.616)--cycle;
\def\gpfillpath{(5.299,0.616)--(5.481,0.616)--(5.481,2.334)--(5.299,2.334)--cycle}
\gpfill{color=gpbgfillcolor} \gpfillpath;
\gpfill{rgb color={0.498,0.498,0.498},gp pattern 7,pattern color=.} \gpfillpath;
\draw[gp path] (5.299,0.616)--(5.299,2.333)--(5.480,2.333)--(5.480,0.616)--cycle;
\def\gpfillpath{(6.564,0.616)--(6.746,0.616)--(6.746,2.486)--(6.564,2.486)--cycle}
\gpfill{color=gpbgfillcolor} \gpfillpath;
\gpfill{rgb color={0.498,0.498,0.498},gp pattern 7,pattern color=.} \gpfillpath;
\draw[gp path] (6.564,0.616)--(6.564,2.485)--(6.745,2.485)--(6.745,0.616)--cycle;
\gpcolor{color=gp lt color border}
\node[gp node right] at (4.576,4.184) {MLlib};
\def\gpfillpath{(4.760,4.107)--(5.676,4.107)--(5.676,4.261)--(4.760,4.261)--cycle}
\gpfill{color=gpbgfillcolor} \gpfillpath;
\gpfill{rgb color={0.294,0.000,0.510},gp pattern 6,pattern color=.} \gpfillpath;
\gpcolor{rgb color={0.294,0.000,0.510}}
\gpsetlinetype{gp lt plot 3}
\draw[gp path] (4.760,4.107)--(5.676,4.107)--(5.676,4.261)--(4.760,4.261)--cycle;
\def\gpfillpath{(2.950,0.616)--(3.131,0.616)--(3.131,2.286)--(2.950,2.286)--cycle}
\gpfill{color=gpbgfillcolor} \gpfillpath;
\gpfill{rgb color={0.294,0.000,0.510},gp pattern 6,pattern color=.} \gpfillpath;
\draw[gp path] (2.950,0.616)--(2.950,2.285)--(3.130,2.285)--(3.130,0.616)--cycle;
\def\gpfillpath{(4.215,0.616)--(4.396,0.616)--(4.396,2.388)--(4.215,2.388)--cycle}
\gpfill{color=gpbgfillcolor} \gpfillpath;
\gpfill{rgb color={0.294,0.000,0.510},gp pattern 6,pattern color=.} \gpfillpath;
\draw[gp path] (4.215,0.616)--(4.215,2.387)--(4.395,2.387)--(4.395,0.616)--cycle;
\def\gpfillpath{(5.480,0.616)--(5.661,0.616)--(5.661,2.591)--(5.480,2.591)--cycle}
\gpfill{color=gpbgfillcolor} \gpfillpath;
\gpfill{rgb color={0.294,0.000,0.510},gp pattern 6,pattern color=.} \gpfillpath;
\draw[gp path] (5.480,0.616)--(5.480,2.590)--(5.660,2.590)--(5.660,0.616)--cycle;
\def\gpfillpath{(6.745,0.616)--(6.926,0.616)--(6.926,2.719)--(6.745,2.719)--cycle}
\gpfill{color=gpbgfillcolor} \gpfillpath;
\gpfill{rgb color={0.294,0.000,0.510},gp pattern 6,pattern color=.} \gpfillpath;
\draw[gp path] (6.745,0.616)--(6.745,2.718)--(6.925,2.718)--(6.925,0.616)--cycle;
\gpcolor{color=gp lt color border}
\node[gp node right] at (6.964,4.492) {Turi};
\def\gpfillpath{(7.148,4.415)--(8.064,4.415)--(8.064,4.569)--(7.148,4.569)--cycle}
\gpfill{color=gpbgfillcolor} \gpfillpath;
\gpfill{rgb color={1.000,0.000,0.000},gp pattern 8,pattern color=.} \gpfillpath;
\gpcolor{rgb color={1.000,0.000,0.000}}
\gpsetlinetype{gp lt plot 4}
\draw[gp path] (7.148,4.415)--(8.064,4.415)--(8.064,4.569)--(7.148,4.569)--cycle;
\def\gpfillpath{(3.130,0.616)--(3.312,0.616)--(3.312,3.194)--(3.130,3.194)--cycle}
\gpfill{color=gpbgfillcolor} \gpfillpath;
\gpfill{rgb color={1.000,0.000,0.000},gp pattern 8,pattern color=.} \gpfillpath;
\draw[gp path] (3.130,0.616)--(3.130,3.193)--(3.311,3.193)--(3.311,0.616)--cycle;
\def\gpfillpath{(4.395,0.616)--(4.577,0.616)--(4.577,3.192)--(4.395,3.192)--cycle}
\gpfill{color=gpbgfillcolor} \gpfillpath;
\gpfill{rgb color={1.000,0.000,0.000},gp pattern 8,pattern color=.} \gpfillpath;
\draw[gp path] (4.395,0.616)--(4.395,3.191)--(4.576,3.191)--(4.576,0.616)--cycle;
\def\gpfillpath{(5.660,0.616)--(5.842,0.616)--(5.842,3.178)--(5.660,3.178)--cycle}
\gpfill{color=gpbgfillcolor} \gpfillpath;
\gpfill{rgb color={1.000,0.000,0.000},gp pattern 8,pattern color=.} \gpfillpath;
\draw[gp path] (5.660,0.616)--(5.660,3.177)--(5.841,3.177)--(5.841,0.616)--cycle;
\def\gpfillpath{(6.925,0.616)--(7.107,0.616)--(7.107,3.184)--(6.925,3.184)--cycle}
\gpfill{color=gpbgfillcolor} \gpfillpath;
\gpfill{rgb color={1.000,0.000,0.000},gp pattern 8,pattern color=.} \gpfillpath;
\draw[gp path] (6.925,0.616)--(6.925,3.183)--(7.106,3.183)--(7.106,0.616)--cycle;
\gpcolor{color=gp lt color border}
\gpsetlinetype{gp lt border}
\draw[gp path] (1.504,3.533)--(1.504,0.616)--(7.829,0.616)--(7.829,3.533)--cycle;
\gpdefrectangularnode{gp plot 1}{\pgfpoint{1.504cm}{0.616cm}}{\pgfpoint{7.829cm}{3.533cm}}
\end{tikzpicture}

%% file: charts/perf.iter.friendster32.tex
\begin{tikzpicture}[gnuplot]
\path (0.000,0.000) rectangle (8.382,4.826);
\gpcolor{color=gp lt color border}
\gpsetlinetype{gp lt border}
\gpsetlinewidth{1.00}
\draw[gp path] (1.504,0.616)--(1.684,0.616);
\draw[gp path] (7.829,0.616)--(7.649,0.616);
\node[gp node right] at (1.320,0.616) { 0.1};
\draw[gp path] (1.504,1.001)--(1.594,1.001);
\draw[gp path] (7.829,1.001)--(7.739,1.001);
\draw[gp path] (1.504,1.227)--(1.594,1.227);
\draw[gp path] (7.829,1.227)--(7.739,1.227);
\draw[gp path] (1.504,1.387)--(1.594,1.387);
\draw[gp path] (7.829,1.387)--(7.739,1.387);
\draw[gp path] (1.504,1.511)--(1.594,1.511);
\draw[gp path] (7.829,1.511)--(7.739,1.511);
\draw[gp path] (1.504,1.612)--(1.594,1.612);
\draw[gp path] (7.829,1.612)--(7.739,1.612);
\draw[gp path] (1.504,1.698)--(1.594,1.698);
\draw[gp path] (7.829,1.698)--(7.739,1.698);
\draw[gp path] (1.504,1.772)--(1.594,1.772);
\draw[gp path] (7.829,1.772)--(7.739,1.772);
\draw[gp path] (1.504,1.838)--(1.594,1.838);
\draw[gp path] (7.829,1.838)--(7.739,1.838);
\draw[gp path] (1.504,1.896)--(1.684,1.896);
\draw[gp path] (7.829,1.896)--(7.649,1.896);
\node[gp node right] at (1.320,1.896) { 1};
\draw[gp path] (1.504,2.282)--(1.594,2.282);
\draw[gp path] (7.829,2.282)--(7.739,2.282);
\draw[gp path] (1.504,2.507)--(1.594,2.507);
\draw[gp path] (7.829,2.507)--(7.739,2.507);
\draw[gp path] (1.504,2.667)--(1.594,2.667);
\draw[gp path] (7.829,2.667)--(7.739,2.667);
\draw[gp path] (1.504,2.791)--(1.594,2.791);
\draw[gp path] (7.829,2.791)--(7.739,2.791);
\draw[gp path] (1.504,2.893)--(1.594,2.893);
\draw[gp path] (7.829,2.893)--(7.739,2.893);
\draw[gp path] (1.504,2.978)--(1.594,2.978);
\draw[gp path] (7.829,2.978)--(7.739,2.978);
\draw[gp path] (1.504,3.053)--(1.594,3.053);
\draw[gp path] (7.829,3.053)--(7.739,3.053);
\draw[gp path] (1.504,3.118)--(1.594,3.118);
\draw[gp path] (7.829,3.118)--(7.739,3.118);
\draw[gp path] (1.504,3.177)--(1.684,3.177);
\draw[gp path] (7.829,3.177)--(7.649,3.177);
\node[gp node right] at (1.320,3.177) { 10};
\draw[gp path] (1.504,3.562)--(1.594,3.562);
\draw[gp path] (7.829,3.562)--(7.739,3.562);
\draw[gp path] (1.504,3.788)--(1.594,3.788);
\draw[gp path] (7.829,3.788)--(7.739,3.788);
\draw[gp path] (1.504,3.948)--(1.594,3.948);
\draw[gp path] (7.829,3.948)--(7.739,3.948);
\draw[gp path] (1.504,4.072)--(1.594,4.072);
\draw[gp path] (7.829,4.072)--(7.739,4.072);
\draw[gp path] (1.504,4.173)--(1.594,4.173);
\draw[gp path] (7.829,4.173)--(7.739,4.173);
\draw[gp path] (1.504,4.259)--(1.594,4.259);
\draw[gp path] (7.829,4.259)--(7.739,4.259);
\draw[gp path] (1.504,4.333)--(1.594,4.333);
\draw[gp path] (7.829,4.333)--(7.739,4.333);
\draw[gp path] (1.504,4.398)--(1.594,4.398);
\draw[gp path] (7.829,4.398)--(7.739,4.398);
\draw[gp path] (1.504,4.457)--(1.684,4.457);
\draw[gp path] (7.829,4.457)--(7.649,4.457);
\node[gp node right] at (1.320,4.457) { 100};
\draw[gp path] (2.769,0.616)--(2.769,0.796);
\draw[gp path] (2.769,4.457)--(2.769,4.277);
\node[gp node center] at (2.769,0.308) {k=10};
\draw[gp path] (4.034,0.616)--(4.034,0.796);
\draw[gp path] (4.034,4.457)--(4.034,4.277);
\node[gp node center] at (4.034,0.308) {k=20};
\draw[gp path] (5.299,0.616)--(5.299,0.796);
\draw[gp path] (5.299,4.457)--(5.299,4.277);
\node[gp node center] at (5.299,0.308) {k=50};
\draw[gp path] (6.564,0.616)--(6.564,0.796);
\draw[gp path] (6.564,4.457)--(6.564,4.277);
\node[gp node center] at (6.564,0.308) {k=100};
\draw[gp path] (1.504,4.457)--(1.504,0.616)--(7.829,0.616)--(7.829,4.457)--cycle;
\node[gp node center,rotate=-270] at (0.246,2.536) {Log Scale Time/iter (sec)};
\gpfill{rgb color={0.000,1.000,0.000},color=.!50} (2.408,0.616)--(2.589,0.616)--(2.589,0.696)--(2.408,0.696)--cycle;
\gpfill{rgb color={0.000,1.000,0.000},color=.!50} (3.673,0.616)--(3.854,0.616)--(3.854,1.017)--(3.673,1.017)--cycle;
\gpfill{rgb color={0.000,1.000,0.000},color=.!50} (4.938,0.616)--(5.119,0.616)--(5.119,1.409)--(4.938,1.409)--cycle;
\gpfill{rgb color={0.000,1.000,0.000},color=.!50} (6.203,0.616)--(6.384,0.616)--(6.384,1.711)--(6.203,1.711)--cycle;
\def\gpfillpath{(2.588,0.616)--(2.770,0.616)--(2.770,2.334)--(2.588,2.334)--cycle}
\gpfill{color=gpbgfillcolor} \gpfillpath;
\gpfill{rgb color={0.373,0.620,0.627},gp pattern 4,pattern color=.} \gpfillpath;
\gpcolor{rgb color={0.373,0.620,0.627}}
\gpsetlinetype{gp lt plot 1}
\draw[gp path] (2.588,0.616)--(2.588,2.333)--(2.769,2.333)--(2.769,0.616)--cycle;
\def\gpfillpath{(3.853,0.616)--(4.035,0.616)--(4.035,2.354)--(3.853,2.354)--cycle}
\gpfill{color=gpbgfillcolor} \gpfillpath;
\gpfill{rgb color={0.373,0.620,0.627},gp pattern 4,pattern color=.} \gpfillpath;
\draw[gp path] (3.853,0.616)--(3.853,2.353)--(4.034,2.353)--(4.034,0.616)--cycle;
\def\gpfillpath{(5.118,0.616)--(5.300,0.616)--(5.300,2.672)--(5.118,2.672)--cycle}
\gpfill{color=gpbgfillcolor} \gpfillpath;
\gpfill{rgb color={0.373,0.620,0.627},gp pattern 4,pattern color=.} \gpfillpath;
\draw[gp path] (5.118,0.616)--(5.118,2.671)--(5.299,2.671)--(5.299,0.616)--cycle;
\def\gpfillpath{(6.383,0.616)--(6.565,0.616)--(6.565,2.860)--(6.383,2.860)--cycle}
\gpfill{color=gpbgfillcolor} \gpfillpath;
\gpfill{rgb color={0.373,0.620,0.627},gp pattern 4,pattern color=.} \gpfillpath;
\draw[gp path] (6.383,0.616)--(6.383,2.859)--(6.564,2.859)--(6.564,0.616)--cycle;
\def\gpfillpath{(2.769,0.616)--(2.951,0.616)--(2.951,2.580)--(2.769,2.580)--cycle}
\gpfill{color=gpbgfillcolor} \gpfillpath;
\gpfill{rgb color={0.498,0.498,0.498},gp pattern 7,pattern color=.} \gpfillpath;
\gpcolor{rgb color={0.498,0.498,0.498}}
\gpsetlinetype{gp lt plot 2}
\draw[gp path] (2.769,0.616)--(2.769,2.579)--(2.950,2.579)--(2.950,0.616)--cycle;
\def\gpfillpath{(4.034,0.616)--(4.216,0.616)--(4.216,2.691)--(4.034,2.691)--cycle}
\gpfill{color=gpbgfillcolor} \gpfillpath;
\gpfill{rgb color={0.498,0.498,0.498},gp pattern 7,pattern color=.} \gpfillpath;
\draw[gp path] (4.034,0.616)--(4.034,2.690)--(4.215,2.690)--(4.215,0.616)--cycle;
\def\gpfillpath{(5.299,0.616)--(5.481,0.616)--(5.481,2.979)--(5.299,2.979)--cycle}
\gpfill{color=gpbgfillcolor} \gpfillpath;
\gpfill{rgb color={0.498,0.498,0.498},gp pattern 7,pattern color=.} \gpfillpath;
\draw[gp path] (5.299,0.616)--(5.299,2.978)--(5.480,2.978)--(5.480,0.616)--cycle;
\def\gpfillpath{(6.564,0.616)--(6.746,0.616)--(6.746,3.216)--(6.564,3.216)--cycle}
\gpfill{color=gpbgfillcolor} \gpfillpath;
\gpfill{rgb color={0.498,0.498,0.498},gp pattern 7,pattern color=.} \gpfillpath;
\draw[gp path] (6.564,0.616)--(6.564,3.215)--(6.745,3.215)--(6.745,0.616)--cycle;
\def\gpfillpath{(2.950,0.616)--(3.131,0.616)--(3.131,2.857)--(2.950,2.857)--cycle}
\gpfill{color=gpbgfillcolor} \gpfillpath;
\gpfill{rgb color={0.294,0.000,0.510},gp pattern 6,pattern color=.} \gpfillpath;
\gpcolor{rgb color={0.294,0.000,0.510}}
\gpsetlinetype{gp lt plot 3}
\draw[gp path] (2.950,0.616)--(2.950,2.856)--(3.130,2.856)--(3.130,0.616)--cycle;
\def\gpfillpath{(4.215,0.616)--(4.396,0.616)--(4.396,3.061)--(4.215,3.061)--cycle}
\gpfill{color=gpbgfillcolor} \gpfillpath;
\gpfill{rgb color={0.294,0.000,0.510},gp pattern 6,pattern color=.} \gpfillpath;
\draw[gp path] (4.215,0.616)--(4.215,3.060)--(4.395,3.060)--(4.395,0.616)--cycle;
\def\gpfillpath{(5.480,0.616)--(5.661,0.616)--(5.661,3.536)--(5.480,3.536)--cycle}
\gpfill{color=gpbgfillcolor} \gpfillpath;
\gpfill{rgb color={0.294,0.000,0.510},gp pattern 6,pattern color=.} \gpfillpath;
\draw[gp path] (5.480,0.616)--(5.480,3.535)--(5.660,3.535)--(5.660,0.616)--cycle;
\def\gpfillpath{(6.745,0.616)--(6.926,0.616)--(6.926,3.847)--(6.745,3.847)--cycle}
\gpfill{color=gpbgfillcolor} \gpfillpath;
\gpfill{rgb color={0.294,0.000,0.510},gp pattern 6,pattern color=.} \gpfillpath;
\draw[gp path] (6.745,0.616)--(6.745,3.846)--(6.925,3.846)--(6.925,0.616)--cycle;
\def\gpfillpath{(3.130,0.616)--(3.312,0.616)--(3.312,4.289)--(3.130,4.289)--cycle}
\gpfill{color=gpbgfillcolor} \gpfillpath;
\gpfill{rgb color={1.000,0.000,0.000},gp pattern 8,pattern color=.} \gpfillpath;
\gpcolor{rgb color={1.000,0.000,0.000}}
\gpsetlinetype{gp lt plot 4}
\draw[gp path] (3.130,0.616)--(3.130,4.288)--(3.311,4.288)--(3.311,0.616)--cycle;
\def\gpfillpath{(4.395,0.616)--(4.577,0.616)--(4.577,4.095)--(4.395,4.095)--cycle}
\gpfill{color=gpbgfillcolor} \gpfillpath;
\gpfill{rgb color={1.000,0.000,0.000},gp pattern 8,pattern color=.} \gpfillpath;
\draw[gp path] (4.395,0.616)--(4.395,4.094)--(4.576,4.094)--(4.576,0.616)--cycle;
\def\gpfillpath{(5.660,0.616)--(5.842,0.616)--(5.842,4.365)--(5.660,4.365)--cycle}
\gpfill{color=gpbgfillcolor} \gpfillpath;
\gpfill{rgb color={1.000,0.000,0.000},gp pattern 8,pattern color=.} \gpfillpath;
\draw[gp path] (5.660,0.616)--(5.660,4.364)--(5.841,4.364)--(5.841,0.616)--cycle;
\def\gpfillpath{(6.925,0.616)--(7.107,0.616)--(7.107,4.311)--(6.925,4.311)--cycle}
\gpfill{color=gpbgfillcolor} \gpfillpath;
\gpfill{rgb color={1.000,0.000,0.000},gp pattern 8,pattern color=.} \gpfillpath;
\draw[gp path] (6.925,0.616)--(6.925,4.310)--(7.106,4.310)--(7.106,0.616)--cycle;
\gpcolor{color=gp lt color border}
\gpsetlinetype{gp lt border}
\draw[gp path] (1.504,4.457)--(1.504,0.616)--(7.829,0.616)--(7.829,4.457)--cycle;
\gpdefrectangularnode{gp plot 1}{\pgfpoint{1.504cm}{0.616cm}}{\pgfpoint{7.829cm}{4.457cm}}
\end{tikzpicture}

%% file: charts/mem.friendster.tex
\begin{tikzpicture}[gnuplot]
\path (0.000,0.000) rectangle (8.382,4.826);
\gpcolor{color=gp lt color border}
\gpsetlinetype{gp lt border}
\gpsetlinewidth{1.00}
\draw[gp path] (1.504,0.616)--(1.684,0.616);
\draw[gp path] (7.829,0.616)--(7.649,0.616);
\node[gp node right] at (1.320,0.616) { 1};
\draw[gp path] (1.504,1.194)--(1.594,1.194);
\draw[gp path] (7.829,1.194)--(7.739,1.194);
\draw[gp path] (1.504,1.532)--(1.594,1.532);
\draw[gp path] (7.829,1.532)--(7.739,1.532);
\draw[gp path] (1.504,1.772)--(1.594,1.772);
\draw[gp path] (7.829,1.772)--(7.739,1.772);
\draw[gp path] (1.504,1.958)--(1.594,1.958);
\draw[gp path] (7.829,1.958)--(7.739,1.958);
\draw[gp path] (1.504,2.110)--(1.594,2.110);
\draw[gp path] (7.829,2.110)--(7.739,2.110);
\draw[gp path] (1.504,2.239)--(1.594,2.239);
\draw[gp path] (7.829,2.239)--(7.739,2.239);
\draw[gp path] (1.504,2.350)--(1.594,2.350);
\draw[gp path] (7.829,2.350)--(7.739,2.350);
\draw[gp path] (1.504,2.449)--(1.594,2.449);
\draw[gp path] (7.829,2.449)--(7.739,2.449);
\draw[gp path] (1.504,2.537)--(1.684,2.537);
\draw[gp path] (7.829,2.537)--(7.649,2.537);
\node[gp node right] at (1.320,2.537) { 10};
\draw[gp path] (1.504,3.115)--(1.594,3.115);
\draw[gp path] (7.829,3.115)--(7.739,3.115);
\draw[gp path] (1.504,3.453)--(1.594,3.453);
\draw[gp path] (7.829,3.453)--(7.739,3.453);
\draw[gp path] (1.504,3.693)--(1.594,3.693);
\draw[gp path] (7.829,3.693)--(7.739,3.693);
\draw[gp path] (1.504,3.879)--(1.594,3.879);
\draw[gp path] (7.829,3.879)--(7.739,3.879);
\draw[gp path] (1.504,4.031)--(1.594,4.031);
\draw[gp path] (7.829,4.031)--(7.739,4.031);
\draw[gp path] (1.504,4.160)--(1.594,4.160);
\draw[gp path] (7.829,4.160)--(7.739,4.160);
\draw[gp path] (1.504,4.271)--(1.594,4.271);
\draw[gp path] (7.829,4.271)--(7.739,4.271);
\draw[gp path] (1.504,4.369)--(1.594,4.369);
\draw[gp path] (7.829,4.369)--(7.739,4.369);
\draw[gp path] (1.504,4.457)--(1.684,4.457);
\draw[gp path] (7.829,4.457)--(7.649,4.457);
\node[gp node right] at (1.320,4.457) { 100};
\draw[gp path] (3.612,0.616)--(3.612,0.796);
\draw[gp path] (3.612,4.457)--(3.612,4.277);
\node[gp node center] at (3.612,0.308) {Friendster-8};
\draw[gp path] (5.721,0.616)--(5.721,0.796);
\draw[gp path] (5.721,4.457)--(5.721,4.277);
\node[gp node center] at (5.721,0.308) {Friendster-32};
\draw[gp path] (1.504,4.457)--(1.504,0.616)--(7.829,0.616)--(7.829,4.457)--cycle;
\node[gp node center,rotate=-270] at (0.246,2.536) {Log Scale Memory (GB)};
\gpfill{rgb color={0.000,1.000,0.000},color=.!50} (3.010,0.616)--(3.312,0.616)--(3.312,1.992)--(3.010,1.992)--cycle;
\gpfill{rgb color={0.000,1.000,0.000},color=.!50} (5.118,0.616)--(5.420,0.616)--(5.420,2.940)--(5.118,2.940)--cycle;
\def\gpfillpath{(3.311,0.616)--(3.613,0.616)--(3.613,1.853)--(3.311,1.853)--cycle}
\gpfill{color=gpbgfillcolor} \gpfillpath;
\gpfill{rgb color={0.373,0.620,0.627},gp pattern 4,pattern color=.} \gpfillpath;
\gpcolor{rgb color={0.373,0.620,0.627}}
\gpsetlinetype{gp lt plot 1}
\draw[gp path] (3.311,0.616)--(3.311,1.852)--(3.612,1.852)--(3.612,0.616)--cycle;
\def\gpfillpath{(5.419,0.616)--(5.722,0.616)--(5.722,1.908)--(5.419,1.908)--cycle}
\gpfill{color=gpbgfillcolor} \gpfillpath;
\gpfill{rgb color={0.373,0.620,0.627},gp pattern 4,pattern color=.} \gpfillpath;
\draw[gp path] (5.419,0.616)--(5.419,1.907)--(5.721,1.907)--(5.721,0.616)--cycle;
\def\gpfillpath{(3.612,0.616)--(3.915,0.616)--(3.915,3.195)--(3.612,3.195)--cycle}
\gpfill{color=gpbgfillcolor} \gpfillpath;
\gpfill{rgb color={0.498,0.498,0.498},gp pattern 7,pattern color=.} \gpfillpath;
\gpcolor{rgb color={0.498,0.498,0.498}}
\gpsetlinetype{gp lt plot 2}
\draw[gp path] (3.612,0.616)--(3.612,3.194)--(3.914,3.194)--(3.914,0.616)--cycle;
\def\gpfillpath{(5.721,0.616)--(6.023,0.616)--(6.023,3.880)--(5.721,3.880)--cycle}
\gpfill{color=gpbgfillcolor} \gpfillpath;
\gpfill{rgb color={0.498,0.498,0.498},gp pattern 7,pattern color=.} \gpfillpath;
\draw[gp path] (5.721,0.616)--(5.721,3.879)--(6.022,3.879)--(6.022,0.616)--cycle;
\def\gpfillpath{(3.914,0.616)--(4.216,0.616)--(4.216,3.481)--(3.914,3.481)--cycle}
\gpfill{color=gpbgfillcolor} \gpfillpath;
\gpfill{rgb color={0.294,0.000,0.510},gp pattern 6,pattern color=.} \gpfillpath;
\gpcolor{rgb color={0.294,0.000,0.510}}
\gpsetlinetype{gp lt plot 3}
\draw[gp path] (3.914,0.616)--(3.914,3.480)--(4.215,3.480)--(4.215,0.616)--cycle;
\def\gpfillpath{(6.022,0.616)--(6.324,0.616)--(6.324,3.773)--(6.022,3.773)--cycle}
\gpfill{color=gpbgfillcolor} \gpfillpath;
\gpfill{rgb color={0.294,0.000,0.510},gp pattern 6,pattern color=.} \gpfillpath;
\draw[gp path] (6.022,0.616)--(6.022,3.772)--(6.323,3.772)--(6.323,0.616)--cycle;
\def\gpfillpath{(4.215,0.616)--(4.517,0.616)--(4.517,2.450)--(4.215,2.450)--cycle}
\gpfill{color=gpbgfillcolor} \gpfillpath;
\gpfill{rgb color={1.000,0.000,0.000},gp pattern 8,pattern color=.} \gpfillpath;
\gpcolor{rgb color={1.000,0.000,0.000}}
\gpsetlinetype{gp lt plot 4}
\draw[gp path] (4.215,0.616)--(4.215,2.449)--(4.516,2.449)--(4.516,0.616)--cycle;
\def\gpfillpath{(6.323,0.616)--(6.625,0.616)--(6.625,3.508)--(6.323,3.508)--cycle}
\gpfill{color=gpbgfillcolor} \gpfillpath;
\gpfill{rgb color={1.000,0.000,0.000},gp pattern 8,pattern color=.} \gpfillpath;
\draw[gp path] (6.323,0.616)--(6.323,3.507)--(6.624,3.507)--(6.624,0.616)--cycle;
\gpcolor{color=gp lt color border}
\gpsetlinetype{gp lt border}
\draw[gp path] (1.504,4.457)--(1.504,0.616)--(7.829,0.616)--(7.829,4.457)--cycle;
\gpdefrectangularnode{gp plot 1}{\pgfpoint{1.504cm}{0.616cm}}{\pgfpoint{7.829cm}{4.457cm}}
\end{tikzpicture}

%% file: charts/perf.iter.rmvm.tex
\begin{tikzpicture}[gnuplot]
\path (0.000,0.000) rectangle (8.382,4.826);
\gpcolor{color=gp lt color border}
\gpsetlinetype{gp lt border}
\gpsetlinewidth{1.00}
\draw[gp path] (1.504,0.616)--(1.684,0.616);
\draw[gp path] (7.829,0.616)--(7.649,0.616);
\node[gp node right] at (1.320,0.616) { 0};
\draw[gp path] (1.504,0.981)--(1.684,0.981);
\draw[gp path] (7.829,0.981)--(7.649,0.981);
\node[gp node right] at (1.320,0.981) { 20};
\draw[gp path] (1.504,1.345)--(1.684,1.345);
\draw[gp path] (7.829,1.345)--(7.649,1.345);
\node[gp node right] at (1.320,1.345) { 40};
\draw[gp path] (1.504,1.710)--(1.684,1.710);
\draw[gp path] (7.829,1.710)--(7.649,1.710);
\node[gp node right] at (1.320,1.710) { 60};
\draw[gp path] (1.504,2.075)--(1.684,2.075);
\draw[gp path] (7.829,2.075)--(7.649,2.075);
\node[gp node right] at (1.320,2.075) { 80};
\draw[gp path] (1.504,2.439)--(1.684,2.439);
\draw[gp path] (7.829,2.439)--(7.649,2.439);
\node[gp node right] at (1.320,2.439) { 100};
\draw[gp path] (1.504,2.804)--(1.684,2.804);
\draw[gp path] (7.829,2.804)--(7.649,2.804);
\node[gp node right] at (1.320,2.804) { 120};
\draw[gp path] (1.504,3.168)--(1.684,3.168);
\draw[gp path] (7.829,3.168)--(7.649,3.168);
\node[gp node right] at (1.320,3.168) { 140};
\draw[gp path] (1.504,3.533)--(1.684,3.533);
\draw[gp path] (7.829,3.533)--(7.649,3.533);
\node[gp node right] at (1.320,3.533) { 160};
\draw[gp path] (3.085,0.616)--(3.085,0.796);
\draw[gp path] (3.085,3.533)--(3.085,3.353);
\node[gp node center] at (3.085,0.308) {RM$_{856M}$};
\draw[gp path] (4.667,0.616)--(4.667,0.796);
\draw[gp path] (4.667,3.533)--(4.667,3.353);
\node[gp node center] at (4.667,0.308) {RM$_{1B}$};
\draw[gp path] (6.248,0.616)--(6.248,0.796);
\draw[gp path] (6.248,3.533)--(6.248,3.353);
\node[gp node center] at (6.248,0.308) {RM$_{2B}$};
\draw[gp path] (1.504,3.533)--(1.504,0.616)--(7.829,0.616)--(7.829,3.533)--cycle;
\node[gp node center,rotate=-270] at (0.246,2.074) {Time/iter (sec)};
\node[gp node right] at (2.280,4.492) {\textsf{knori}};
\gpfill{rgb color={0.000,1.000,0.000},color=.!50} (2.464,4.415)--(3.380,4.415)--(3.380,4.569)--(2.464,4.569)--cycle;
\gpfill{rgb color={0.000,1.000,0.000},color=.!50} (2.633,0.616)--(2.860,0.616)--(2.860,0.672)--(2.633,0.672)--cycle;
\gpfill{rgb color={0.000,1.000,0.000},color=.!50} (4.215,0.616)--(4.442,0.616)--(4.442,0.760)--(4.215,0.760)--cycle;
\node[gp node right] at (2.280,4.184) {\textsf{knors}};
\def\gpfillpath{(2.464,4.107)--(3.380,4.107)--(3.380,4.261)--(2.464,4.261)--cycle}
\gpfill{color=gpbgfillcolor} \gpfillpath;
\gpfill{rgb color={0.373,0.620,0.627},gp pattern 4,pattern color=.} \gpfillpath;
\gpcolor{rgb color={0.373,0.620,0.627}}
\gpsetlinetype{gp lt plot 1}
\draw[gp path] (2.464,4.107)--(3.380,4.107)--(3.380,4.261)--(2.464,4.261)--cycle;
\def\gpfillpath{(2.859,0.616)--(3.086,0.616)--(3.086,0.678)--(2.859,0.678)--cycle}
\gpfill{color=gpbgfillcolor} \gpfillpath;
\gpfill{rgb color={0.373,0.620,0.627},gp pattern 4,pattern color=.} \gpfillpath;
\draw[gp path] (2.859,0.616)--(2.859,0.677)--(3.085,0.677)--(3.085,0.616)--cycle;
\def\gpfillpath{(4.441,0.616)--(4.668,0.616)--(4.668,1.130)--(4.441,1.130)--cycle}
\gpfill{color=gpbgfillcolor} \gpfillpath;
\gpfill{rgb color={0.373,0.620,0.627},gp pattern 4,pattern color=.} \gpfillpath;
\draw[gp path] (4.441,0.616)--(4.441,1.129)--(4.667,1.129)--(4.667,0.616)--cycle;
\def\gpfillpath{(6.022,0.616)--(6.249,0.616)--(6.249,2.890)--(6.022,2.890)--cycle}
\gpfill{color=gpbgfillcolor} \gpfillpath;
\gpfill{rgb color={0.373,0.620,0.627},gp pattern 4,pattern color=.} \gpfillpath;
\draw[gp path] (6.022,0.616)--(6.022,2.889)--(6.248,2.889)--(6.248,0.616)--cycle;
\gpcolor{color=gp lt color border}
\node[gp node right] at (4.484,4.492) {H$_2$O};
\def\gpfillpath{(4.668,4.415)--(5.584,4.415)--(5.584,4.569)--(4.668,4.569)--cycle}
\gpfill{color=gpbgfillcolor} \gpfillpath;
\gpfill{rgb color={0.498,0.498,0.498},gp pattern 7,pattern color=.} \gpfillpath;
\gpcolor{rgb color={0.498,0.498,0.498}}
\gpsetlinetype{gp lt plot 2}
\draw[gp path] (4.668,4.415)--(5.584,4.415)--(5.584,4.569)--(4.668,4.569)--cycle;
\def\gpfillpath{(3.085,0.616)--(3.312,0.616)--(3.312,0.978)--(3.085,0.978)--cycle}
\gpfill{color=gpbgfillcolor} \gpfillpath;
\gpfill{rgb color={0.498,0.498,0.498},gp pattern 7,pattern color=.} \gpfillpath;
\draw[gp path] (3.085,0.616)--(3.085,0.977)--(3.311,0.977)--(3.311,0.616)--cycle;
\def\gpfillpath{(4.667,0.616)--(4.893,0.616)--(4.893,1.750)--(4.667,1.750)--cycle}
\gpfill{color=gpbgfillcolor} \gpfillpath;
\gpfill{rgb color={0.498,0.498,0.498},gp pattern 7,pattern color=.} \gpfillpath;
\draw[gp path] (4.667,0.616)--(4.667,1.749)--(4.892,1.749)--(4.892,0.616)--cycle;
\gpcolor{color=gp lt color border}
\node[gp node right] at (4.484,4.184) {MlLib};
\def\gpfillpath{(4.668,4.107)--(5.584,4.107)--(5.584,4.261)--(4.668,4.261)--cycle}
\gpfill{color=gpbgfillcolor} \gpfillpath;
\gpfill{rgb color={0.294,0.000,0.510},gp pattern 6,pattern color=.} \gpfillpath;
\gpcolor{rgb color={0.294,0.000,0.510}}
\gpsetlinetype{gp lt plot 3}
\draw[gp path] (4.668,4.107)--(5.584,4.107)--(5.584,4.261)--(4.668,4.261)--cycle;
\def\gpfillpath{(3.311,0.616)--(3.538,0.616)--(3.538,1.116)--(3.311,1.116)--cycle}
\gpfill{color=gpbgfillcolor} \gpfillpath;
\gpfill{rgb color={0.294,0.000,0.510},gp pattern 6,pattern color=.} \gpfillpath;
\draw[gp path] (3.311,0.616)--(3.311,1.115)--(3.537,1.115)--(3.537,0.616)--cycle;
\def\gpfillpath{(4.892,0.616)--(5.119,0.616)--(5.119,3.337)--(4.892,3.337)--cycle}
\gpfill{color=gpbgfillcolor} \gpfillpath;
\gpfill{rgb color={0.294,0.000,0.510},gp pattern 6,pattern color=.} \gpfillpath;
\draw[gp path] (4.892,0.616)--(4.892,3.336)--(5.118,3.336)--(5.118,0.616)--cycle;
\gpcolor{color=gp lt color border}
\node[gp node right] at (6.688,4.492) {Turi};
\def\gpfillpath{(6.872,4.415)--(7.788,4.415)--(7.788,4.569)--(6.872,4.569)--cycle}
\gpfill{color=gpbgfillcolor} \gpfillpath;
\gpfill{rgb color={1.000,0.000,0.000},gp pattern 8,pattern color=.} \gpfillpath;
\gpcolor{rgb color={1.000,0.000,0.000}}
\gpsetlinetype{gp lt plot 4}
\draw[gp path] (6.872,4.415)--(7.788,4.415)--(7.788,4.569)--(6.872,4.569)--cycle;
\def\gpfillpath{(3.537,0.616)--(3.764,0.616)--(3.764,1.425)--(3.537,1.425)--cycle}
\gpfill{color=gpbgfillcolor} \gpfillpath;
\gpfill{rgb color={1.000,0.000,0.000},gp pattern 8,pattern color=.} \gpfillpath;
\draw[gp path] (3.537,0.616)--(3.537,1.424)--(3.763,1.424)--(3.763,0.616)--cycle;
\gpcolor{color=gp lt color border}
\gpsetlinetype{gp lt border}
\draw[gp path] (1.504,3.533)--(1.504,0.616)--(7.829,0.616)--(7.829,3.533)--cycle;
\gpdefrectangularnode{gp plot 1}{\pgfpoint{1.504cm}{0.616cm}}{\pgfpoint{7.829cm}{3.533cm}}
\end{tikzpicture}

%% file: charts/mem.rmvm.tex
\begin{tikzpicture}[gnuplot]
\path (0.000,0.000) rectangle (8.382,4.826);
\gpcolor{color=gp lt color border}
\gpsetlinetype{gp lt border}
\gpsetlinewidth{1.00}
\draw[gp path] (1.504,0.616)--(1.684,0.616);
\draw[gp path] (7.829,0.616)--(7.649,0.616);
\node[gp node right] at (1.320,0.616) { 0};
\draw[gp path] (1.504,1.165)--(1.684,1.165);
\draw[gp path] (7.829,1.165)--(7.649,1.165);
\node[gp node right] at (1.320,1.165) { 50};
\draw[gp path] (1.504,1.713)--(1.684,1.713);
\draw[gp path] (7.829,1.713)--(7.649,1.713);
\node[gp node right] at (1.320,1.713) { 100};
\draw[gp path] (1.504,2.262)--(1.684,2.262);
\draw[gp path] (7.829,2.262)--(7.649,2.262);
\node[gp node right] at (1.320,2.262) { 150};
\draw[gp path] (1.504,2.811)--(1.684,2.811);
\draw[gp path] (7.829,2.811)--(7.649,2.811);
\node[gp node right] at (1.320,2.811) { 200};
\draw[gp path] (1.504,3.360)--(1.684,3.360);
\draw[gp path] (7.829,3.360)--(7.649,3.360);
\node[gp node right] at (1.320,3.360) { 250};
\draw[gp path] (1.504,3.908)--(1.684,3.908);
\draw[gp path] (7.829,3.908)--(7.649,3.908);
\node[gp node right] at (1.320,3.908) { 300};
\draw[gp path] (1.504,4.457)--(1.684,4.457);
\draw[gp path] (7.829,4.457)--(7.649,4.457);
\node[gp node right] at (1.320,4.457) { 350};
\draw[gp path] (3.085,0.616)--(3.085,0.796);
\draw[gp path] (3.085,4.457)--(3.085,4.277);
\node[gp node center] at (3.085,0.308) {RM$_{856M}$};
\draw[gp path] (4.667,0.616)--(4.667,0.796);
\draw[gp path] (4.667,4.457)--(4.667,4.277);
\node[gp node center] at (4.667,0.308) {RM$_{1B}$};
\draw[gp path] (6.248,0.616)--(6.248,0.796);
\draw[gp path] (6.248,4.457)--(6.248,4.277);
\node[gp node center] at (6.248,0.308) {RM$_{2B}$};
\draw[gp path] (1.504,4.457)--(1.504,0.616)--(7.829,0.616)--(7.829,4.457)--cycle;
\node[gp node center,rotate=-270] at (0.246,2.536) {Memory (GB)};
\gpfill{rgb color={0.000,1.000,0.000},color=.!50} (2.633,0.616)--(2.860,0.616)--(2.860,1.835)--(2.633,1.835)--cycle;
\gpfill{rgb color={0.000,1.000,0.000},color=.!50} (4.215,0.616)--(4.442,0.616)--(4.442,3.437)--(4.215,3.437)--cycle;
\def\gpfillpath{(2.859,0.616)--(3.086,0.616)--(3.086,0.891)--(2.859,0.891)--cycle}
\gpfill{color=gpbgfillcolor} \gpfillpath;
\gpfill{rgb color={0.373,0.620,0.627},gp pattern 4,pattern color=.} \gpfillpath;
\gpcolor{rgb color={0.373,0.620,0.627}}
\gpsetlinetype{gp lt plot 1}
\draw[gp path] (2.859,0.616)--(2.859,0.890)--(3.085,0.890)--(3.085,0.616)--cycle;
\def\gpfillpath{(4.441,0.616)--(4.668,0.616)--(4.668,0.946)--(4.441,0.946)--cycle}
\gpfill{color=gpbgfillcolor} \gpfillpath;
\gpfill{rgb color={0.373,0.620,0.627},gp pattern 4,pattern color=.} \gpfillpath;
\draw[gp path] (4.441,0.616)--(4.441,0.945)--(4.667,0.945)--(4.667,0.616)--cycle;
\def\gpfillpath{(6.022,0.616)--(6.249,0.616)--(6.249,1.188)--(6.022,1.188)--cycle}
\gpfill{color=gpbgfillcolor} \gpfillpath;
\gpfill{rgb color={0.373,0.620,0.627},gp pattern 4,pattern color=.} \gpfillpath;
\draw[gp path] (6.022,0.616)--(6.022,1.187)--(6.248,1.187)--(6.248,0.616)--cycle;
\def\gpfillpath{(3.085,0.616)--(3.312,0.616)--(3.312,1.813)--(3.085,1.813)--cycle}
\gpfill{color=gpbgfillcolor} \gpfillpath;
\gpfill{rgb color={0.498,0.498,0.498},gp pattern 7,pattern color=.} \gpfillpath;
\gpcolor{rgb color={0.498,0.498,0.498}}
\gpsetlinetype{gp lt plot 2}
\draw[gp path] (3.085,0.616)--(3.085,1.812)--(3.311,1.812)--(3.311,0.616)--cycle;
\def\gpfillpath{(4.667,0.616)--(4.893,0.616)--(4.893,3.920)--(4.667,3.920)--cycle}
\gpfill{color=gpbgfillcolor} \gpfillpath;
\gpfill{rgb color={0.498,0.498,0.498},gp pattern 7,pattern color=.} \gpfillpath;
\draw[gp path] (4.667,0.616)--(4.667,3.919)--(4.892,3.919)--(4.892,0.616)--cycle;
\def\gpfillpath{(3.311,0.616)--(3.538,0.616)--(3.538,3.953)--(3.311,3.953)--cycle}
\gpfill{color=gpbgfillcolor} \gpfillpath;
\gpfill{rgb color={0.294,0.000,0.510},gp pattern 6,pattern color=.} \gpfillpath;
\gpcolor{rgb color={0.294,0.000,0.510}}
\gpsetlinetype{gp lt plot 3}
\draw[gp path] (3.311,0.616)--(3.311,3.952)--(3.537,3.952)--(3.537,0.616)--cycle;
\def\gpfillpath{(4.892,0.616)--(5.119,0.616)--(5.119,4.118)--(4.892,4.118)--cycle}
\gpfill{color=gpbgfillcolor} \gpfillpath;
\gpfill{rgb color={0.294,0.000,0.510},gp pattern 6,pattern color=.} \gpfillpath;
\draw[gp path] (4.892,0.616)--(4.892,4.117)--(5.118,4.117)--(5.118,0.616)--cycle;
\def\gpfillpath{(3.537,0.616)--(3.764,0.616)--(3.764,2.055)--(3.537,2.055)--cycle}
\gpfill{color=gpbgfillcolor} \gpfillpath;
\gpfill{rgb color={1.000,0.000,0.000},gp pattern 8,pattern color=.} \gpfillpath;
\gpcolor{rgb color={1.000,0.000,0.000}}
\gpsetlinetype{gp lt plot 4}
\draw[gp path] (3.537,0.616)--(3.537,2.054)--(3.763,2.054)--(3.763,0.616)--cycle;
\gpcolor{color=gp lt color border}
\gpsetlinetype{gp lt border}
\draw[gp path] (1.504,4.457)--(1.504,0.616)--(7.829,0.616)--(7.829,4.457)--cycle;
\gpdefrectangularnode{gp plot 1}{\pgfpoint{1.504cm}{0.616cm}}{\pgfpoint{7.829cm}{4.457cm}}
\end{tikzpicture}

%% file: charts/numa.speedup.dist.friendster32.tex
\begin{tikzpicture}[gnuplot]
\path (0.000,0.000) rectangle (8.382,5.334);
\gpcolor{color=gp lt color border}
\gpsetlinetype{gp lt border}
\gpsetlinewidth{1.00}
\draw[gp path] (1.136,1.725)--(1.316,1.725);
\draw[gp path] (7.829,1.725)--(7.649,1.725);
\node[gp node right] at (0.952,1.725) {24};
\draw[gp path] (1.136,2.497)--(1.316,2.497);
\draw[gp path] (7.829,2.497)--(7.649,2.497);
\node[gp node right] at (0.952,2.497) {48};
\draw[gp path] (1.136,4.041)--(1.316,4.041);
\draw[gp path] (7.829,4.041)--(7.649,4.041);
\node[gp node right] at (0.952,4.041) {96};
\draw[gp path] (2.756,0.985)--(2.756,1.165);
\draw[gp path] (2.756,4.041)--(2.756,3.861);
\node[gp node center] at (2.756,0.677) {24};
\draw[gp path] (4.447,0.985)--(4.447,1.165);
\draw[gp path] (4.447,4.041)--(4.447,3.861);
\node[gp node center] at (4.447,0.677) {48};
\draw[gp path] (7.829,0.985)--(7.829,1.165);
\draw[gp path] (7.829,4.041)--(7.829,3.861);
\node[gp node center] at (7.829,0.677) {96};
\draw[gp path] (1.136,4.041)--(1.136,0.985)--(7.829,0.985)--(7.829,4.041)--cycle;
\node[gp node center,rotate=-270] at (0.246,2.513) {Relative Performance};
\node[gp node center] at (4.482,0.215) {No. of Threads};
\node[gp node right] at (3.198,5.000) {MLlib-EC2};
\gpcolor{rgb color={0.294,0.000,0.510}}
\gpsetlinetype{gp lt plot 0}
\draw[gp path] (3.382,5.000)--(4.298,5.000);
\draw[gp path] (1.136,0.985)--(2.756,1.725)--(4.447,2.343)--(7.829,3.703);
\gpsetpointsize{4.00}
\gppoint{gp mark 1}{(1.136,0.985)}
\gppoint{gp mark 1}{(2.756,1.725)}
\gppoint{gp mark 1}{(4.447,2.343)}
\gppoint{gp mark 1}{(7.829,3.703)}
\gppoint{gp mark 1}{(3.840,5.000)}
\gpcolor{color=gp lt color border}
\node[gp node right] at (3.198,4.692) {MPI};
\gpcolor{rgb color={1.000,0.412,0.706}}
\draw[gp path] (3.382,4.692)--(4.298,4.692);
\draw[gp path] (1.136,0.985)--(2.756,1.725)--(4.447,2.292)--(7.829,3.168);
\gppoint{gp mark 1}{(1.136,0.985)}
\gppoint{gp mark 1}{(2.756,1.725)}
\gppoint{gp mark 1}{(4.447,2.292)}
\gppoint{gp mark 1}{(7.829,3.168)}
\gppoint{gp mark 1}{(3.840,4.692)}
\gpcolor{color=gp lt color border}
\node[gp node right] at (7.058,5.000) {\textsf{knord}};
\gpcolor{rgb color={0.000,1.000,0.000}}
\draw[gp path] (7.242,5.000)--(8.158,5.000);
\draw[gp path] (1.136,0.985)--(2.756,1.725)--(4.447,2.361)--(7.829,3.369);
\gppoint{gp mark 1}{(1.136,0.985)}
\gppoint{gp mark 1}{(2.756,1.725)}
\gppoint{gp mark 1}{(4.447,2.361)}
\gppoint{gp mark 1}{(7.829,3.369)}
\gppoint{gp mark 1}{(7.700,5.000)}
\gpcolor{color=gp lt color border}
\node[gp node right] at (7.058,4.692) {Linear (Ideal)};
\gpcolor{rgb color={0.000,0.749,1.000}}
\gpsetlinetype{gp lt plot 1}
\draw[gp path] (7.242,4.692)--(8.158,4.692);
\draw[gp path] (1.136,0.985)--(2.756,1.725)--(4.447,2.497)--(7.829,4.041);
\gppoint{gp mark 2}{(1.136,0.985)}
\gppoint{gp mark 2}{(2.756,1.725)}
\gppoint{gp mark 2}{(4.447,2.497)}
\gppoint{gp mark 2}{(7.829,4.041)}
\gppoint{gp mark 2}{(7.700,4.692)}
\gpcolor{color=gp lt color border}
\gpsetlinetype{gp lt border}
\draw[gp path] (1.136,4.041)--(1.136,0.985)--(7.829,0.985)--(7.829,4.041)--cycle;
\gpdefrectangularnode{gp plot 1}{\pgfpoint{1.136cm}{0.985cm}}{\pgfpoint{7.829cm}{4.041cm}}
\end{tikzpicture}

%% file: charts/numa.speedup.dist.rm1b.tex
\begin{tikzpicture}[gnuplot]
\path (0.000,0.000) rectangle (8.382,5.334);
\gpcolor{color=gp lt color border}
\gpsetlinetype{gp lt border}
\gpsetlinewidth{1.00}
\draw[gp path] (1.320,1.741)--(1.500,1.741);
\draw[gp path] (7.829,1.741)--(7.649,1.741);
\node[gp node right] at (1.136,1.741) {72};
\draw[gp path] (1.320,2.508)--(1.500,2.508);
\draw[gp path] (7.829,2.508)--(7.649,2.508);
\node[gp node right] at (1.136,2.508) {144};
\draw[gp path] (1.320,4.041)--(1.500,4.041);
\draw[gp path] (7.829,4.041)--(7.649,4.041);
\node[gp node right] at (1.136,4.041) {288};
\draw[gp path] (2.930,0.985)--(2.930,1.165);
\draw[gp path] (2.930,4.041)--(2.930,3.861);
\node[gp node center] at (2.930,0.677) {72};
\draw[gp path] (4.563,0.985)--(4.563,1.165);
\draw[gp path] (4.563,4.041)--(4.563,3.861);
\node[gp node center] at (4.563,0.677) {144};
\draw[gp path] (7.829,0.985)--(7.829,1.165);
\draw[gp path] (7.829,4.041)--(7.829,3.861);
\node[gp node center] at (7.829,0.677) {288};
\draw[gp path] (1.320,4.041)--(1.320,0.985)--(7.829,0.985)--(7.829,4.041)--cycle;
\node[gp node center,rotate=-270] at (0.246,2.513) {Relative Performance};
\node[gp node center] at (4.574,0.215) {No. of Threads};
\node[gp node right] at (3.290,5.000) {MLlib-EC2};
\gpcolor{rgb color={0.294,0.000,0.510}}
\gpsetlinetype{gp lt plot 0}
\draw[gp path] (3.474,5.000)--(4.390,5.000);
\draw[gp path] (1.320,0.985)--(2.930,1.741)--(4.563,2.305)--(7.829,3.821);
\gpsetpointsize{4.00}
\gppoint{gp mark 1}{(1.320,0.985)}
\gppoint{gp mark 1}{(2.930,1.741)}
\gppoint{gp mark 1}{(4.563,2.305)}
\gppoint{gp mark 1}{(7.829,3.821)}
\gppoint{gp mark 1}{(3.932,5.000)}
\gpcolor{color=gp lt color border}
\node[gp node right] at (3.290,4.692) {MPI};
\gpcolor{rgb color={1.000,0.412,0.706}}
\draw[gp path] (3.474,4.692)--(4.390,4.692);
\draw[gp path] (1.320,0.985)--(2.930,1.741)--(4.563,2.231)--(7.829,3.278);
\gppoint{gp mark 1}{(1.320,0.985)}
\gppoint{gp mark 1}{(2.930,1.741)}
\gppoint{gp mark 1}{(4.563,2.231)}
\gppoint{gp mark 1}{(7.829,3.278)}
\gppoint{gp mark 1}{(3.932,4.692)}
\gpcolor{color=gp lt color border}
\node[gp node right] at (7.150,5.000) {\textsf{knord}};
\gpcolor{rgb color={0.000,1.000,0.000}}
\draw[gp path] (7.334,5.000)--(8.250,5.000);
\draw[gp path] (1.320,0.985)--(2.930,1.741)--(4.563,2.380)--(7.829,3.478);
\gppoint{gp mark 1}{(1.320,0.985)}
\gppoint{gp mark 1}{(2.930,1.741)}
\gppoint{gp mark 1}{(4.563,2.380)}
\gppoint{gp mark 1}{(7.829,3.478)}
\gppoint{gp mark 1}{(7.792,5.000)}
\gpcolor{color=gp lt color border}
\node[gp node right] at (7.150,4.692) {Linear (Ideal)};
\gpcolor{rgb color={0.000,0.749,1.000}}
\gpsetlinetype{gp lt plot 1}
\draw[gp path] (7.334,4.692)--(8.250,4.692);
\draw[gp path] (1.320,0.985)--(2.930,1.741)--(4.563,2.508)--(7.829,4.041);
\gppoint{gp mark 2}{(1.320,0.985)}
\gppoint{gp mark 2}{(2.930,1.741)}
\gppoint{gp mark 2}{(4.563,2.508)}
\gppoint{gp mark 2}{(7.829,4.041)}
\gppoint{gp mark 2}{(7.792,4.692)}
\gpcolor{color=gp lt color border}
\gpsetlinetype{gp lt border}
\draw[gp path] (1.320,4.041)--(1.320,0.985)--(7.829,0.985)--(7.829,4.041)--cycle;
\gpdefrectangularnode{gp plot 1}{\pgfpoint{1.320cm}{0.985cm}}{\pgfpoint{7.829cm}{4.041cm}}
\end{tikzpicture}

%% file: charts/titles.tex
\begin{tikzpicture}[gnuplot]
\gpcolor{color=gp lt color border}
\gpsetlinetype{gp lt border}
\gpsetlinewidth{1.00}
\node[gp node right] at (3.198,4.492) {\textsf{knord}};
\gpfill{rgb color={0.000,1.000,0.000},color=.!50} (3.382,4.415)--(4.298,4.415)--(4.298,4.569)--(3.382,4.569)--cycle;
\node[gp node right] at (3.198,4.184) {MPI};
\gpfill{rgb color={1.000,0.412,0.706},color=.!50} (3.382,4.107)--(4.298,4.107)--(4.298,4.261)--(3.382,4.261)--cycle;
\node[gp node right] at (3.198,3.876) {\textsf{knord-}};
\def\gpfillpath{(3.382,3.799)--(4.298,3.799)--(4.298,3.953)--(3.382,3.953)--cycle}
\gpfill{color=gpbgfillcolor} \gpfillpath;
\gpfill{rgb color={0.498,1.000,0.831},gp pattern 3,pattern color=.} \gpfillpath;
\node[gp node right] at (6.138,4.492) {MPI-};
\gpfill{rgb color={1.000,0.000,1.000},color=.!50} (6.322,4.415)--(7.238,4.415)--(7.238,4.569)--(6.322,4.569)--cycle;
\node[gp node right] at (6.138,4.184) {MLlib-EC2};
\def\gpfillpath{(6.322,4.107)--(7.238,4.107)--(7.238,4.261)--(6.322,4.261)--cycle}
\gpfill{color=gpbgfillcolor} \gpfillpath;
\gpfill{rgb color={0.294,0.000,0.510},gp pattern 6,pattern color=.} \gpfillpath;
\gpcolor{rgb color={0.294,0.000,0.510}}
\gpsetlinetype{gp lt plot 4}
\draw[gp path] (6.322,4.107)--(7.238,4.107)--(7.238,4.261)--(6.322,4.261)--cycle;
\gpcolor{color=gp lt color border}
\end{tikzpicture}

%% file: charts/perf.iter.dist.friendsterX.tex
\begin{tikzpicture}[gnuplot]
\path (0.000,0.000) rectangle (8.382,4.826);
\gpcolor{color=gp lt color border}
\gpsetlinetype{gp lt border}
\gpsetlinewidth{1.00}
\draw[gp path] (1.136,0.985)--(1.316,0.985);
\draw[gp path] (3.638,0.985)--(3.458,0.985);
\node[gp node right] at (0.952,0.985) { 0};
\draw[gp path] (1.136,1.419)--(1.316,1.419);
\draw[gp path] (3.638,1.419)--(3.458,1.419);
\node[gp node right] at (0.952,1.419) { 1};
\draw[gp path] (1.136,1.853)--(1.316,1.853);
\draw[gp path] (3.638,1.853)--(3.458,1.853);
\node[gp node right] at (0.952,1.853) { 2};
\draw[gp path] (1.136,2.287)--(1.316,2.287);
\draw[gp path] (3.638,2.287)--(3.458,2.287);
\node[gp node right] at (0.952,2.287) { 3};
\draw[gp path] (1.136,2.721)--(1.316,2.721);
\draw[gp path] (3.638,2.721)--(3.458,2.721);
\node[gp node right] at (0.952,2.721) { 4};
\draw[gp path] (1.136,3.155)--(1.316,3.155);
\draw[gp path] (3.638,3.155)--(3.458,3.155);
\node[gp node right] at (0.952,3.155) { 5};
\draw[gp path] (1.136,3.589)--(1.316,3.589);
\draw[gp path] (3.638,3.589)--(3.458,3.589);
\node[gp node right] at (0.952,3.589) { 6};
\draw[gp path] (1.136,4.023)--(1.316,4.023);
\draw[gp path] (3.638,4.023)--(3.458,4.023);
\node[gp node right] at (0.952,4.023) { 7};
\draw[gp path] (1.136,4.457)--(1.316,4.457);
\draw[gp path] (3.638,4.457)--(3.458,4.457);
\node[gp node right] at (0.952,4.457) { 8};
\draw[gp path] (1.970,0.985)--(1.970,1.165);
\draw[gp path] (1.970,4.457)--(1.970,4.277);
\node[gp node center] at (1.970,0.677) {48};
\draw[gp path] (2.804,0.985)--(2.804,1.165);
\draw[gp path] (2.804,4.457)--(2.804,4.277);
\node[gp node center] at (2.804,0.677) {64};
\draw[gp path] (1.136,4.457)--(1.136,0.985)--(3.638,0.985)--(3.638,4.457)--cycle;
\node[gp node center,rotate=-270] at (0.246,2.721) {Time/iter (sec)};
\node[gp node center] at (2.387,0.215) {No. Cores};
\gpfill{rgb color={0.000,1.000,0.000},color=.!50} (1.732,0.985)--(1.852,0.985)--(1.852,1.187)--(1.732,1.187)--cycle;
\gpfill{rgb color={0.000,1.000,0.000},color=.!50} (2.566,0.985)--(2.686,0.985)--(2.686,1.126)--(2.566,1.126)--cycle;
\gpfill{rgb color={1.000,0.412,0.706},color=.!50} (1.851,0.985)--(1.971,0.985)--(1.971,1.333)--(1.851,1.333)--cycle;
\gpfill{rgb color={1.000,0.412,0.706},color=.!50} (2.685,0.985)--(2.805,0.985)--(2.805,1.294)--(2.685,1.294)--cycle;
\def\gpfillpath{(1.970,0.985)--(2.090,0.985)--(2.090,1.831)--(1.970,1.831)--cycle}
\gpfill{color=gpbgfillcolor} \gpfillpath;
\gpfill{rgb color={0.498,1.000,0.831},gp pattern 3,pattern color=.} \gpfillpath;
\gpcolor{rgb color={0.498,1.000,0.831}}
\gpsetlinetype{gp lt plot 2}
\draw[gp path] (1.970,0.985)--(1.970,1.830)--(2.089,1.830)--(2.089,0.985)--cycle;
\def\gpfillpath{(2.804,0.985)--(2.924,0.985)--(2.924,1.690)--(2.804,1.690)--cycle}
\gpfill{color=gpbgfillcolor} \gpfillpath;
\gpfill{rgb color={0.498,1.000,0.831},gp pattern 3,pattern color=.} \gpfillpath;
\draw[gp path] (2.804,0.985)--(2.804,1.689)--(2.923,1.689)--(2.923,0.985)--cycle;
\gpfill{rgb color={1.000,0.000,1.000},color=.!50} (2.089,0.985)--(2.209,0.985)--(2.209,2.355)--(2.089,2.355)--cycle;
\gpfill{rgb color={1.000,0.000,1.000},color=.!50} (2.923,0.985)--(3.043,0.985)--(3.043,2.081)--(2.923,2.081)--cycle;
\def\gpfillpath{(2.208,0.985)--(2.328,0.985)--(2.328,4.075)--(2.208,4.075)--cycle}
\gpfill{color=gpbgfillcolor} \gpfillpath;
\gpfill{rgb color={0.294,0.000,0.510},gp pattern 6,pattern color=.} \gpfillpath;
\gpcolor{rgb color={0.294,0.000,0.510}}
\gpsetlinetype{gp lt plot 4}
\draw[gp path] (2.208,0.985)--(2.208,4.074)--(2.327,4.074)--(2.327,0.985)--cycle;
\def\gpfillpath{(3.042,0.985)--(3.162,0.985)--(3.162,3.721)--(3.042,3.721)--cycle}
\gpfill{color=gpbgfillcolor} \gpfillpath;
\gpfill{rgb color={0.294,0.000,0.510},gp pattern 6,pattern color=.} \gpfillpath;
\draw[gp path] (3.042,0.985)--(3.042,3.720)--(3.161,3.720)--(3.161,0.985)--cycle;
\gpcolor{color=gp lt color border}
\gpsetlinetype{gp lt border}
\draw[gp path] (1.136,4.457)--(1.136,0.985)--(3.638,0.985)--(3.638,4.457)--cycle;
\gpdefrectangularnode{gp plot 1}{\pgfpoint{1.136cm}{0.985cm}}{\pgfpoint{3.638cm}{4.457cm}}
\draw[gp path] (5.203,0.985)--(5.383,0.985);
\draw[gp path] (7.829,0.985)--(7.649,0.985);
\node[gp node right] at (5.019,0.985) { 0};
\draw[gp path] (5.203,1.332)--(5.383,1.332);
\draw[gp path] (7.829,1.332)--(7.649,1.332);
\node[gp node right] at (5.019,1.332) { 1};
\draw[gp path] (5.203,1.679)--(5.383,1.679);
\draw[gp path] (7.829,1.679)--(7.649,1.679);
\node[gp node right] at (5.019,1.679) { 2};
\draw[gp path] (5.203,2.027)--(5.383,2.027);
\draw[gp path] (7.829,2.027)--(7.649,2.027);
\node[gp node right] at (5.019,2.027) { 3};
\draw[gp path] (5.203,2.374)--(5.383,2.374);
\draw[gp path] (7.829,2.374)--(7.649,2.374);
\node[gp node right] at (5.019,2.374) { 4};
\draw[gp path] (5.203,2.721)--(5.383,2.721);
\draw[gp path] (7.829,2.721)--(7.649,2.721);
\node[gp node right] at (5.019,2.721) { 5};
\draw[gp path] (5.203,3.068)--(5.383,3.068);
\draw[gp path] (7.829,3.068)--(7.649,3.068);
\node[gp node right] at (5.019,3.068) { 6};
\draw[gp path] (5.203,3.415)--(5.383,3.415);
\draw[gp path] (7.829,3.415)--(7.649,3.415);
\node[gp node right] at (5.019,3.415) { 7};
\draw[gp path] (5.203,3.763)--(5.383,3.763);
\draw[gp path] (7.829,3.763)--(7.649,3.763);
\node[gp node right] at (5.019,3.763) { 8};
\draw[gp path] (5.203,4.110)--(5.383,4.110);
\draw[gp path] (7.829,4.110)--(7.649,4.110);
\node[gp node right] at (5.019,4.110) { 9};
\draw[gp path] (5.203,4.457)--(5.383,4.457);
\draw[gp path] (7.829,4.457)--(7.649,4.457);
\node[gp node right] at (5.019,4.457) { 10};
\draw[gp path] (5.860,0.985)--(5.860,1.165);
\draw[gp path] (5.860,4.457)--(5.860,4.277);
\node[gp node center] at (5.860,0.677) {48};
\draw[gp path] (6.516,0.985)--(6.516,1.165);
\draw[gp path] (6.516,4.457)--(6.516,4.277);
\node[gp node center] at (6.516,0.677) {96};
\draw[gp path] (7.173,0.985)--(7.173,1.165);
\draw[gp path] (7.173,4.457)--(7.173,4.277);
\node[gp node center] at (7.173,0.677) {126};
\draw[gp path] (5.203,4.457)--(5.203,0.985)--(7.829,0.985)--(7.829,4.457)--cycle;
\node[gp node center] at (6.516,0.215) {No. Cores};
\gpfill{rgb color={0.000,1.000,0.000},color=.!50} (5.672,0.985)--(5.767,0.985)--(5.767,1.175)--(5.672,1.175)--cycle;
\gpfill{rgb color={0.000,1.000,0.000},color=.!50} (6.328,0.985)--(6.423,0.985)--(6.423,1.104)--(6.328,1.104)--cycle;
\gpfill{rgb color={0.000,1.000,0.000},color=.!50} (6.985,0.985)--(7.080,0.985)--(7.080,1.065)--(6.985,1.065)--cycle;
\gpfill{rgb color={1.000,0.412,0.706},color=.!50} (5.766,0.985)--(5.861,0.985)--(5.861,1.190)--(5.766,1.190)--cycle;
\gpfill{rgb color={1.000,0.412,0.706},color=.!50} (6.422,0.985)--(6.517,0.985)--(6.517,1.116)--(6.422,1.116)--cycle;
\gpfill{rgb color={1.000,0.412,0.706},color=.!50} (7.079,0.985)--(7.174,0.985)--(7.174,1.094)--(7.079,1.094)--cycle;
\def\gpfillpath{(5.860,0.985)--(5.954,0.985)--(5.954,2.781)--(5.860,2.781)--cycle}
\gpfill{color=gpbgfillcolor} \gpfillpath;
\gpfill{rgb color={0.498,1.000,0.831},gp pattern 3,pattern color=.} \gpfillpath;
\gpcolor{rgb color={0.498,1.000,0.831}}
\gpsetlinetype{gp lt plot 2}
\draw[gp path] (5.860,0.985)--(5.860,2.780)--(5.953,2.780)--(5.953,0.985)--cycle;
\def\gpfillpath{(6.516,0.985)--(6.611,0.985)--(6.611,2.032)--(6.516,2.032)--cycle}
\gpfill{color=gpbgfillcolor} \gpfillpath;
\gpfill{rgb color={0.498,1.000,0.831},gp pattern 3,pattern color=.} \gpfillpath;
\draw[gp path] (6.516,0.985)--(6.516,2.031)--(6.610,2.031)--(6.610,0.985)--cycle;
\def\gpfillpath{(7.173,0.985)--(7.267,0.985)--(7.267,1.823)--(7.173,1.823)--cycle}
\gpfill{color=gpbgfillcolor} \gpfillpath;
\gpfill{rgb color={0.498,1.000,0.831},gp pattern 3,pattern color=.} \gpfillpath;
\draw[gp path] (7.173,0.985)--(7.173,1.822)--(7.266,1.822)--(7.266,0.985)--cycle;
\gpfill{rgb color={1.000,0.000,1.000},color=.!50} (5.953,0.985)--(6.048,0.985)--(6.048,3.674)--(5.953,3.674)--cycle;
\gpfill{rgb color={1.000,0.000,1.000},color=.!50} (6.610,0.985)--(6.705,0.985)--(6.705,2.611)--(6.610,2.611)--cycle;
\gpfill{rgb color={1.000,0.000,1.000},color=.!50} (7.266,0.985)--(7.361,0.985)--(7.361,2.079)--(7.266,2.079)--cycle;
\def\gpfillpath{(6.047,0.985)--(6.142,0.985)--(6.142,4.114)--(6.047,4.114)--cycle}
\gpfill{color=gpbgfillcolor} \gpfillpath;
\gpfill{rgb color={0.294,0.000,0.510},gp pattern 6,pattern color=.} \gpfillpath;
\gpcolor{rgb color={0.294,0.000,0.510}}
\gpsetlinetype{gp lt plot 4}
\draw[gp path] (6.047,0.985)--(6.047,4.113)--(6.141,4.113)--(6.141,0.985)--cycle;
\def\gpfillpath{(6.704,0.985)--(6.798,0.985)--(6.798,2.567)--(6.704,2.567)--cycle}
\gpfill{color=gpbgfillcolor} \gpfillpath;
\gpfill{rgb color={0.294,0.000,0.510},gp pattern 6,pattern color=.} \gpfillpath;
\draw[gp path] (6.704,0.985)--(6.704,2.566)--(6.797,2.566)--(6.797,0.985)--cycle;
\def\gpfillpath{(7.360,0.985)--(7.455,0.985)--(7.455,2.344)--(7.360,2.344)--cycle}
\gpfill{color=gpbgfillcolor} \gpfillpath;
\gpfill{rgb color={0.294,0.000,0.510},gp pattern 6,pattern color=.} \gpfillpath;
\draw[gp path] (7.360,0.985)--(7.360,2.343)--(7.454,2.343)--(7.454,0.985)--cycle;
\gpcolor{color=gp lt color border}
\gpsetlinetype{gp lt border}
\draw[gp path] (5.203,4.457)--(5.203,0.985)--(7.829,0.985)--(7.829,4.457)--cycle;
\gpdefrectangularnode{gp plot 2}{\pgfpoint{5.203cm}{0.985cm}}{\pgfpoint{7.829cm}{4.457cm}}
\end{tikzpicture}

%% file: charts/perf.iter.dist.rand.tex
\begin{tikzpicture}[gnuplot]
\path (0.000,0.000) rectangle (8.382,4.826);
\gpcolor{color=gp lt color border}
\gpsetlinetype{gp lt border}
\gpsetlinewidth{1.00}
\draw[gp path] (1.136,0.985)--(1.316,0.985);
\draw[gp path] (3.638,0.985)--(3.458,0.985);
\node[gp node right] at (0.952,0.985) {0};
\draw[gp path] (1.136,1.481)--(1.316,1.481);
\draw[gp path] (3.638,1.481)--(3.458,1.481);
\node[gp node right] at (0.952,1.481) {10};
\draw[gp path] (1.136,1.977)--(1.316,1.977);
\draw[gp path] (3.638,1.977)--(3.458,1.977);
\node[gp node right] at (0.952,1.977) {20};
\draw[gp path] (1.136,2.473)--(1.316,2.473);
\draw[gp path] (3.638,2.473)--(3.458,2.473);
\node[gp node right] at (0.952,2.473) {30};
\draw[gp path] (1.136,2.969)--(1.316,2.969);
\draw[gp path] (3.638,2.969)--(3.458,2.969);
\node[gp node right] at (0.952,2.969) {40};
\draw[gp path] (1.136,3.465)--(1.316,3.465);
\draw[gp path] (3.638,3.465)--(3.458,3.465);
\node[gp node right] at (0.952,3.465) {50};
\draw[gp path] (1.136,3.961)--(1.316,3.961);
\draw[gp path] (3.638,3.961)--(3.458,3.961);
\node[gp node right] at (0.952,3.961) {60};
\draw[gp path] (1.136,4.457)--(1.316,4.457);
\draw[gp path] (3.638,4.457)--(3.458,4.457);
\node[gp node right] at (0.952,4.457) {70};
\draw[gp path] (1.762,0.985)--(1.762,1.165);
\draw[gp path] (1.762,4.457)--(1.762,4.277);
\node[gp node center] at (1.762,0.677) {72};
\draw[gp path] (2.387,0.985)--(2.387,1.165);
\draw[gp path] (2.387,4.457)--(2.387,4.277);
\node[gp node center] at (2.387,0.677) {144};
\draw[gp path] (3.013,0.985)--(3.013,1.165);
\draw[gp path] (3.013,4.457)--(3.013,4.277);
\node[gp node center] at (3.013,0.677) {288};
\draw[gp path] (1.136,4.457)--(1.136,0.985)--(3.638,0.985)--(3.638,4.457)--cycle;
\node[gp node center,rotate=-270] at (0.246,2.721) {Time/iter (sec)};
\node[gp node center] at (2.387,0.215) {No. Cores};
\gpfill{rgb color={0.000,1.000,0.000},color=.!50} (1.583,0.985)--(1.673,0.985)--(1.673,1.409)--(1.583,1.409)--cycle;
\gpfill{rgb color={0.000,1.000,0.000},color=.!50} (2.208,0.985)--(2.299,0.985)--(2.299,1.204)--(2.208,1.204)--cycle;
\gpfill{rgb color={0.000,1.000,0.000},color=.!50} (2.834,0.985)--(2.924,0.985)--(2.924,1.121)--(2.834,1.121)--cycle;
\gpfill{rgb color={1.000,0.412,0.706},color=.!50} (1.672,0.985)--(1.763,0.985)--(1.763,1.654)--(1.672,1.654)--cycle;
\gpfill{rgb color={1.000,0.412,0.706},color=.!50} (2.298,0.985)--(2.388,0.985)--(2.388,1.575)--(2.298,1.575)--cycle;
\gpfill{rgb color={1.000,0.412,0.706},color=.!50} (2.923,0.985)--(3.014,0.985)--(3.014,1.436)--(2.923,1.436)--cycle;
\def\gpfillpath{(1.762,0.985)--(1.852,0.985)--(1.852,1.320)--(1.762,1.320)--cycle}
\gpfill{color=gpbgfillcolor} \gpfillpath;
\gpfill{rgb color={0.498,1.000,0.831},gp pattern 3,pattern color=.} \gpfillpath;
\gpcolor{rgb color={0.498,1.000,0.831}}
\gpsetlinetype{gp lt plot 2}
\draw[gp path] (1.762,0.985)--(1.762,1.319)--(1.851,1.319)--(1.851,0.985)--cycle;
\def\gpfillpath{(2.387,0.985)--(2.477,0.985)--(2.477,1.171)--(2.387,1.171)--cycle}
\gpfill{color=gpbgfillcolor} \gpfillpath;
\gpfill{rgb color={0.498,1.000,0.831},gp pattern 3,pattern color=.} \gpfillpath;
\draw[gp path] (2.387,0.985)--(2.387,1.170)--(2.476,1.170)--(2.476,0.985)--cycle;
\def\gpfillpath{(3.013,0.985)--(3.103,0.985)--(3.103,1.081)--(3.013,1.081)--cycle}
\gpfill{color=gpbgfillcolor} \gpfillpath;
\gpfill{rgb color={0.498,1.000,0.831},gp pattern 3,pattern color=.} \gpfillpath;
\draw[gp path] (3.013,0.985)--(3.013,1.080)--(3.102,1.080)--(3.102,0.985)--cycle;
\gpfill{rgb color={1.000,0.000,1.000},color=.!50} (1.851,0.985)--(1.941,0.985)--(1.941,1.694)--(1.851,1.694)--cycle;
\gpfill{rgb color={1.000,0.000,1.000},color=.!50} (2.476,0.985)--(2.567,0.985)--(2.567,1.530)--(2.476,1.530)--cycle;
\gpfill{rgb color={1.000,0.000,1.000},color=.!50} (3.102,0.985)--(3.192,0.985)--(3.192,1.445)--(3.102,1.445)--cycle;
\def\gpfillpath{(1.940,0.985)--(2.031,0.985)--(2.031,4.271)--(1.940,4.271)--cycle}
\gpfill{color=gpbgfillcolor} \gpfillpath;
\gpfill{rgb color={0.294,0.000,0.510},gp pattern 6,pattern color=.} \gpfillpath;
\gpcolor{rgb color={0.294,0.000,0.510}}
\gpsetlinetype{gp lt plot 4}
\draw[gp path] (1.940,0.985)--(1.940,4.270)--(2.030,4.270)--(2.030,0.985)--cycle;
\def\gpfillpath{(2.566,0.985)--(2.656,0.985)--(2.656,2.424)--(2.566,2.424)--cycle}
\gpfill{color=gpbgfillcolor} \gpfillpath;
\gpfill{rgb color={0.294,0.000,0.510},gp pattern 6,pattern color=.} \gpfillpath;
\draw[gp path] (2.566,0.985)--(2.566,2.423)--(2.655,2.423)--(2.655,0.985)--cycle;
\def\gpfillpath{(3.191,0.985)--(3.282,0.985)--(3.282,1.658)--(3.191,1.658)--cycle}
\gpfill{color=gpbgfillcolor} \gpfillpath;
\gpfill{rgb color={0.294,0.000,0.510},gp pattern 6,pattern color=.} \gpfillpath;
\draw[gp path] (3.191,0.985)--(3.191,1.657)--(3.281,1.657)--(3.281,0.985)--cycle;
\gpcolor{color=gp lt color border}
\gpsetlinetype{gp lt border}
\draw[gp path] (1.136,4.457)--(1.136,0.985)--(3.638,0.985)--(3.638,4.457)--cycle;
\gpdefrectangularnode{gp plot 1}{\pgfpoint{1.136cm}{0.985cm}}{\pgfpoint{3.638cm}{4.457cm}}
\draw[gp path] (5.019,0.985)--(5.199,0.985);
\draw[gp path] (7.829,0.985)--(7.649,0.985);
\node[gp node right] at (4.835,0.985) {0};
\draw[gp path] (5.019,1.564)--(5.199,1.564);
\draw[gp path] (7.829,1.564)--(7.649,1.564);
\node[gp node right] at (4.835,1.564) {5};
\draw[gp path] (5.019,2.142)--(5.199,2.142);
\draw[gp path] (7.829,2.142)--(7.649,2.142);
\node[gp node right] at (4.835,2.142) {10};
\draw[gp path] (5.019,2.721)--(5.199,2.721);
\draw[gp path] (7.829,2.721)--(7.649,2.721);
\node[gp node right] at (4.835,2.721) {15};
\draw[gp path] (5.019,3.300)--(5.199,3.300);
\draw[gp path] (7.829,3.300)--(7.649,3.300);
\node[gp node right] at (4.835,3.300) {20};
\draw[gp path] (5.019,3.878)--(5.199,3.878);
\draw[gp path] (7.829,3.878)--(7.649,3.878);
\node[gp node right] at (4.835,3.878) {25};
\draw[gp path] (5.019,4.457)--(5.199,4.457);
\draw[gp path] (7.829,4.457)--(7.649,4.457);
\node[gp node right] at (4.835,4.457) {30};
\draw[gp path] (5.956,0.985)--(5.956,1.165);
\draw[gp path] (5.956,4.457)--(5.956,4.277);
\node[gp node center] at (5.956,0.677) {144};
\draw[gp path] (6.892,0.985)--(6.892,1.165);
\draw[gp path] (6.892,4.457)--(6.892,4.277);
\node[gp node center] at (6.892,0.677) {288};
\draw[gp path] (5.019,4.457)--(5.019,0.985)--(7.829,0.985)--(7.829,4.457)--cycle;
\node[gp node center] at (6.424,0.215) {No. Cores};
\gpfill{rgb color={0.000,1.000,0.000},color=.!50} (5.688,0.985)--(5.823,0.985)--(5.823,1.129)--(5.688,1.129)--cycle;
\gpfill{rgb color={0.000,1.000,0.000},color=.!50} (6.625,0.985)--(6.760,0.985)--(6.760,1.087)--(6.625,1.087)--cycle;
\gpfill{rgb color={1.000,0.412,0.706},color=.!50} (5.822,0.985)--(5.957,0.985)--(5.957,1.176)--(5.822,1.176)--cycle;
\gpfill{rgb color={1.000,0.412,0.706},color=.!50} (6.759,0.985)--(6.893,0.985)--(6.893,1.134)--(6.759,1.134)--cycle;
\def\gpfillpath{(5.956,0.985)--(6.090,0.985)--(6.090,1.475)--(5.956,1.475)--cycle}
\gpfill{color=gpbgfillcolor} \gpfillpath;
\gpfill{rgb color={0.498,1.000,0.831},gp pattern 3,pattern color=.} \gpfillpath;
\gpcolor{rgb color={0.498,1.000,0.831}}
\gpsetlinetype{gp lt plot 2}
\draw[gp path] (5.956,0.985)--(5.956,1.474)--(6.089,1.474)--(6.089,0.985)--cycle;
\def\gpfillpath{(6.892,0.985)--(7.027,0.985)--(7.027,1.307)--(6.892,1.307)--cycle}
\gpfill{color=gpbgfillcolor} \gpfillpath;
\gpfill{rgb color={0.498,1.000,0.831},gp pattern 3,pattern color=.} \gpfillpath;
\draw[gp path] (6.892,0.985)--(6.892,1.306)--(7.026,1.306)--(7.026,0.985)--cycle;
\gpfill{rgb color={1.000,0.000,1.000},color=.!50} (6.089,0.985)--(6.224,0.985)--(6.224,2.411)--(6.089,2.411)--cycle;
\gpfill{rgb color={1.000,0.000,1.000},color=.!50} (7.026,0.985)--(7.161,0.985)--(7.161,1.763)--(7.026,1.763)--cycle;
\def\gpfillpath{(6.223,0.985)--(6.358,0.985)--(6.358,4.025)--(6.223,4.025)--cycle}
\gpfill{color=gpbgfillcolor} \gpfillpath;
\gpfill{rgb color={0.294,0.000,0.510},gp pattern 6,pattern color=.} \gpfillpath;
\gpcolor{rgb color={0.294,0.000,0.510}}
\gpsetlinetype{gp lt plot 4}
\draw[gp path] (6.223,0.985)--(6.223,4.024)--(6.357,4.024)--(6.357,0.985)--cycle;
\def\gpfillpath{(7.160,0.985)--(7.295,0.985)--(7.295,2.689)--(7.160,2.689)--cycle}
\gpfill{color=gpbgfillcolor} \gpfillpath;
\gpfill{rgb color={0.294,0.000,0.510},gp pattern 6,pattern color=.} \gpfillpath;
\draw[gp path] (7.160,0.985)--(7.160,2.688)--(7.294,2.688)--(7.294,0.985)--cycle;
\gpcolor{color=gp lt color border}
\gpsetlinetype{gp lt border}
\draw[gp path] (5.019,4.457)--(5.019,0.985)--(7.829,0.985)--(7.829,4.457)--cycle;
\gpdefrectangularnode{gp plot 2}{\pgfpoint{5.019cm}{0.985cm}}{\pgfpoint{7.829cm}{4.457cm}}
\end{tikzpicture}

%% file: charts/perf.iter.sem-ec2.tex
\begin{tikzpicture}[gnuplot]
\path (0.000,0.000) rectangle (8.382,6.096);
\gpcolor{color=gp lt color border}
\gpsetlinetype{gp lt border}
\gpsetlinewidth{1.00}
\draw[gp path] (1.504,1.845)--(1.684,1.845);
\draw[gp path] (7.829,1.845)--(7.649,1.845);
\node[gp node right] at (1.320,1.845) { 0.1};
\draw[gp path] (1.504,2.142)--(1.594,2.142);
\draw[gp path] (7.829,2.142)--(7.739,2.142);
\draw[gp path] (1.504,2.315)--(1.594,2.315);
\draw[gp path] (7.829,2.315)--(7.739,2.315);
\draw[gp path] (1.504,2.439)--(1.594,2.439);
\draw[gp path] (7.829,2.439)--(7.739,2.439);
\draw[gp path] (1.504,2.534)--(1.594,2.534);
\draw[gp path] (7.829,2.534)--(7.739,2.534);
\draw[gp path] (1.504,2.612)--(1.594,2.612);
\draw[gp path] (7.829,2.612)--(7.739,2.612);
\draw[gp path] (1.504,2.678)--(1.594,2.678);
\draw[gp path] (7.829,2.678)--(7.739,2.678);
\draw[gp path] (1.504,2.735)--(1.594,2.735);
\draw[gp path] (7.829,2.735)--(7.739,2.735);
\draw[gp path] (1.504,2.786)--(1.594,2.786);
\draw[gp path] (7.829,2.786)--(7.739,2.786);
\draw[gp path] (1.504,2.831)--(1.684,2.831);
\draw[gp path] (7.829,2.831)--(7.649,2.831);
\node[gp node right] at (1.320,2.831) { 1};
\draw[gp path] (1.504,3.128)--(1.594,3.128);
\draw[gp path] (7.829,3.128)--(7.739,3.128);
\draw[gp path] (1.504,3.301)--(1.594,3.301);
\draw[gp path] (7.829,3.301)--(7.739,3.301);
\draw[gp path] (1.504,3.425)--(1.594,3.425);
\draw[gp path] (7.829,3.425)--(7.739,3.425);
\draw[gp path] (1.504,3.520)--(1.594,3.520);
\draw[gp path] (7.829,3.520)--(7.739,3.520);
\draw[gp path] (1.504,3.598)--(1.594,3.598);
\draw[gp path] (7.829,3.598)--(7.739,3.598);
\draw[gp path] (1.504,3.664)--(1.594,3.664);
\draw[gp path] (7.829,3.664)--(7.739,3.664);
\draw[gp path] (1.504,3.721)--(1.594,3.721);
\draw[gp path] (7.829,3.721)--(7.739,3.721);
\draw[gp path] (1.504,3.772)--(1.594,3.772);
\draw[gp path] (7.829,3.772)--(7.739,3.772);
\draw[gp path] (1.504,3.817)--(1.684,3.817);
\draw[gp path] (7.829,3.817)--(7.649,3.817);
\node[gp node right] at (1.320,3.817) { 10};
\draw[gp path] (1.504,4.114)--(1.594,4.114);
\draw[gp path] (7.829,4.114)--(7.739,4.114);
\draw[gp path] (1.504,4.287)--(1.594,4.287);
\draw[gp path] (7.829,4.287)--(7.739,4.287);
\draw[gp path] (1.504,4.411)--(1.594,4.411);
\draw[gp path] (7.829,4.411)--(7.739,4.411);
\draw[gp path] (1.504,4.506)--(1.594,4.506);
\draw[gp path] (7.829,4.506)--(7.739,4.506);
\draw[gp path] (1.504,4.584)--(1.594,4.584);
\draw[gp path] (7.829,4.584)--(7.739,4.584);
\draw[gp path] (1.504,4.650)--(1.594,4.650);
\draw[gp path] (7.829,4.650)--(7.739,4.650);
\draw[gp path] (1.504,4.707)--(1.594,4.707);
\draw[gp path] (7.829,4.707)--(7.739,4.707);
\draw[gp path] (1.504,4.758)--(1.594,4.758);
\draw[gp path] (7.829,4.758)--(7.739,4.758);
\draw[gp path] (1.504,4.803)--(1.684,4.803);
\draw[gp path] (7.829,4.803)--(7.649,4.803);
\node[gp node right] at (1.320,4.803) { 100};
\draw[gp path] (2.769,1.845)--(2.769,2.025);
\draw[gp path] (2.769,4.803)--(2.769,4.623);
\node[gp node left,rotate=-40] at (2.769,1.661) {Friendster-8};
\draw[gp path] (4.034,1.845)--(4.034,2.025);
\draw[gp path] (4.034,4.803)--(4.034,4.623);
\node[gp node left,rotate=-40] at (4.034,1.661) {Friendster-32};
\draw[gp path] (5.299,1.845)--(5.299,2.025);
\draw[gp path] (5.299,4.803)--(5.299,4.623);
\node[gp node left,rotate=-40] at (5.299,1.661) {RM$_{856}$};
\draw[gp path] (6.564,1.845)--(6.564,2.025);
\draw[gp path] (6.564,4.803)--(6.564,4.623);
\node[gp node left,rotate=-40] at (6.564,1.661) {RU$_{1B}$};
\draw[gp path] (1.504,4.803)--(1.504,1.845)--(7.829,1.845)--(7.829,4.803)--cycle;
\node[gp node center,rotate=-270] at (0.246,3.324) {Log Time/iter (sec)};
\node[gp node right] at (3.382,5.762) {\textsf{knors}};
\def\gpfillpath{(3.566,5.685)--(4.482,5.685)--(4.482,5.839)--(3.566,5.839)--cycle}
\gpfill{color=gpbgfillcolor} \gpfillpath;
\gpfill{rgb color={0.373,0.620,0.627},gp pattern 4,pattern color=.} \gpfillpath;
\gpcolor{rgb color={0.373,0.620,0.627}}
\gpsetlinetype{gp lt plot 0}
\draw[gp path] (3.566,5.685)--(4.482,5.685)--(4.482,5.839)--(3.566,5.839)--cycle;
\def\gpfillpath{(2.453,1.845)--(2.665,1.845)--(2.665,2.654)--(2.453,2.654)--cycle}
\gpfill{color=gpbgfillcolor} \gpfillpath;
\gpfill{rgb color={0.373,0.620,0.627},gp pattern 4,pattern color=.} \gpfillpath;
\draw[gp path] (2.453,1.845)--(2.453,2.653)--(2.664,2.653)--(2.664,1.845)--cycle;
\def\gpfillpath{(3.718,1.845)--(3.930,1.845)--(3.930,3.065)--(3.718,3.065)--cycle}
\gpfill{color=gpbgfillcolor} \gpfillpath;
\gpfill{rgb color={0.373,0.620,0.627},gp pattern 4,pattern color=.} \gpfillpath;
\draw[gp path] (3.718,1.845)--(3.718,3.064)--(3.929,3.064)--(3.929,1.845)--cycle;
\def\gpfillpath{(4.983,1.845)--(5.195,1.845)--(5.195,4.085)--(4.983,4.085)--cycle}
\gpfill{color=gpbgfillcolor} \gpfillpath;
\gpfill{rgb color={0.373,0.620,0.627},gp pattern 4,pattern color=.} \gpfillpath;
\draw[gp path] (4.983,1.845)--(4.983,4.084)--(5.194,4.084)--(5.194,1.845)--cycle;
\def\gpfillpath{(6.248,1.845)--(6.460,1.845)--(6.460,4.487)--(6.248,4.487)--cycle}
\gpfill{color=gpbgfillcolor} \gpfillpath;
\gpfill{rgb color={0.373,0.620,0.627},gp pattern 4,pattern color=.} \gpfillpath;
\draw[gp path] (6.248,1.845)--(6.248,4.486)--(6.459,4.486)--(6.459,1.845)--cycle;
\gpcolor{color=gp lt color border}
\node[gp node right] at (3.382,5.454) {MLlib-EC2};
\def\gpfillpath{(3.566,5.377)--(4.482,5.377)--(4.482,5.531)--(3.566,5.531)--cycle}
\gpfill{color=gpbgfillcolor} \gpfillpath;
\gpfill{rgb color={0.294,0.000,0.510},gp pattern 6,pattern color=.} \gpfillpath;
\gpcolor{rgb color={0.294,0.000,0.510}}
\gpsetlinetype{gp lt plot 1}
\draw[gp path] (3.566,5.377)--(4.482,5.377)--(4.482,5.531)--(3.566,5.531)--cycle;
\def\gpfillpath{(2.664,1.845)--(2.875,1.845)--(2.875,3.620)--(2.664,3.620)--cycle}
\gpfill{color=gpbgfillcolor} \gpfillpath;
\gpfill{rgb color={0.294,0.000,0.510},gp pattern 6,pattern color=.} \gpfillpath;
\draw[gp path] (2.664,1.845)--(2.664,3.619)--(2.874,3.619)--(2.874,1.845)--cycle;
\def\gpfillpath{(3.929,1.845)--(4.140,1.845)--(4.140,3.493)--(3.929,3.493)--cycle}
\gpfill{color=gpbgfillcolor} \gpfillpath;
\gpfill{rgb color={0.294,0.000,0.510},gp pattern 6,pattern color=.} \gpfillpath;
\draw[gp path] (3.929,1.845)--(3.929,3.492)--(4.139,3.492)--(4.139,1.845)--cycle;
\def\gpfillpath{(5.194,1.845)--(5.405,1.845)--(5.405,4.628)--(5.194,4.628)--cycle}
\gpfill{color=gpbgfillcolor} \gpfillpath;
\gpfill{rgb color={0.294,0.000,0.510},gp pattern 6,pattern color=.} \gpfillpath;
\draw[gp path] (5.194,1.845)--(5.194,4.627)--(5.404,4.627)--(5.404,1.845)--cycle;
\def\gpfillpath{(6.459,1.845)--(6.670,1.845)--(6.670,4.231)--(6.459,4.231)--cycle}
\gpfill{color=gpbgfillcolor} \gpfillpath;
\gpfill{rgb color={0.294,0.000,0.510},gp pattern 6,pattern color=.} \gpfillpath;
\draw[gp path] (6.459,1.845)--(6.459,4.230)--(6.669,4.230)--(6.669,1.845)--cycle;
\gpcolor{color=gp lt color border}
\node[gp node right] at (6.322,5.762) {\textsf{knord}};
\gpfill{rgb color={0.000,1.000,0.000},color=.!50} (6.506,5.685)--(7.422,5.685)--(7.422,5.839)--(6.506,5.839)--cycle;
\gpfill{rgb color={0.000,1.000,0.000},color=.!50} (2.874,1.845)--(3.086,1.845)--(3.086,2.349)--(2.874,2.349)--cycle;
\gpfill{rgb color={0.000,1.000,0.000},color=.!50} (4.139,1.845)--(4.351,1.845)--(4.351,2.572)--(4.139,2.572)--cycle;
\gpfill{rgb color={0.000,1.000,0.000},color=.!50} (5.404,1.845)--(5.616,1.845)--(5.616,3.750)--(5.404,3.750)--cycle;
\gpfill{rgb color={0.000,1.000,0.000},color=.!50} (6.669,1.845)--(6.881,1.845)--(6.881,2.924)--(6.669,2.924)--cycle;
\node[gp node right] at (6.322,5.454) {MPI};
\gpfill{rgb color={1.000,0.412,0.706},color=.!50} (6.506,5.377)--(7.422,5.377)--(7.422,5.531)--(6.506,5.531)--cycle;
\gpfill{rgb color={1.000,0.412,0.706},color=.!50} (3.085,1.845)--(3.297,1.845)--(3.297,2.685)--(3.085,2.685)--cycle;
\gpfill{rgb color={1.000,0.412,0.706},color=.!50} (4.350,1.845)--(4.562,1.845)--(4.562,2.603)--(4.350,2.603)--cycle;
\gpfill{rgb color={1.000,0.412,0.706},color=.!50} (5.615,1.845)--(5.827,1.845)--(5.827,3.945)--(5.615,3.945)--cycle;
\gpfill{rgb color={1.000,0.412,0.706},color=.!50} (6.880,1.845)--(7.092,1.845)--(7.092,3.044)--(6.880,3.044)--cycle;
\gpsetlinetype{gp lt border}
\draw[gp path] (1.504,4.803)--(1.504,1.845)--(7.829,1.845)--(7.829,4.803)--cycle;
\gpdefrectangularnode{gp plot 1}{\pgfpoint{1.504cm}{1.845cm}}{\pgfpoint{7.829cm}{4.803cm}}
\end{tikzpicture}

%% file: future_discuss.tex
\section{Future Work and Discussion}

We intend to expand the distributed aspects of \textsf{knor} into a
generalized programming framework for distributed machine learning on NUMA
machines.
We observe that NUMA architectures are prevalent today in cloud
computing. A large contingency of
machines available through Unix-based cloud service providers such as Amazon
EC$2$, IBM Cloud \cite{ibmcloud} and Google Cloud Platform \cite{gcp} are
indeed NUMA systems.

We aim to build a suite of machine learning algorithms on top of
the generalized derivative framework. We aim to demonstrate that
the NUMA optimizations we deploy for \textsf{knor} are applicable to a variety of
compute-intensive applications. The initial phase will target
other variants of k-means like spherical k-means \cite{skmeans}, semi-supervised k-means++
\cite{sskmeanspp} etc. Later phases will target machine learning algorithms like
GMM \cite{gmm}, agglomerative clustering \cite{aggclust} and
k-nearest neighbors \cite{knn}. Finally, we aim to provide a C++ interface upon
which users may implement custom algorithms and benefit from our NUMA and
external memory optimizations.

We are an open source project available at \\
\href{https://github.com/flashxio/knor}{https://github.com/flashxio/knor}

%% file: conclusion.tex
\section{Conclusion}

We accelerate k-means by over an order of
magnitude by rethinking Lloyd's algorithm for modern multiprocessor
NUMA architectures through the minimization of critical regions.
We demonstrate that our modifications to Lloyd's are relevant to
both in-memory (\textsf{knori}), distributed memory (\textsf{knord})
and semi-external memory (\textsf{knors}) applications as we outperform
state-of-the-art frameworks running the exact same algorithms.

We formulate a minimal triangle inequality pruning technique (MTI)
that further boosts the performance of k-means on real-world billion point datasets
by over $100$x when compared to some popular frameworks. MTI does so without
significantly increasing memory consumption.

The addition of a row caching layer yields performance improvements
over vanilla SEM implementations. We demonstrate that k-means in SEM
performs only a small constant factor slower than in-memory algorithms for large scale datasets
and scales beyond the limits of memory at which point in-memory algorithms fail.

Finally, we demonstrate that there are large performance benefits
associated with NUMA-targeted optimizations. We show
that data locality optimizations such as NUMA-node thread binding,
NUMA-aware task scheduling, and NUMA-aware memory allocation schemes
provide several times speedup for k-means on NUMA hardware.

%% file: hpdc17.bbl
\begin{thebibliography}{10}

\bibitem{Abello98}
J.~Abello, A.~L. Buchsbaum, and J.~R. Westbrook.
\newblock A functional approach to external graph algorithms.
\newblock In {\em Algorithmica}, pages 332--343. Springer-Verlag, 1998.

\bibitem{Ang14}
J.~Ang, R.~F. Barrett, R.~Benner, D.~Burke, C.~Chan, J.~Cook, D.~Donofrio,
  S.~D. Hammond, K.~S. Hemmert, S.~Kelly, et~al.
\newblock Abstract machine models and proxy architectures for exascale
  computing.
\newblock In {\em Hardware-Software Co-Design for High Performance Computing
  (Co-HPC), 2014}, pages 25--32. IEEE, 2014.

\bibitem{netflixprize}
J.~Bennett and S.~Lanning.
\newblock The netflix prize.
\newblock In {\em Proceedings of KDD cup and workshop}, volume 2007, page~35,
  2007.

\bibitem{conn1}
N.~Binkiewicz, J.~T. Vogelstein, and K.~Rohe.
\newblock Covariate assisted spectral clustering.
\newblock {\em arXiv preprint arXiv:1411.2158}, 2014.

\bibitem{mlpack}
R.~R. Curtin, J.~R. Cline, N.~P. Slagle, W.~B. March, P.~Ram, N.~A. Mehta, and
  A.~G. Gray.
\newblock Mlpack: A scalable c++ machine learning library.
\newblock {\em Journal of Machine Learning Research}, 14(Mar):801--805, 2013.

\bibitem{cfgoogle}
A.~S. Das, M.~Datar, A.~Garg, and S.~Rajaram.
\newblock Google news personalization: scalable online collaborative filtering.
\newblock In {\em Proceedings of the 16th international conference on World
  Wide Web}, pages 271--280. ACM, 2007.

\bibitem{mapreduce}
J.~Dean and S.~Ghemawat.
\newblock {MapReduce}: Simplified data processing on large clusters.
\newblock In {\em Proceedings of the 6th Conference on Symposium on Opearting
  Systems Design \& Implementation - Volume 6}, 2004.

\bibitem{EM}
A.~P. Dempster, N.~M. Laird, and D.~B. Rubin.
\newblock Maximum likelihood from incomplete data via the em algorithm.
\newblock {\em Journal of the royal statistical society. Series B
  (methodological)}, pages 1--38, 1977.

\bibitem{ding2015yinyang}
Y.~Ding, Y.~Zhao, X.~Shen, M.~Musuvathi, and T.~Mytkowicz.
\newblock Yinyang k-means: A drop-in replacement of the classic k-means with
  consistent speedup.
\newblock In {\em Proceedings of the 32nd International Conference on Machine
  Learning (ICML-15)}, pages 579--587, 2015.

\bibitem{knn}
R.~O. Duda, P.~E. Hart, et~al.
\newblock {\em Pattern classification and scene analysis}, volume~3.
\newblock Wiley New York, 1973.

\bibitem{triineq}
C.~Elkan.
\newblock Using the triangle inequality to accelerate k-means.
\newblock In {\em ICML}, volume~3, pages 147--153, 2003.

\bibitem{MPI}
M.~P. Forum.
\newblock Mpi: A message-passing interface standard.
\newblock Technical report, Knoxville, TN, USA, 1994.

\bibitem{friendster}
Frienster graph.
\newblock \url{https://archive.org/download/friendster-dataset-201107},
  Accessed 4/18/2014.

\bibitem{cache-obl}
M.~Frigo, C.~E. Leiserson, H.~Prokop, and S.~Ramachandran.
\newblock Cache-oblivious algorithms.
\newblock In {\em Foundations of Computer Science, 1999. 40th Annual Symposium
  on}, pages 285--297. IEEE, 1999.

\bibitem{gmm}
C.~F. Gauss.
\newblock {\em Theory of the motion of the heavenly bodies moving about the sun
  in conic sections: a translation of Carl Frdr. Gauss" Theoria motus": With an
  appendix. By Ch. H. Davis}.
\newblock Little, Brown and Comp., 1857.

\bibitem{h2o}
h2o.
\newblock h2o.
\newblock \url{http://h2o.ai/}, 2005--2015.

\bibitem{skmeans}
K.~Hornik, I.~Feinerer, M.~Kober, and C.~Buchta.
\newblock Spherical k-means clustering.
\newblock {\em Journal of Statistical Software}, 50(10):1--22, 2012.

\bibitem{aws}
A.~Inc.
\newblock Amazon web services.

\bibitem{gcp}
G.~Inc.
\newblock Google cloud platform.

\bibitem{genetic1}
L.~B. Jorde and S.~P. Wooding.
\newblock Genetic variation, classification and'race'.
\newblock {\em Nature genetics}, 36:S28--S33, 2004.

\bibitem{BLAS}
C.~L. Lawson, R.~J. Hanson, D.~R. Kincaid, and F.~T. Krogh.
\newblock Basic linear algebra subprograms for fortran usage.
\newblock {\em ACM Transactions on Mathematical Software (TOMS)},
  5(3):308--323, 1979.

\bibitem{lloyds}
S.~P. Lloyd.
\newblock Least squares quantization in pcm.
\newblock {\em Information Theory, IEEE Transactions on}, 28(2):129--137, 1982.

\bibitem{graphlab}
Y.~Low, J.~E. Gonzalez, A.~Kyrola, D.~Bickson, C.~E. Guestrin, and
  J.~Hellerstein.
\newblock Graphlab: A new framework for parallel machine learning.
\newblock {\em arXiv preprint arXiv:1408.2041}, 2014.

\bibitem{conn2}
V.~Lyzinski, D.~L. Sussman, D.~E. Fishkind, H.~Pao, L.~Chen, J.~T. Vogelstein,
  Y.~Park, and C.~E. Priebe.
\newblock Spectral clustering for divide-and-conquer graph matching.
\newblock {\em Parallel Computing}, 2015.

\bibitem{conn3}
V.~Lyzinski, M.~Tang, A.~Athreya, Y.~Park, and C.~E. Priebe.
\newblock Community detection and classification in hierarchical stochastic
  blockmodels.
\newblock {\em arXiv preprint arXiv:1503.02115}, 2015.

\bibitem{matlab}
MATLAB.
\newblock {\em version 7.10.0 (R2010a)}.
\newblock The MathWorks Inc., Natick, Massachusetts, 2010.

\bibitem{McSherry15}
F.~McSherry, M.~Isard, and D.~G. Murray.
\newblock Scalability! but at what cost?
\newblock In {\em 15th Workshop on Hot Topics in Operating Systems (HotOS XV)},
  2015.

\bibitem{mllib}
X.~Meng, J.~Bradley, B.~Yavuz, E.~Sparks, S.~Venkataraman, D.~Liu, J.~Freeman,
  D.~Tsai, M.~Amde, S.~Owen, et~al.
\newblock Mllib: Machine learning in apache spark.
\newblock {\em arXiv preprint arXiv:1505.06807}, 2015.

\bibitem{mahout}
S.~Owen, R.~Anil, T.~Dunning, and E.~Friedman.
\newblock {\em Mahout in action}.
\newblock Manning Shelter Island, 2011.

\bibitem{genetic2}
N.~Patterson, A.~L. Price, and D.~Reich.
\newblock Population structure and eigenanalysis.
\newblock 2006.

\bibitem{Pearce10}
R.~Pearce, M.~Gokhale, and N.~M. Amato.
\newblock Multithreaded asynchronous graph traversal for in-memory and
  semi-external memory.
\newblock In {\em Proceedings of the 2010 ACM/IEEE International Conference for
  High Performance Computing, Networking, Storage and Analysis}, 2010.

\bibitem{sklearn}
F.~Pedregosa, G.~Varoquaux, A.~Gramfort, V.~Michel, B.~Thirion, O.~Grisel,
  M.~Blondel, P.~Prettenhofer, R.~Weiss, V.~Dubourg, J.~Vanderplas, A.~Passos,
  D.~Cournapeau, M.~Brucher, M.~Perrot, and E.~Duchesnay.
\newblock Scikit-learn: Machine learning in {P}ython.
\newblock {\em Journal of Machine Learning Research}, 12:2825--2830, 2011.

\bibitem{rpackage}
{R Core Team}.
\newblock {\em R: A Language and Environment for Statistical Computing}.
\newblock R Foundation for Statistical Computing, Vienna, Austria, 2015.

\bibitem{aggclust}
L.~Rokach and O.~Maimon.
\newblock Clustering methods.
\newblock In {\em Data mining and knowledge discovery handbook}, pages
  321--352. Springer, 2005.

\bibitem{McSherryBlog}
Scalability! but at what cost?
\newblock
  \url{http://www.frankmcsherry.org/graph/scalability/cost/2015/01/15/COST.html},
  Accessed 9/3/2016.

\bibitem{sculley2010web-scale}
D.~Sculley.
\newblock Web-scale k-means clustering.
\newblock In {\em ACM Digital library}, pages 1177--1178, 2010.

\bibitem{shindler2011fast}
M.~Shindler, A.~Wong, and A.~W. Meyerson.
\newblock Fast and accurate k-means for large datasets.
\newblock In {\em Advances in neural information processing systems}, pages
  2375--2383, 2011.

\bibitem{facebookanatomy}
J.~Ugander, B.~Karrer, L.~Backstrom, and C.~Marlow.
\newblock The anatomy of the facebook social graph.
\newblock {\em arXiv preprint arXiv:1111.4503}, 2011.

\bibitem{behavioromics}
J.~T. Vogelstein, Y.~Park, T.~Ohyama, R.~A. Kerr, J.~W. Truman, C.~E. Priebe,
  and M.~Zlatic.
\newblock Discovery of brainwide neural-behavioral maps via multiscale
  unsupervised structure learning.
\newblock {\em Science}, 344(6182):386--392, 2014.

\bibitem{sskmeanspp}
J.~Yoder and C.~E. Priebe.
\newblock Semi-supervised k-means++.
\newblock {\em arXiv preprint arXiv:1602.00360}, 2016.

\bibitem{rdd}
M.~Zaharia, M.~Chowdhury, T.~Das, A.~Dave, J.~Ma, M.~McCauley, M.~J. Franklin,
  S.~Shenker, and I.~Stoica.
\newblock Resilient distributed datasets: A fault-tolerant abstraction for
  in-memory cluster computing.
\newblock In {\em Proceedings of the 9th USENIX conference on Networked Systems
  Design and Implementation}, pages 2--2. USENIX Association, 2012.

\bibitem{spark}
M.~Zaharia, M.~Chowdhury, M.~J. Franklin, S.~Shenker, and I.~Stoica.
\newblock Spark: Cluster computing with working sets.
\newblock {\em HotCloud}, 10:10--10, 2010.

\bibitem{zhao2009parallel}
W.~Zhao, H.~Ma, and Q.~He.
\newblock Parallel k-means clustering based on mapreduce.
\newblock In {\em Cloud Computing}, pages 674--679. Springer, 2009.

\bibitem{safs}
D.~Zheng, R.~Burns, and A.~S. Szalay.
\newblock Toward millions of file system {IOPS} on low-cost, commodity
  hardware.
\newblock In {\em Proceedings of the International Conference on High
  Performance Computing, Networking, Storage and Analysis}, 2013.

\bibitem{flashgraph}
D.~Zheng, D.~Mhembere, R.~Burns, J.~Vogelstein, C.~E. Priebe, and A.~S. Szalay.
\newblock {FlashGraph}: Processing billion-node graphs on an array of commodity
  {SSDs}.
\newblock In {\em 13th USENIX Conference on File and Storage Technologies (FAST
  15)}, 2015.

\bibitem{ibmcloud}
J.~Zhu, X.~Fang, Z.~Guo, M.~H. Niu, F.~Cao, S.~Yue, and Q.~Y. Liu.
\newblock Ibm cloud computing powering a smarter planet.
\newblock In {\em IEEE International Conference on Cloud Computing}, pages
  621--625. Springer, 2009.

\end{thebibliography}
